\documentclass[lineno]{jfm}
\usepackage{graphicx}
\usepackage{newtxtext}
\usepackage{newtxmath}
\usepackage{natbib}
\usepackage{orcidlink}
\usepackage{hyperref}
\hypersetup{
	colorlinks = true,
	urlcolor   = blue,
	citecolor  = blue,
	linkcolor  = blue,
	filecolor  = blue,
}

\newcommand{\RomanNumeralCaps}[1]
\linenumbers

\usepackage{float}
\usepackage{caption}
\usepackage{subcaption}
\usepackage{amssymb}
\usepackage{booktabs}
\usepackage{multirow}
\usepackage[font=normalsize,justification=justified]{caption}
\usepackage{soul}
\title{Comparison of moving and stationary clapping bodies}

\author{Suyog V Mahulkar\aff{1}
	\corresp{\email{suyogm@iisc.ac.in}},
	Jaywant H Arakeri{\aff{1,}\aff{2,}\aff{3}}}

\affiliation{\aff{1} Department of Mechanical Engineering, Indian Institute of Science, Bangalore, 560012, India \aff{2} Engineering Mechanics Unit, Jawaharlal Nehru Centre For Advanced Scientific Research, Bangalore, 560064, India
	\aff{3} Department of Mechanical Engineering, Indian Institute of Technology, Jodhpur, 342030, India}

\begin{document}
\maketitle
\begin{abstract}
	We report differences in the flow dynamics and in the body kinematics between a clapping body that is allowed to propel forward freely (Dynamic) and one that is constrained from moving forward (Stationary). The experiments were done in quiescent water. The body consists of two interconnected plates hinged at one end with a ‘torsion’ spring. Initially, a thread loop holds the plates apart at an interplate angle of 60 degrees. Cutting of the thread initiates the clapping motion, and if allowed, the body propels forward a certain distance. Experiments have been performed for three values of $d^*$(= depth/length): 1.5, 1.0, and 0.5. In both cases, vortex loops initially develop along the three edges of each plate, which reconnect by the end of the clapping motion resulting in the formation of an elliptical vortex loop in the wake for $d^*$ = 1.5, 1.0 bodies, and multiple connected rings for $d^*$ = 0.5 bodies. Three main and unexpected differences are observed between the ‘Dynamic’ and ‘Stationary’ bodies. In the dynamic case, the clapping action is faster compared to the stationary case, with the maximum angular plate velocity being twice as high. The mean thrust coefficient, $\overline{C_T}$, based on plate tip velocity, is higher for the stationary. The value of  $\overline{C_T}$ and the circulation in the starting vortices is almost independent of $d^*$ for the dynamic case, whereas it increases with $d^*$ for the stationary case. The core separation of starting vortices closely matches in both stationary and dynamic cases although circulation varies. 
\end{abstract}

\begin{section}{Introduction}
	One class of marine animals propel themselves by generating pulsed jets, for example, jellyfish and squids. A squid contracts its body cavity to produce the jet. Two types of jetting patterns were observed by Bartol et al.\cite{Bartol2D09} in the wakes of the squid species, {\it Lolliguncula brevis}, first one is a short pulse jet consisting of isolated rings, and the second is a long pulse jet where the vortex ring is followed by trailing jets. Bartol et al.\cite{Bartol3D16} further analyzed the 3-D flowfield developed in the complex interaction between the animal fins and ejected pulse jets. Similarlly, jellyfish generate thrust by contracting their subumbrellar cavity and a starting vortex forms in the wake that is associated with the thrust production phase (Dabiri et al.\cite{Dabiri05}, \cite{Dabiri06}). Swimming performances of a few other aquatic animals that use pulse jet propulsion is discussed in the review by Gemmell et al.\cite{Gemmell21}. In recent years, some aquatic robots have been developed that use pulsed jet propulsion (J. Nichols et al.\cite{robosquid10}, M. Krieg et al.\cite{Krieg08}), and T. Bujard et al.\cite{Bujard21}). The clapping motion in some insects also produces pulse jets. It may be observed in butterflies where the clapping phase of the wing motion produces a transient jet (Brodsky\cite{Brodsky91}, L. C. Johansson et al.\cite{Johanassan21}).\par
	
	Motivated by the need to understand clapping propulsion, Kim et al.\cite{Kim13} studied  the flow dynamics associated with two clapping plates submerged in quiescent water. They reported that the impulse generated during the clapping motion and circulation of starting vortices available in the wake increases with the height of the plate, with the length kept constant. In a later study (Martin et al.\cite{Martin17}), it was reported that the clapping mode of pulse jet propulsion produces high thrust compared to the flapping type of propulsion observed in fishes with undulating tail, although the latter mode of propulsion is more efficient. Das et al. (\cite{Das13} and \cite{Das18}) studied the effect of flexible flaps, often found in marine animals, at the exit of a 2-D channel through which a transient jet was produced. In these laboratory studies, the ‘body’ was constrained from moving forward. The natural question is, what are the differences between the flows and forces produced when the body is free to move due to the clapping action, and when it is constrained from translating? In a companion paper (Mahulkar and Arakeri\cite{Mahulkar23}), we report the body and flow dynamics of a simple clapping body that is allowed to move forward freely. In the present work, we report the differences that arise when the same clapping body is constrained from moving forward and when it is allowed to propel forward. \par
	
	The body consists of two rigid plates hinged together at the front with a ‘torsion’ spring see (figures \hyperref[fig:ProbSchematic_StatDyn]{1a, c}). Initially, the plates are pulled apart by a thin threaded loop. There is an effective initial torque, $T_o$, on the  plates. Clapping motion of the plates is initiated by cutting of the thread resulting in a transient jet of fluid ejected from the interplate cavity. In the forward motion-constrained case, the body was held at the hinge point, and only the plates were allowed to come together due to the torsion spring action (figure \hyperref[fig:ProbSchematic_StatDyn]{1b}). In the free-to-translate case, the body was propelled forward under the clapping action of the plates (figure \hyperref[fig:ProbSchematic_StatDyn]{1d}). In the latter case, we had to ensure that the body was neutrally buoyant and the centers of mass and buoyancy coincided to obtain linear horizontal motion.\par
		
	The details of clapping body construction and analysis methods is provided in \S \hyperref[sec:ExptSetup_StatDyn]{2}. The comparative analysis of body kinematics is discussed in \S \hyperref[sec:Kinematics_StatDyn]{3.1} and flow characteristics in \S \hyperref[sec:WakeDynamics_StatDyn]{3.2}. A discussion on the differences in wake momentum and circulation is provided in \S \hyperref[sec:WakeMom_Gamma_StatDyn]{3.2.4} followed by the concluding remarks in \S \hyperref[sec:conclusion_StatDyn]{4}.
\begin{figure}
	\centering\
	\begin{subfigure}[b]{0.35\textwidth}
		\includegraphics[width=\textwidth]{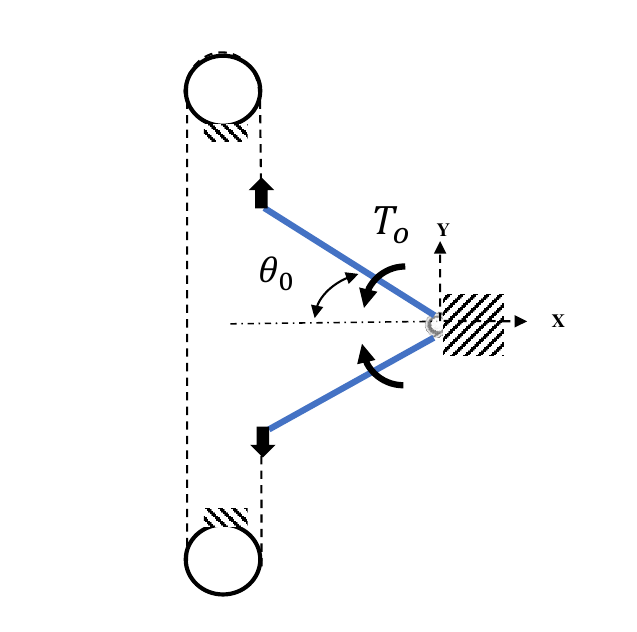}
		\caption{}
		\label{fig:Stat_to}
	\end{subfigure}\hspace{10mm}
	\begin{subfigure}[b]{0.35\textwidth}
		\includegraphics[width=\textwidth]{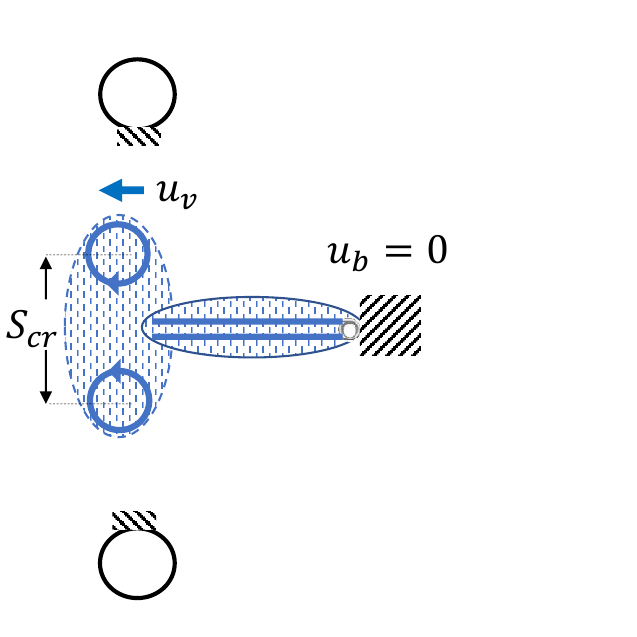}
		\caption{}
		\label{fig:Stat_tf}
	\end{subfigure}\hspace{10mm}	
	
	\begin{subfigure}[b]{0.35\textwidth}
		\includegraphics[width=\textwidth]{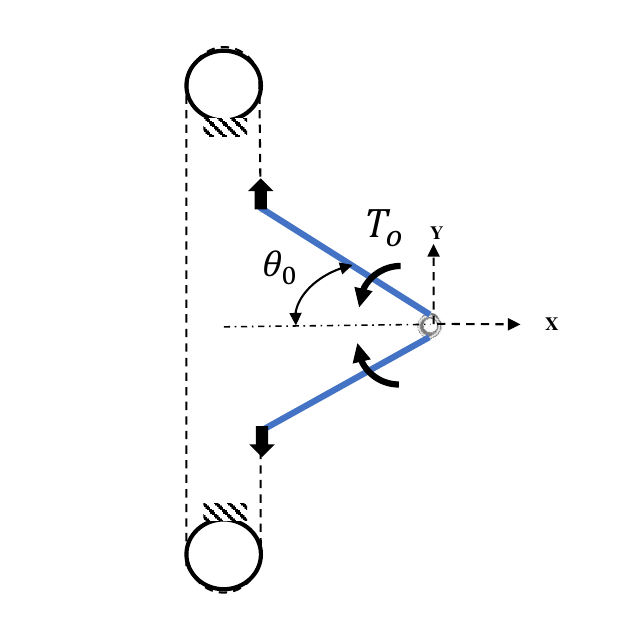}
		\caption{}
		\label{fig:Dyn_to}
	\end{subfigure}\hspace{10mm}		
	\begin{subfigure}[b]{0.35\textwidth}
		\includegraphics[width=\textwidth]{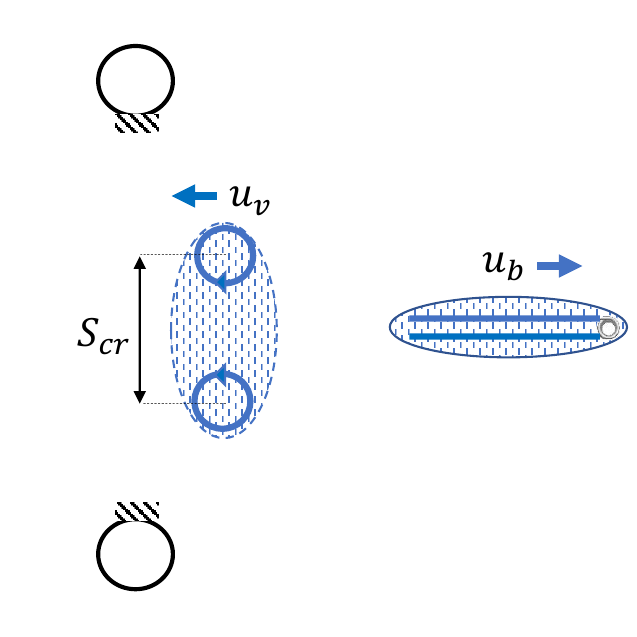}
		\caption{}
		\label{fig:Dyn_tf}
	\end{subfigure}\hspace{10mm}	
	\caption{(a) Schematic of the clapping body for the stationary case. It consists of two rigid plates with an initial interplate angle 2$\theta_o$. An initial torque $T_o$ is applied to keep the  plates apart using a thread loop mounted on a release stand consisting of two circular rods. The lab reference-based origin of the XY coordinate system is located at the hinge point; the mid-depth of the body lies on the Z = 0 plane. The body is clamped at the hinge point to prevent  its forward motion. (b) Cutting of the thread initiates clapping and generation of a starting vortex with core separation $S_{cr}$. (c) For the dynamic case, the initial configuration is same as for the stationary case, but the body is allowed to freely propel forward. (d) Clapping motion caused cutting of the thread results in the body translating to the right with velocity, $u_b$, and the starting vortex moving to the left in the -ve X direction with velocity, $u_v$. The blue dashed lines show fluid moving with a clapping body and with starting vortices.}\label{fig:ProbSchematic_StatDyn}
\end{figure}

\end{section}
%%%%%%%%%%%%%%%%%%%%%%%%%%%%%%%%%%%%%%%%%%%%%%%%%%%%%%%%%%%%%%%%%%%%%%%%%%%%%%%%%%%%%%%%
\begin{section}{Experimental apparatus}
\label{sec:ExptSetup_StatDyn}

	The clapping body consists of two plates (figure \hyperref[fig:Fabricated clapping body_StatDyn]{2a}) held together by a ‘torsion’ spring. The torsion spring action comes from two thin steel plates. For the self-propelling body, it was necessary that the body must have a neutrally buoyant configuration and that it should be configured to move in a straight line in the horizontal (X) direction. This was ensured by using a combination of materials with different specific gravity (SG) values:  steel (SG 8.09 gm/cc), plastic (SG 0.89 gm/cc), balsa wood (SG 0.22 gm/cc), and Bond-Tite adhesive (SG 1.05 gm/cc). The coincidence of the center of mass COM and center of buoyancy COB removes the unbalanced torque, ensuring rectilinear translation. Each plate (figure \hyperref[fig:Fabricated clapping body_StatDyn]{2a}) has an aero-foil shape balsa  at the leading edge (LE), followed by steel plate of length, $L_e$ = 45 mm, and thickness 0.14 mm, and at the trailing edge, a rigid plastic sheet of length, $L_{Plastic}$ = 30 mm, and thickness 0.7 mm. On the outside of the plastic sheet, a second piece of balsa is attached (figure \hyperref[fig:Fabricated clapping body_StatDyn]{2b, c}). The two steel plates are glued together at the leading edge using ‘Bond-Tite’ glue to form the clapping body (see figure \hyperref[fig:Fabricated clapping body_StatDyn]{2b}). Tiny bubbles, entrapped in between the two front aerofoils, disturbed the neutrally buoyant equilibrium condition and were removed using a jet of water produced by a syringe. Small thin steel pieces were added to achieve the neutrally buoyant condition and coincidence of COM and COB. The red circle in the figure \hyperref[fig:Fabricated clapping body_StatDyn]{2c} shows such an additional mass. \par
	
	A fishing thread (Caperlan) of diameter 0.25 mm was looped around two acrylic rods that were used to keep the trailing edges of the two plates apart at an angle 2$\theta_o$ (figure \hyperref[fig:Dyn_to]{1a, c}). Cutting of the thread initiates the clapping action. All the experiments are performed in quiescent water kept in a tank measuring 80 cm $\times$ 80 cm $\times$ 30 cm. The clapping body was held about 15 cm below the water surface. A laparoscopic scissor was used to cut the thread and had the advantage of causing a minimum disturbance in the fluid while cutting. \par
	
	\begin{figure}
		\centering
		\begin{subfigure}[b]{0.7\textwidth}
			\includegraphics[width=\textwidth]{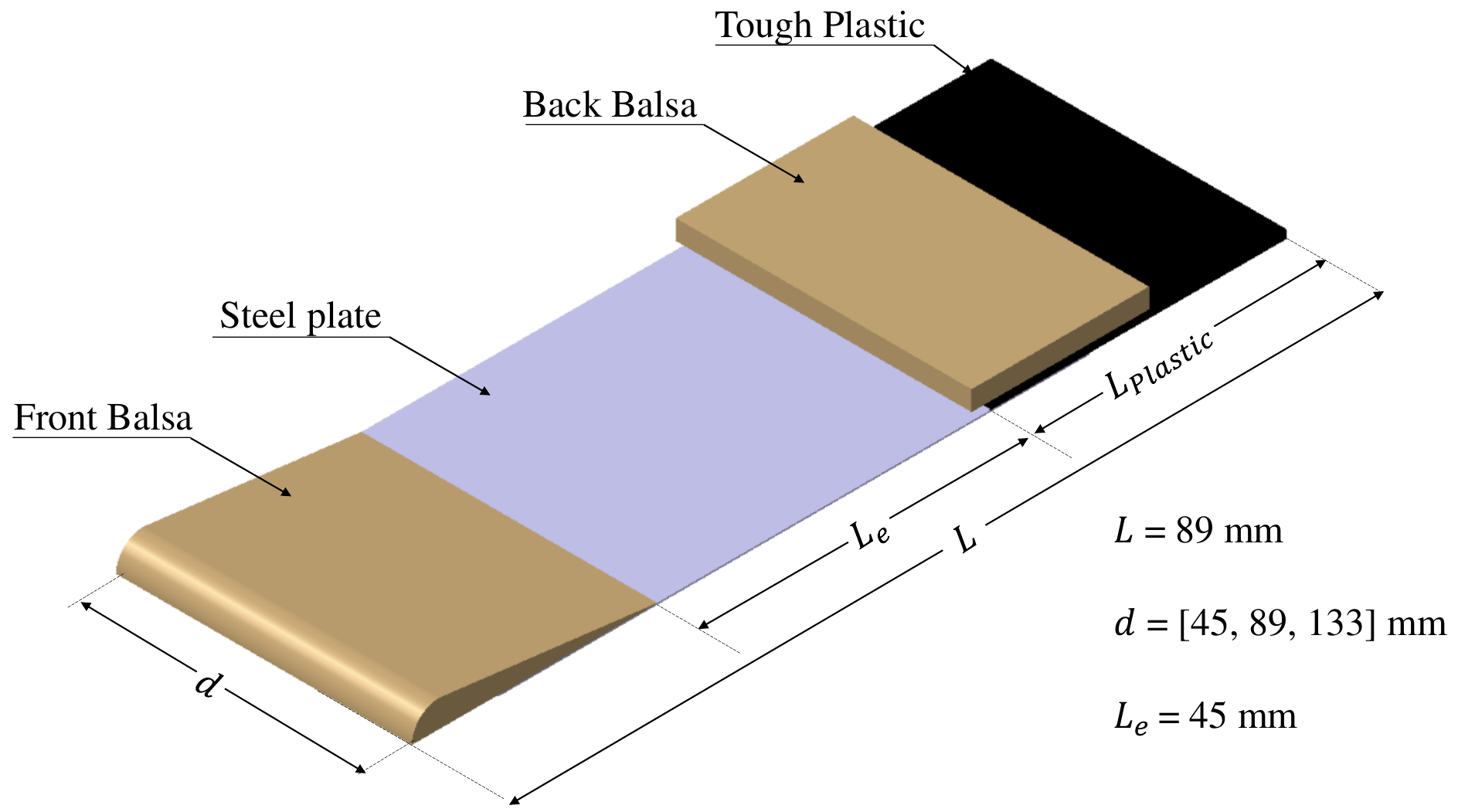}
			\caption{}
			\label{fig:IsoApparatus_StatDyn}
		\end{subfigure}\vspace{4mm}	
		\begin{subfigure}[b]{0.47\textwidth}
			\includegraphics[width=\textwidth]{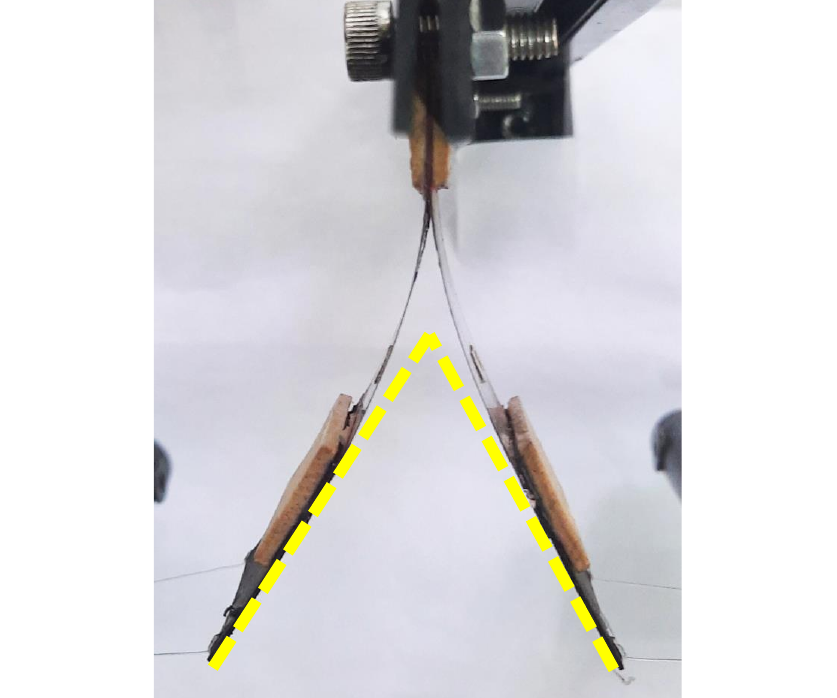}
			\caption{}
			\label{fig:SuperImp_Schematic_StatDyn}
		\end{subfigure}
		\begin{subfigure}[b]{0.5\textwidth}
			\includegraphics[width=\textwidth]{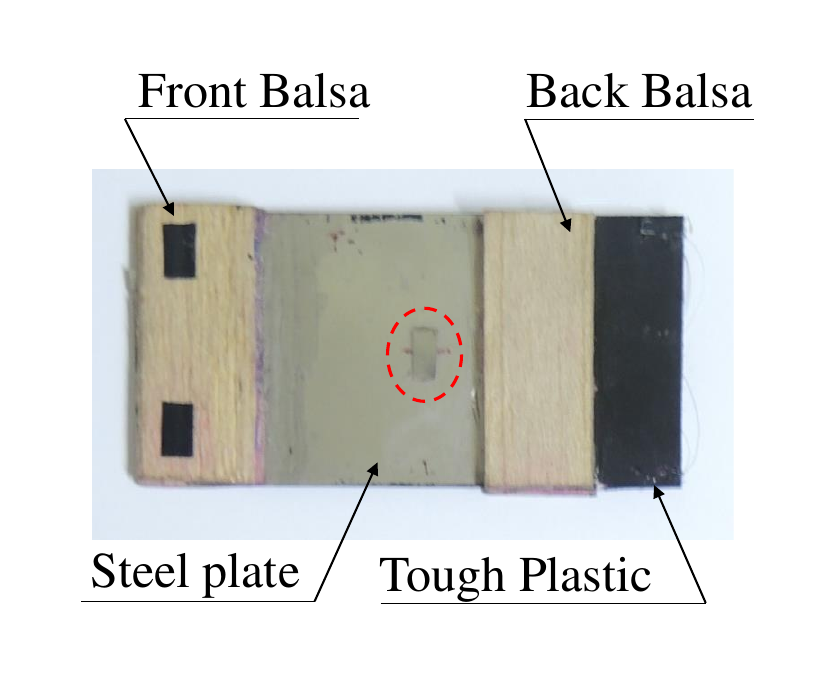}
			\caption{}
			\label{fig:ClappingBody_StatDyn}
		\end{subfigure}

		\caption{(a) An iso-metric view of one of the plates in the clapping body with $d^*=$ 0.5. The body is made by gluing two such plates at the front end. (b) Top view of the initial configuration  of the clapping body held by a clamp at the front end to restrict forward motion. The yellow lines are tangents to the rear portions of the clapping plates meeting at a virtual point. (c) Side view of the clapping body with $d^*=$ 0.5. The small mass used for balancing is encircled by a dashed red line.}\label{fig:Fabricated clapping body_StatDyn}
	\end{figure}

   The total body length $L$ is 89 mm, the maximum body thickness is 6 mm, and the initial interplate angle, $2 \theta_o$, was kept at 60 deg. Experiments were done with three different depth values of the body, $d$, = 45 mm, 89 mm, and 133 mm, giving the non-dimensional depths $d^* (=d/L)$ as 0.5, 1.0, and 1.5. The spring stiffness, $\kappa$, of steel plate of length, $L_e$, is given as the ratio of the strain energy, ($SE$), in the steel plates to the square of net angular deflection. The experimentally measure $\kappa$ values are 114.4 mJ/rad$^2$ for $d^*$ = 1.5, 99.3 mJ/rad$^2$ for $d^*$ = 1.0, and 40.5 mJ/rad$^2$ for $d^*$ = 0.5. Stiffness per unit depth, $Kt$ (= $\kappa /d$), is approximately constant (0.8-1.1\  mJ/mm.rad$^2$). The parametric space consists of cases with the three $d^*$ values for constrained forward motion and self-propelling conditions resulting in six independent experiments, each of which was repeated three times.\par
	
	\begin{figure}
		\centering
		\begin{subfigure}[b]{0.8\textwidth}
			\includegraphics[width=\textwidth]{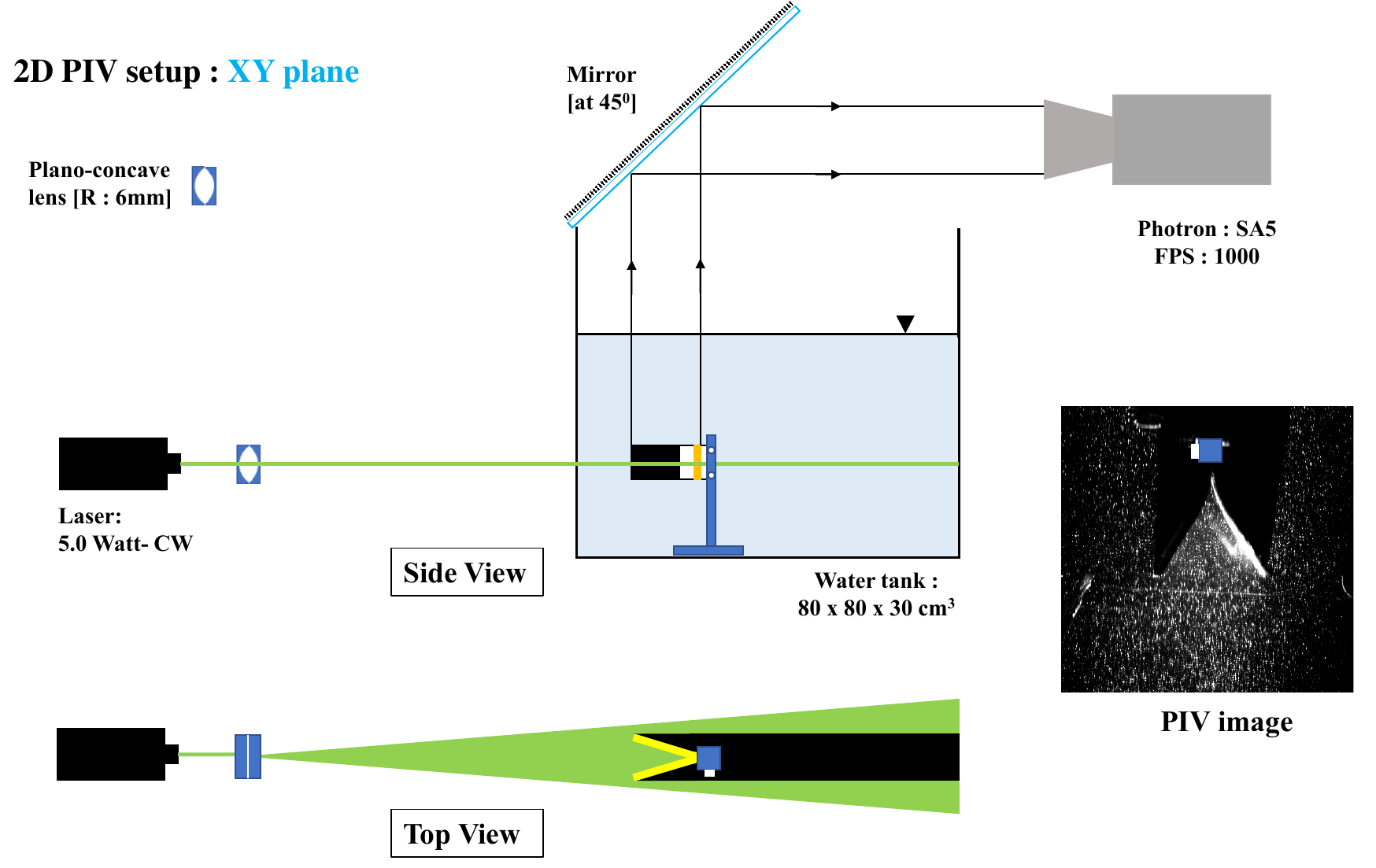}	
			\label{fig:PIVSetup_StatDyn}
		\end{subfigure}	
	\caption{Schematic showing the optical arrangment for PIV and flow visualization. The image captured by a high-speed camera on the right side shows the tracer particles illuminated by a laser light sheet for the stationary body.}
	\end{figure}
	
	The flow fields were measured using 2-D particle image velocimetry (PIV). The PIV system consists of a high-speed camera (Photron SA5), a CW-laser of 5W power and 532 nm wavelength (CNI laser, MGL-N-532-A), and sheet-making optics that created a 2-3 mm thick laser light sheet; the tracer particles were 10-15$\mu$m diameter hollow glass spheres (CONDUCT-O-FIL, Potters Inc). The camera was run at 1000 FPS. The PIV measurements were done in two mutually perpendicular planes: XY (Z=0), termed as a top view, and XZ (Y=0), termed as a side view. An overhead mirror inclined at 45 degrees was used for the XY plane measurements. See figure \hyperref[fig:PIVSetup_StatDyn]{3}. The region of interest (ROI) for top view PIV is 80-100 mm $\times$ 80-100 mm, where 1Pixel $\approx$ 0.3 mm and the interrogation window is 24 $\times$ 24 Pixel$^2$. In the case of side view PIV, the camera can directly see the laser sheet illuminated along the XZ plane. In the side view PIV, ROI is 100-120 mm $\times$ 120-150 mm, where 1 Pixel $\approx$ 0.23 mm, and the interrogation window is the same as in the case of top view PIV. \par
	
	Circulation, $\Gamma$, in the starting vortices present in the wake of clapping bodies is calculated using the following equation: 
	\begin{gather}\label{eq:gamma_stat}
	{\Gamma}=\int{\omega} \ dA_c\ , 
	\end{gather}
	where $\omega$ is the Z-component of vorticity, and integration is done around the vortex over an area where $\omega \geq 0.05\omega_{max}$.  \par
	
	Flow visualization was done using the planar laser-induced fluorescence (PLIF) technique with Rhodamine B dye, which emits light at 625 nm when excited with the light of 532 nm. Based on the dye coating technique given by David et al.\cite{JeemReeves18}, a dye paste is prepared using slow-drying gel (Daler-Rowney), honey, and Rhodamine-B. The dye paste coat was applied on the inside surfaces at the trailing edge. The system components were the same as for the PIV measurements (figure \hyperref[fig:PIVSetup_StatDyn]{3}). 
\end{section}

%%%%%%%%%%%%%%%%%%%%%%%%%%%%%%%%%%%%%%%%%%%%%%%%%%%%%%%%%%%%%%%%%%%%%%%%%%%%%%%%%%%%%%%%
\begin{section}{Results and discussion}
	\label{sec:Results_StatDyn}
	This section presents a detailed analysis of the body kinematics and of the flow fields for the two types of experiments: the self-propelling case, which we term 'Dynamic', and for the case where the body is constrained from translating, which we term as 'Stationary'. The errors estimates are obtained from the average of standard deviations from all the data of the particular variable in the three experiments. \par

	\begin{subsection}{Body kinematics}
	\label{sec:Kinematics_StatDyn}
	Figures \hyperref[fig:SuperImp_Stat_StatDyn]{4a} and \hyperref[fig:SuperImp_Dyn_StatDyn]{4b} show the plate positions at different time instants, 6 ms apart, for the stationary and dynamic cases, respectively. The lines in the figures correspond to the ‘straight’ portions of the plates extrapolated to a point as indicated by yellow dashed lines in figure \hyperref[fig:SuperImp_Schematic_StatDyn]{2b}. For the stationary body, the plates have pure rotational motion, whereas, for the dynamic body, the plates have both rotational and translational motions. An important difference between the dynamic and stationary cases is that the freedom to move forward in the former results in a quicker closing of the two plates (figure \hyperref[fig:AngVelo_Disp_StatDyn]{5b}) with nearly twice the angular velocity (figure \hyperref[fig:AngVelo_AR067_StatDyn]{5a}). The clapping action ends at around 60 - 80 ms for the dynamic cases and around 100 - 120 ms for the stationary cases. The angular velocity, $\dot{\theta}$, in both cases increases rapidly till a maximum, followed by a gradual reduction (figure \hyperref[fig:AngVelo_AR067_StatDyn]{5a}). The velocity curve is a  polynomial fit (5th degree) to the data points obtained from image analysis. The average of standard deviations over time in $\dot{\theta}$ is less than 4\% of the maximum angular velocity and the variation in $\theta_o$ is less than 3 degrees for both stationary and dynamic conditions. The maximum angular velocity in the stationary cases, $\dot{\theta}_{ms}$, range between 5.6 - 7.9 rad/s, and in the dynamic cases, $\dot{\theta}_{md}$, range between 11 - 13.1 rad/s (Table \hyperref[tab:maximaAngular]{1}). The variations in maximum angular velocity across $d^*$ values are partly caused by slight differences in the spring stiffness values resulting from unavoidable variations in the construction details of the bodies.\par
	
	The equation for the rotational motion of each plate in both the stationary and dynamic cases can be written as
	\begin{gather}\label{eq:rot_eqmStat}
	T= (I_{add} + I_{b})\ \ddot{\theta} + T_f(\dot{\theta}, \theta ) \ ,
	\end{gather}
	where $T$ is the applied torque by the spring ($= 2\ \kappa \ \theta$), $I_{add}$ and $I_b$ are the moments inertia due to fluid mass and the plate mass, and $T_f$ is unknown fluid torque that depends on $\dot{\theta}$ and $\theta$. $T_f$ accounts for the torque due to drag forces and wake history forces. In the dynamic clapping cases, there is an additional torque due to the inertial force ($m\dot{u}_b$) acting at the COM of the plate. This contribution is negligible because of the close proximity of the COM ($\approx$ 35 mm from the LE) to the center of rotation ($\approx$ 32 mm from the LE). The center of rotation is a virtual point formed due to the intersection of tangents to both plates, as indicated by yellow dashed lines in figure \hyperref[fig:SuperImp_Schematic_StatDyn]{2b}. In the initial phase of the clapping motion, the angular acceleration of the plate $\ddot{\theta}$, as indicated by the slope of $\dot{\theta}-t$ curve, is significantly lower for the stationary body (figure \hyperref[fig:AngVelo_AR067_StatDyn]{5a}). This implies an increase in $I_{add}$ in the stationary case, as $I_b$ and initial torque $T_o$ ($= 2\ \kappa \ \theta_o$) are the same in stationary and dynamic cases. Note that $T_f$ is negligible in this phase of the clapping motion.\par
	\begin{figure}
		\centering
		\begin{subfigure}[b]{0.45\textwidth}
			\includegraphics[width=\textwidth]{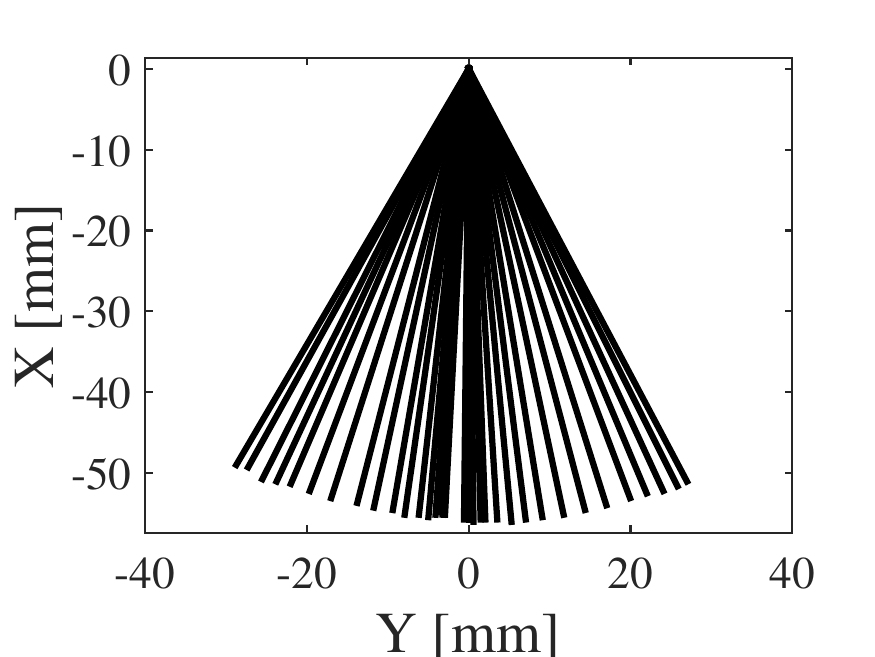}
			\caption{}
			\label{fig:SuperImp_Stat_StatDyn}
		\end{subfigure}\hspace{2mm}
		\begin{subfigure}[b]{0.45\textwidth}
			\includegraphics[width=\textwidth]{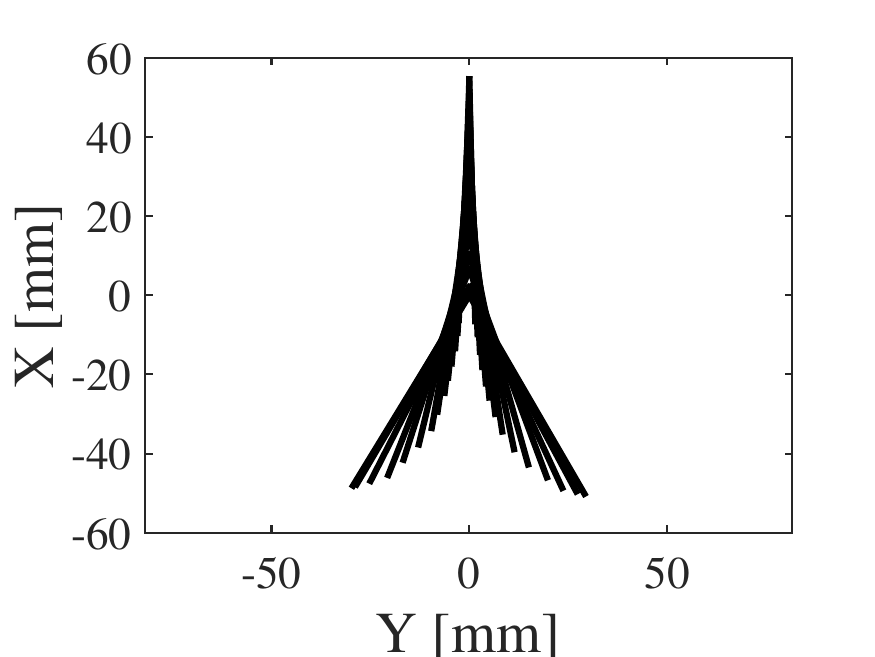}
			\caption{}
			\label{fig:SuperImp_Dyn_StatDyn}
		\end{subfigure}
		\caption{Superimposed positions of the clapping plates for (a) the stationary case and (b) the dynamic case. The black lines represent the tangents, as shown by the yellow lines in figure \hyperref[fig:SuperImp_Schematic_StatDyn]{2b}. Both figures show  superimposed positions at 6 ms intervals, covering a total time of around 90 ms for the $d^*$ = 0.5 case.
		}\label{fig:Sumperimposed_clapping_body_StatDyn}
	\end{figure}
		
	\begin{figure}
		\centering
		\begin{subfigure}[b]{0.45\textwidth}
			\includegraphics[width=\textwidth]{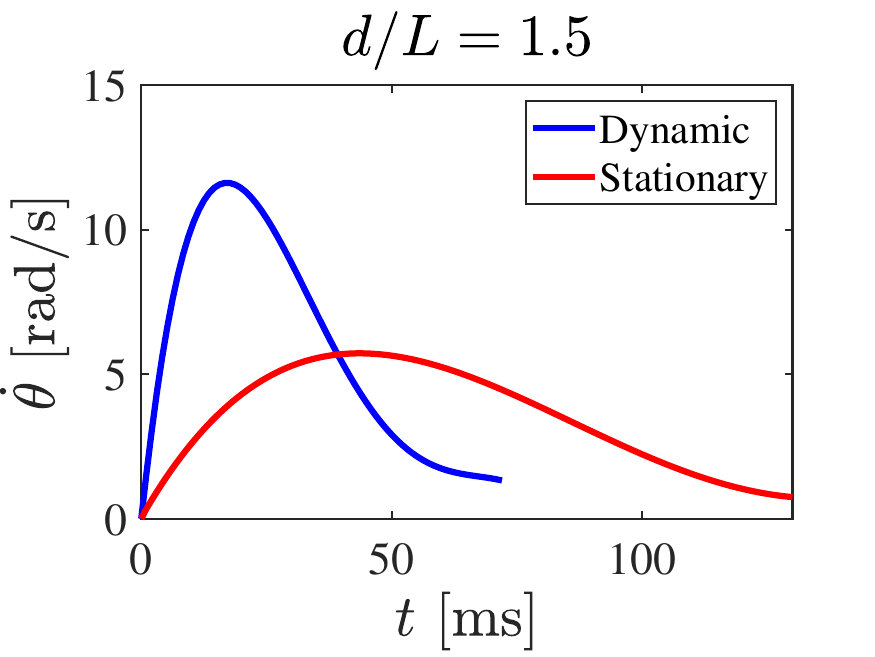}
			\caption{}
			\label{fig:AngVelo_AR067_StatDyn}
		\end{subfigure}\hspace{2mm}
		\begin{subfigure}[b]{0.45\textwidth}
			\includegraphics[width=\textwidth]{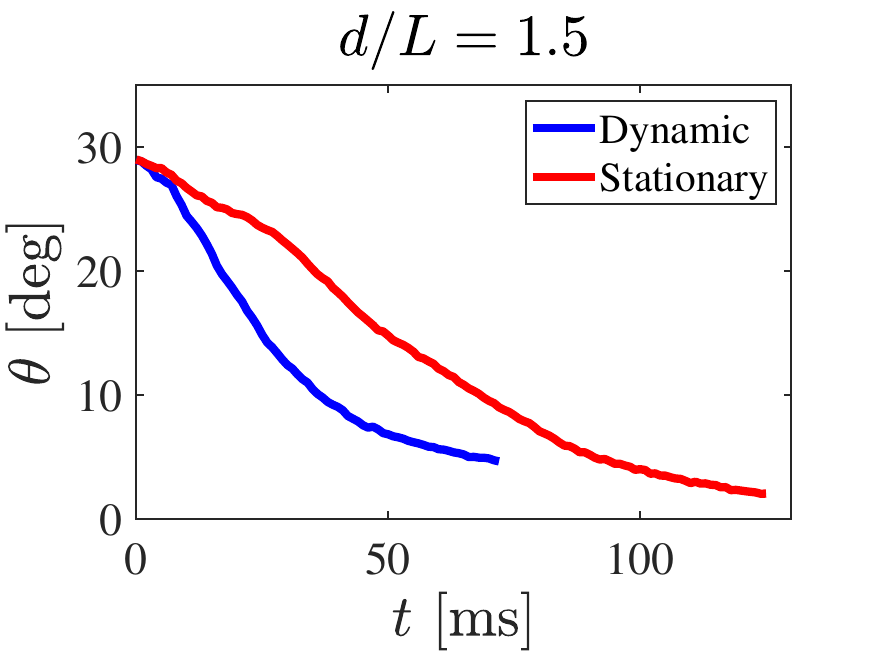}
			\caption{}
			\label{fig:AngDisp_AR067_StatDyn}
		\end{subfigure}
		\caption{The time variation in (a) angular velocity, $\dot{\theta}$, and (b) semi-clapping angle, $\theta$, of the clapping plate for $d^* =$ 1.5, in both freely moving (dynamic) and forward motion-constrained (stationary) cases.} \label{fig:AngVelo_Disp_StatDyn}
	\end{figure}

% Table generated by Excel2LaTeX from sheet 
	\begin{table}
		\centering
		\begin{tabular}{ccccrccrcc}
			\cmidrule{1-1}\cmidrule{3-4}\cmidrule{6-7}\cmidrule{9-10}    \multirow{2}[4]{*}{\text{$d^*$}} &       & \multicolumn{2}{c}{$\dot{\theta}_m$ [rad/s]} &       & \multicolumn{2}{c}{$t_{\dot{\theta}m}$ [ms]} &       & $u_m$ [m/s] & $\ \ t_{um}$ [ms] \\
			\cmidrule{3-4}\cmidrule{6-7}\cmidrule{9-10}          &       & Stationary  & Dynamic &       & Stationary  & Dynamic &       & \multicolumn{2}{c}{Dynamic} \\
			\cmidrule{1-1}\cmidrule{3-4}\cmidrule{6-7}\cmidrule{9-10}    \text{1.5} &       & 5.70  & 11.61 &       & 44    & 17    &       & 0.73  & 47 \\
			\text{1.0} &       & 5.55  & 11.04 &       & 48    & 20    &       & 0.69  & 49 \\
			\text{0.5} &       & 7.93  & 13.05 &       & 38    & 18    &       & 0.71  & 49 \\
			\cmidrule{1-1}\cmidrule{3-4}\cmidrule{6-7}\cmidrule{9-10}    \end{tabular}%
		\caption{Maximum values of angular velocities ($\dot{\theta}_m$) and times ($t_{\dot{\theta}m}$) when the maximum angular velocity is observed for the stationary and dynamic cases. The third column lists maximum translation velocities ($u_m$) of the body for the dynamic case and the times ($t_{um}$)  corresponding to $u_m$.}
		\label{tab:maximaAngular}%
	\end{table}%
	The Reynolds number is high for the cases in the present study, implying the dominance of inertia forces. The Reynolds number based on the maximum tip-velocity, $Re_{\dot{\theta} m} (=\nu  R_c^2 {\dot{\theta}}_m)$, is in the range 1.8 - 2.6 $\times$ 10$^4$ for the stationary clapping case, and for the dynamic clapping case the range is 3.7 - 4.4 $\times$ 10$^4$; $\nu$ is a kinematic viscosity of water and the radius of rotation, $R_c$, is shown by yellow dashed line in figure \hyperref[fig:SuperImp_Schematic_StatDyn]{2b}. The times at which maximum angular velocity is reached in the stationary cases, $t_{\dot{\theta}_{ms}}$, and in the dynamics cases, 
	$t_{\dot{\theta}_{md}}$, show slight variation across the $d^*$ range (Table \hyperref[tab:maximaAngular]{1}), but $t_{\dot{\theta}_{ms}} \ \approx 2 \ t_{\dot{\theta}_{md}}$ , corresponding to the faster clapping in the dynamic case.\par
	
	Interestingly, all the $\dot{\theta}$ versus time curves during the acceleration phase for both dynamic and stationary cases and for the three $d^*$ values collapse when normalized by $\dot{\theta}_m$ for velocity and by $t_{\dot{\theta}m}$ for time (figure \hyperref[fig:AngVelo_nonDim_StatDyn]{6a}). A collapse is also observed during the deceleration phase, although it is not good in the tail portion of the curves (figure \hyperref[fig:AngVelo_nonDim_StatDyn]{6b}); here, time is scaled by the time when the angular speed is half the maximum value. The implication of the collapse in the two phases is that the overall nature of the angular velocity curve is independent of variations in the flow field resulting from constraining the forward motion of the self-propelled clapping body. \par
	
	\begin{figure}
		\centering
		\begin{subfigure}[b]{0.45\textwidth}
			\includegraphics[width=\textwidth]{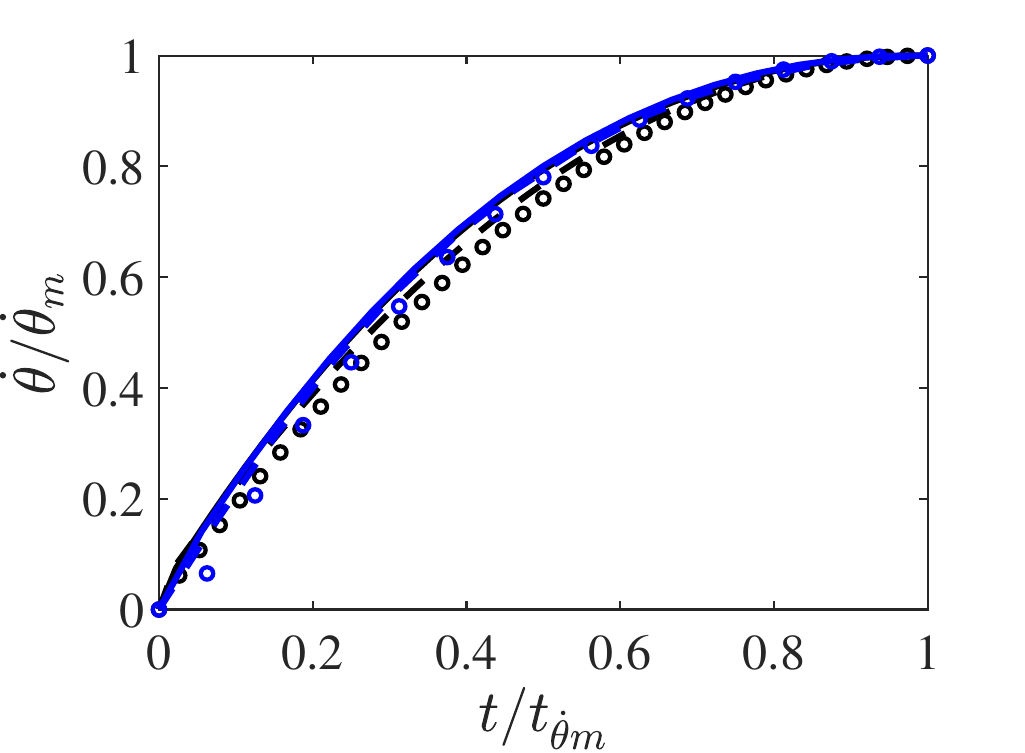}
			\caption{}
			\label{fig:AngVelo_Acc_nonDim_StatDyn}
		\end{subfigure}\hspace{10mm}
		\begin{subfigure}[b]{0.45\textwidth}
			\includegraphics[width=\textwidth]{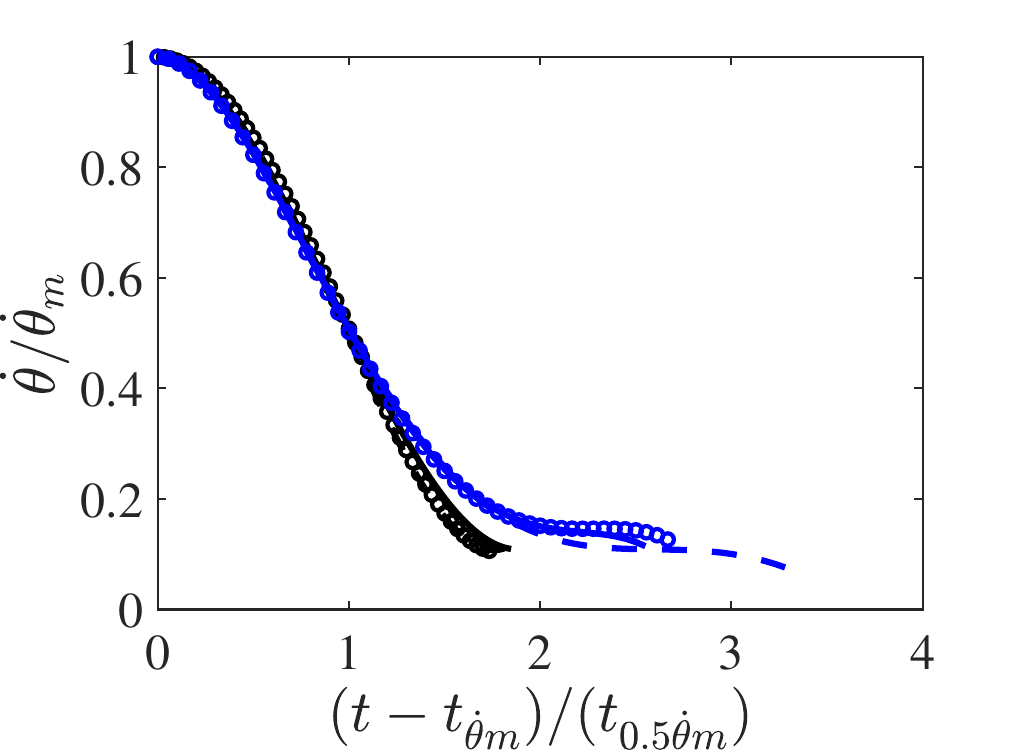}
			\caption{}
			\label{fig:AngVelo_Dacc_nonDim_StatDyn}
		\end{subfigure}
		\begin{subfigure}[b]{0.45\textwidth}
		\includegraphics[width=\textwidth]{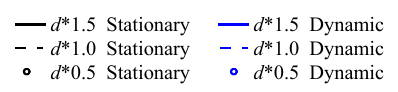}
		\label{fig:AngVelo_Dacc_legend}
		\end{subfigure}
		\caption{Plot of normalized angular velocity versus normalized time for the clapping body in stationary and dynamic cases. (a) acceleration phase (b) retardation phase. $\dot{\theta}_m$ is the maximum angular velocity of a clapping plate. $t_{\dot{\theta}m}$ and $t_{0.5\dot{\theta}m}$ are the times at which angular velocity reaches the maximum value and half of the maximum value, respectively.}\label{fig:AngVelo_nonDim_StatDyn}
	\end{figure}
	\begin{figure}
	\centering
	\begin{subfigure}[b]{0.5\textwidth}
		\includegraphics[width=\textwidth]{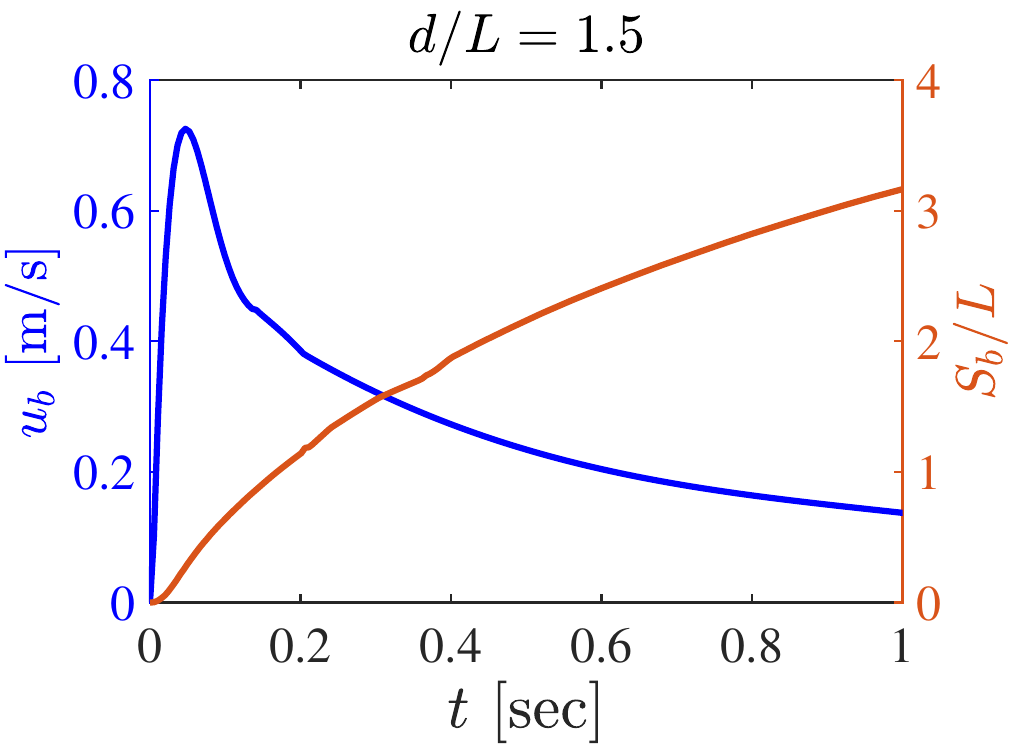}	
		\label{fig:ub_disp_StatDyn}
	\end{subfigure}	
	\caption{Time variation of translational velocity, $u_b$, and displacement, $S_b$, of the freely moving clapping body for $d^* =$ 1.5. The distance traveled by the body $S_b$ is given in terms of body length $L$. }
	\end{figure}

	The translational velocity of the clapping body, $u_b$, in the self-propelling condition increases rapidly and reaches a maximum value approximately at the end of the clapping action; during this linear acceleration phase, the body experiences net thrust force (figure \hyperref[fig:ub_disp_StatDyn]{7}). Once the clapping action is over and the plates are together, the body slowly decelerates due to the drag force acting on the body. The average of standard deviations in $u_b$ is less than 2\% of the maximum body velocity, $u_m$. For all the three $d^*$ configurations, the time scale corresponding to the acceleration phase, $t_{um}$,  is approximately 50 ms, at which the body attains the maximum translational velocity, $u_m$, of 0.7m/s (Table \hyperref[tab:maximaAngular]{1}), and the retardation phase continues for more than 1 sec where the body traverses a distance of approximately $3L$, see figure \hyperref[fig:ub_disp_StatDyn]{7}. A detailed analysis of the results corresponding to a freely moving clapping body is given in Mahulkar and Arakeri\cite{Mahulkar23}. \par

	\end{subsection}
%%%%%%%%%%%%%%%%%%%%%%%%%%%%%%%%%%%%%%%%%%%%%%%%%%%%%%%%%%%%%%%%%%%%%%%%%%%%%%%%%%%%%%%%

	\begin{subsection}{Flow characteristics}
	\label{sec:WakeDynamics_StatDyn}
 	The large difference in the kinematics between the dynamic and stationary cases is reflected in the differences in flow characteristics of the two cases, which we discuss in the following sections.
	
	\begin{subsubsection}{Flow fields in the XY plane}
	\label{sec:vorticityZ_StatDyn}
		\begin{figure}
		\centering
		\begin{subfigure}[b]{0.30\textwidth}
			\includegraphics[width=\textwidth]{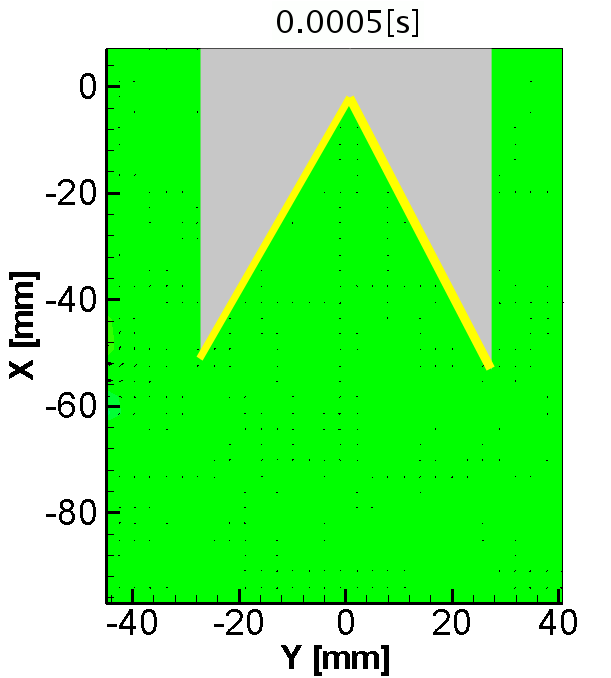}
			\caption{}
		\end{subfigure}\hspace{5mm}
		\begin{subfigure}[b]{0.30\textwidth}
			\includegraphics[width=\textwidth]{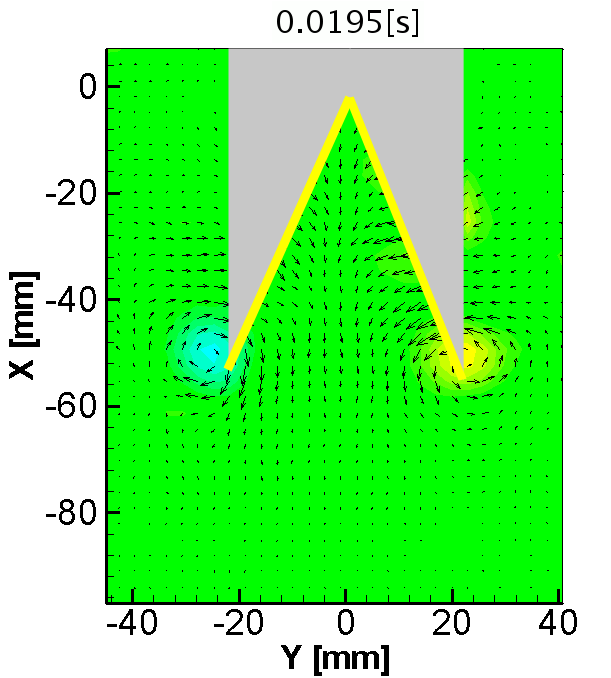}
			\caption{}
		\end{subfigure}\hspace{5mm}
		\begin{subfigure}[b]{0.30\textwidth}
			\includegraphics[width=\textwidth]{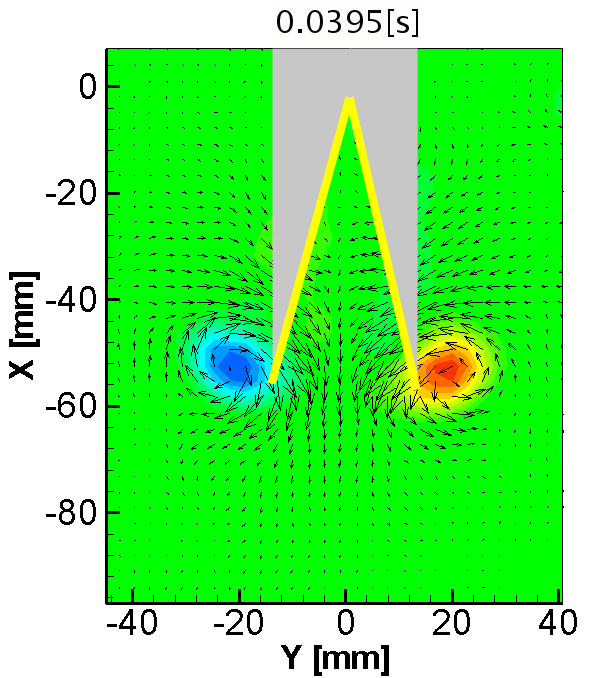}
			\caption{}
		\end{subfigure}\vspace{10mm}
		\begin{subfigure}[b]{0.30\textwidth}
			\includegraphics[width=\textwidth]{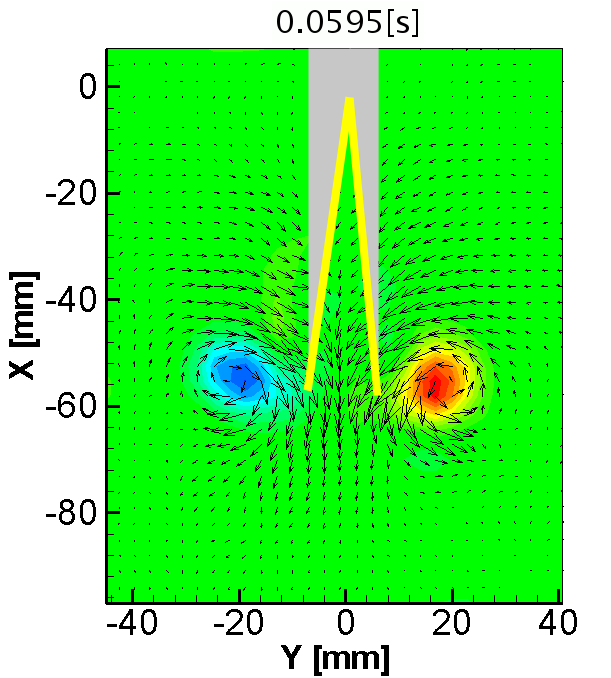}
			\caption{}
		\end{subfigure}\hspace{5mm}
		\begin{subfigure}[b]{0.30\textwidth}
			\includegraphics[width=\textwidth]{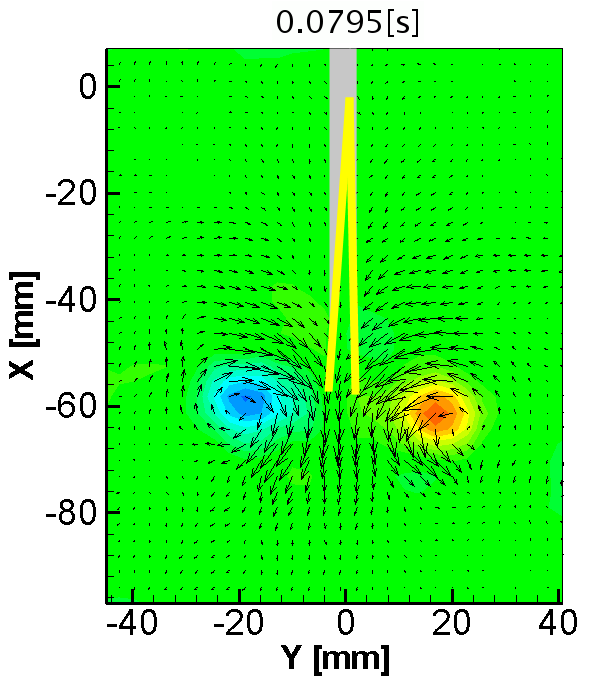}
			\caption{}
		\end{subfigure}\hspace{5mm}
		\begin{subfigure}[b]{0.30\textwidth}
			\includegraphics[width=\textwidth]{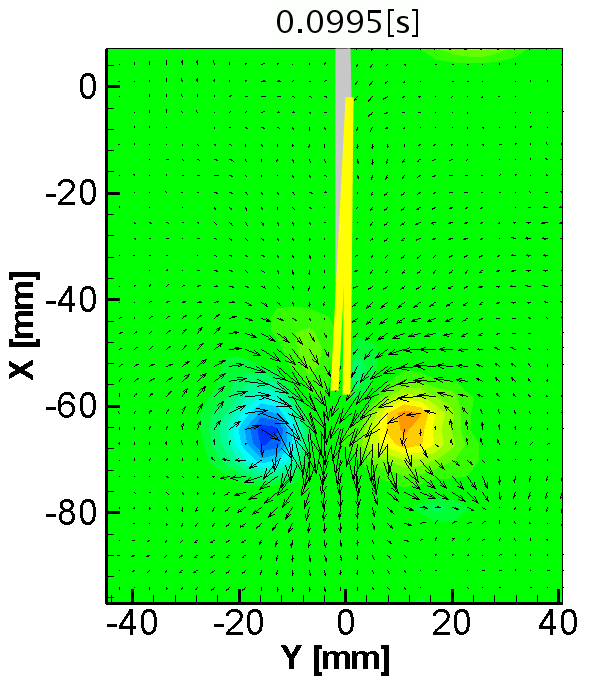}
			\caption{}
		\end{subfigure}\vspace{10mm}
		\begin{subfigure}[b]{0.30\textwidth}
			\includegraphics[width=\textwidth]{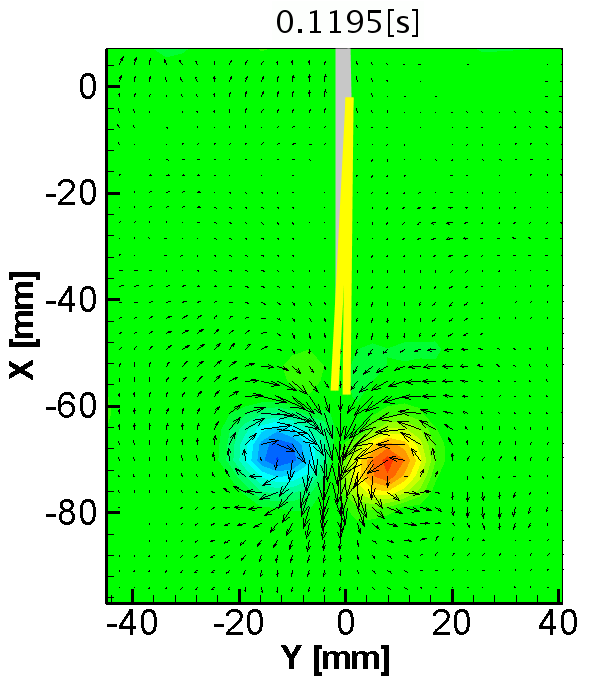}
			\caption{}
		\end{subfigure}\hspace{5mm}
		\begin{subfigure}[b]{0.30\textwidth}
			\includegraphics[width=\textwidth]{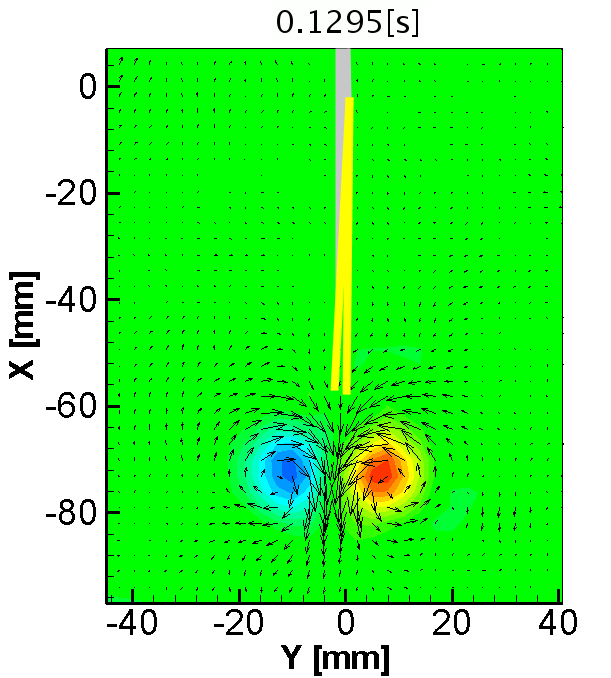}
			\caption{}
		\end{subfigure}\hspace{5mm}
		\begin{subfigure}[b]{0.30\textwidth}
			\includegraphics[width=\textwidth]{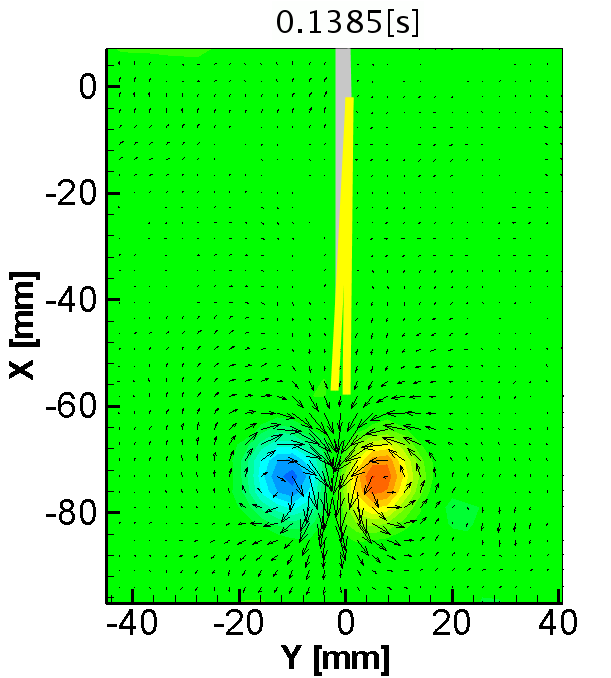}
			\caption{}
		\end{subfigure}\vspace{5mm}
		\begin{subfigure}[b]{0.4\textwidth}
			\includegraphics[width=\textwidth]{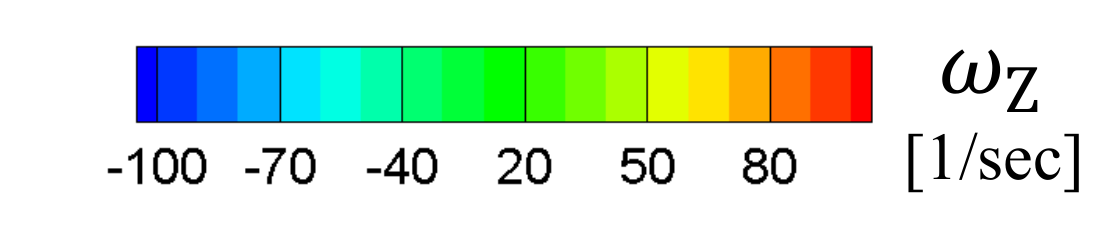}
		\end{subfigure}
		\caption{The Z-component vorticity field for the stationary case for $d^*$ = 0.5.}\label{fig:VortZ_S_StatDyn}
	\end{figure}
	\begin{figure}
		\centering
		\begin{subfigure}[b]{0.28\textwidth}
			\includegraphics[width=\textwidth]{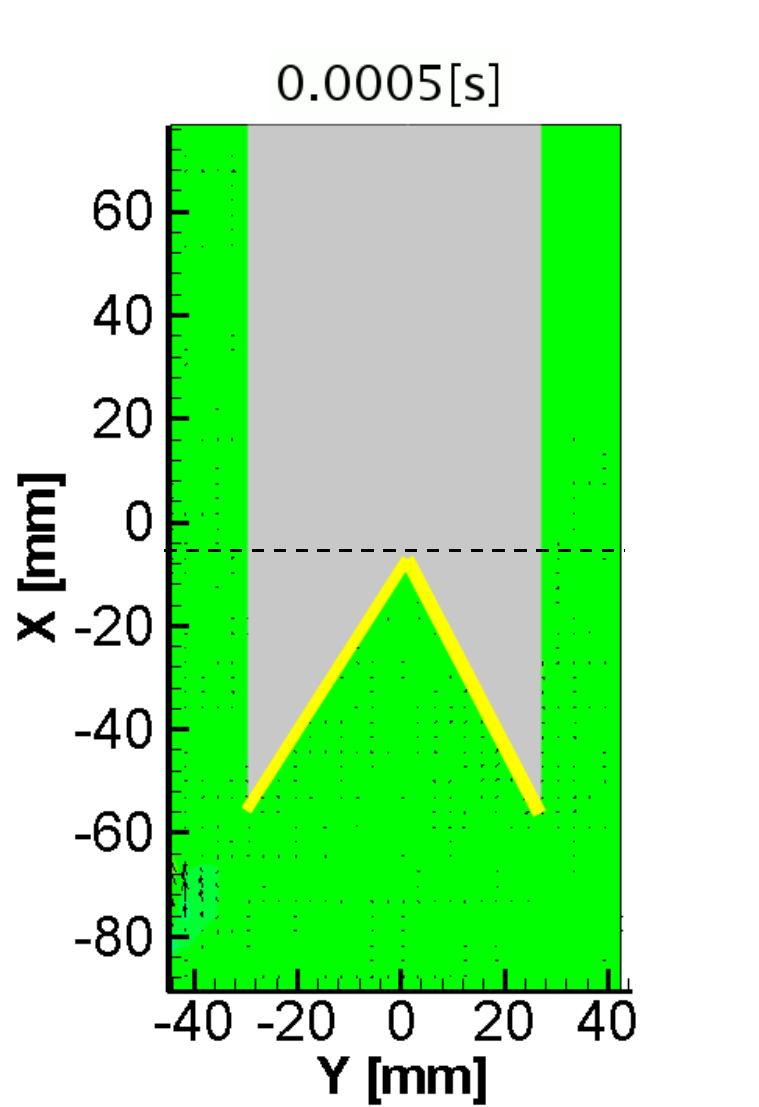}
			\caption{}
		\end{subfigure}\hspace{8mm}	
		\begin{subfigure}[b]{0.28\textwidth}
			\includegraphics[width=\textwidth]{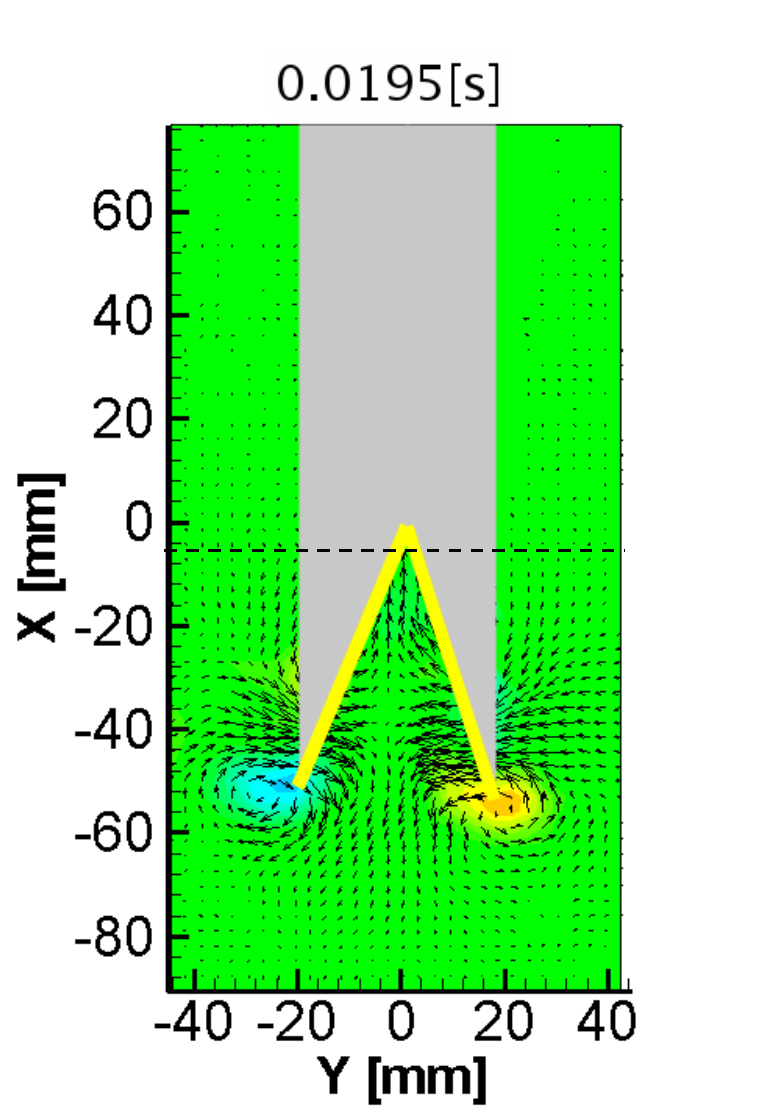}
			\caption{}
		\end{subfigure}\hspace{8mm}
		\begin{subfigure}[b]{0.28\textwidth}
			\includegraphics[width=\textwidth]{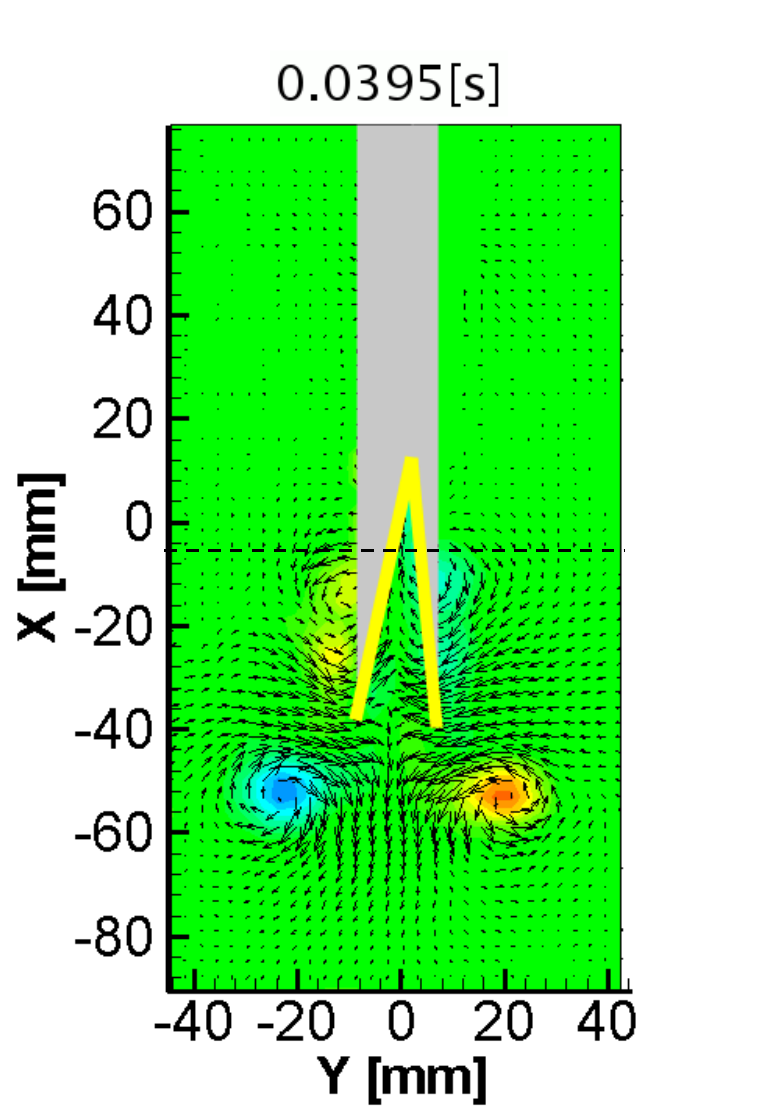}
			\caption{}
		\end{subfigure}\vspace{1mm}
		 \begin{subfigure}[b]{0.28\textwidth}
			\includegraphics[width=\textwidth]{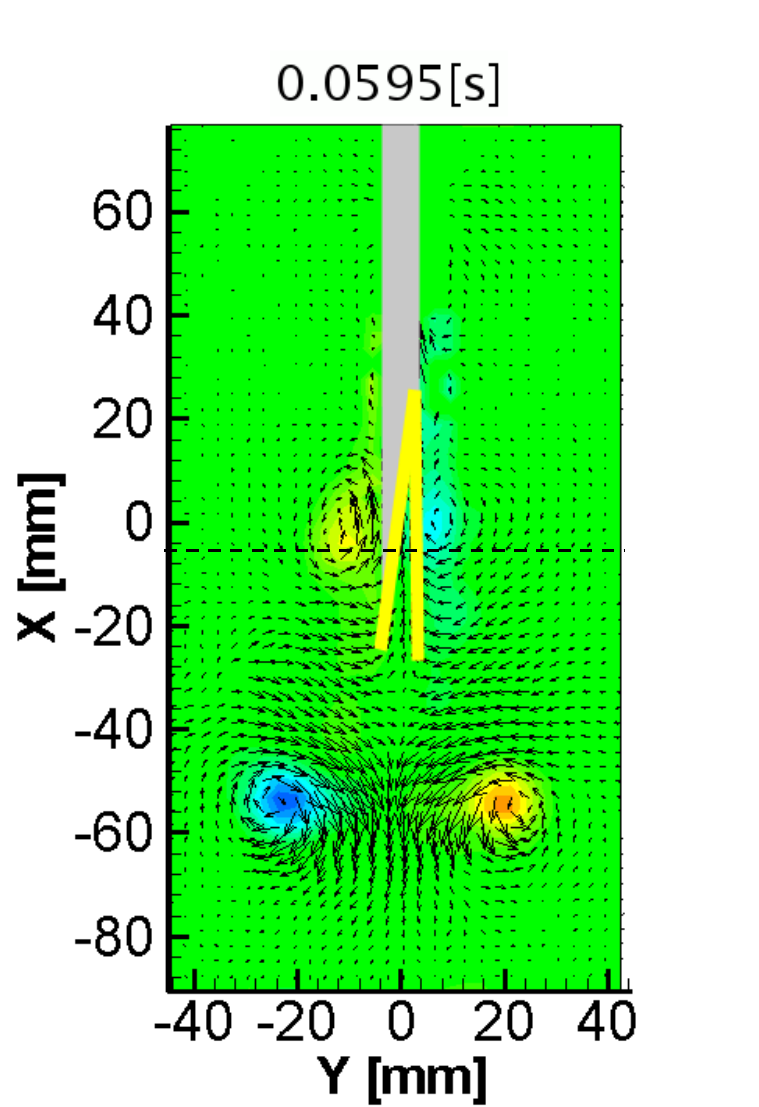}
			\caption{}
		\end{subfigure}\hspace{8mm}
		\begin{subfigure}[b]{0.28\textwidth}
			\includegraphics[width=\textwidth]{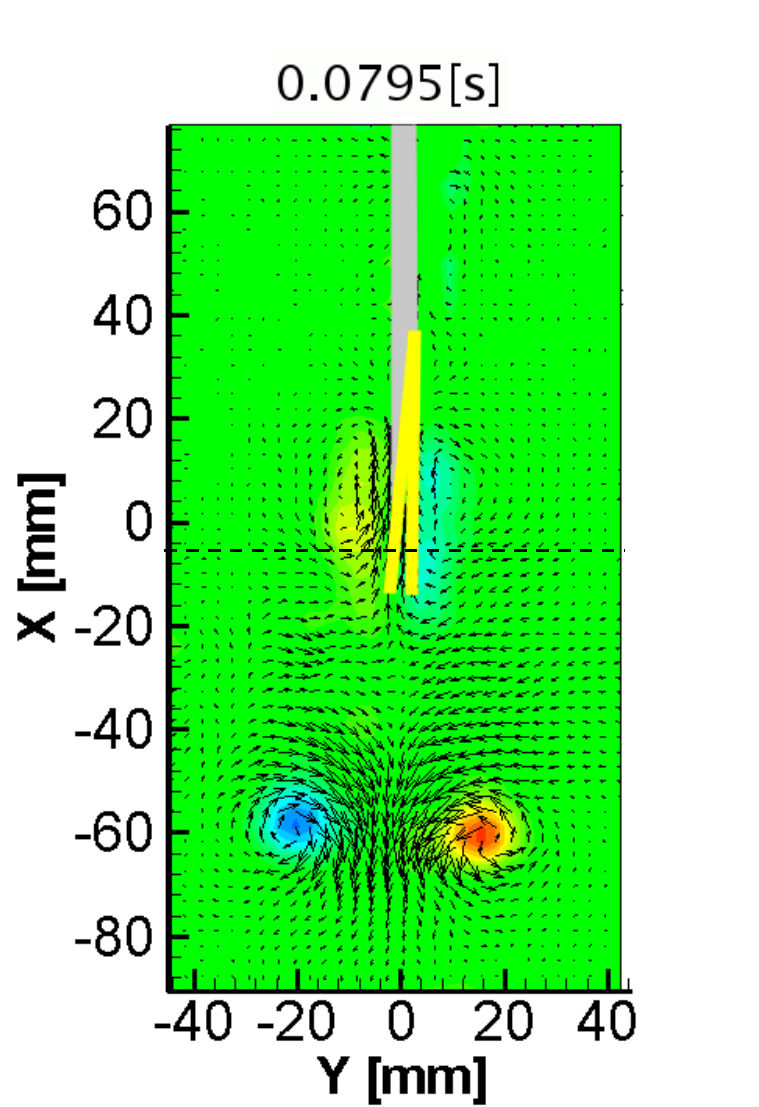}
			\caption{}
		\end{subfigure}\hspace{8mm}
		\begin{subfigure}[b]{0.28\textwidth}
			\includegraphics[width=\textwidth]{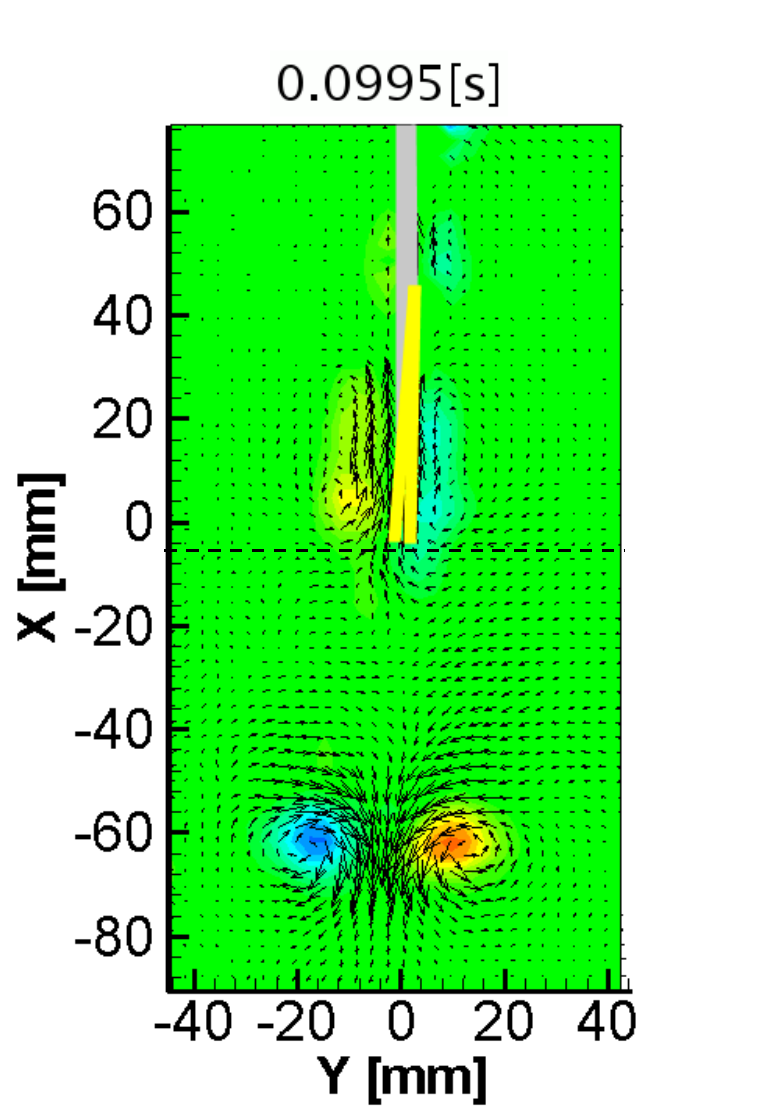}
			\caption{}
		\end{subfigure}\vspace{1mm}
		\begin{subfigure}[b]{0.28\textwidth}
			\includegraphics[width=\textwidth]{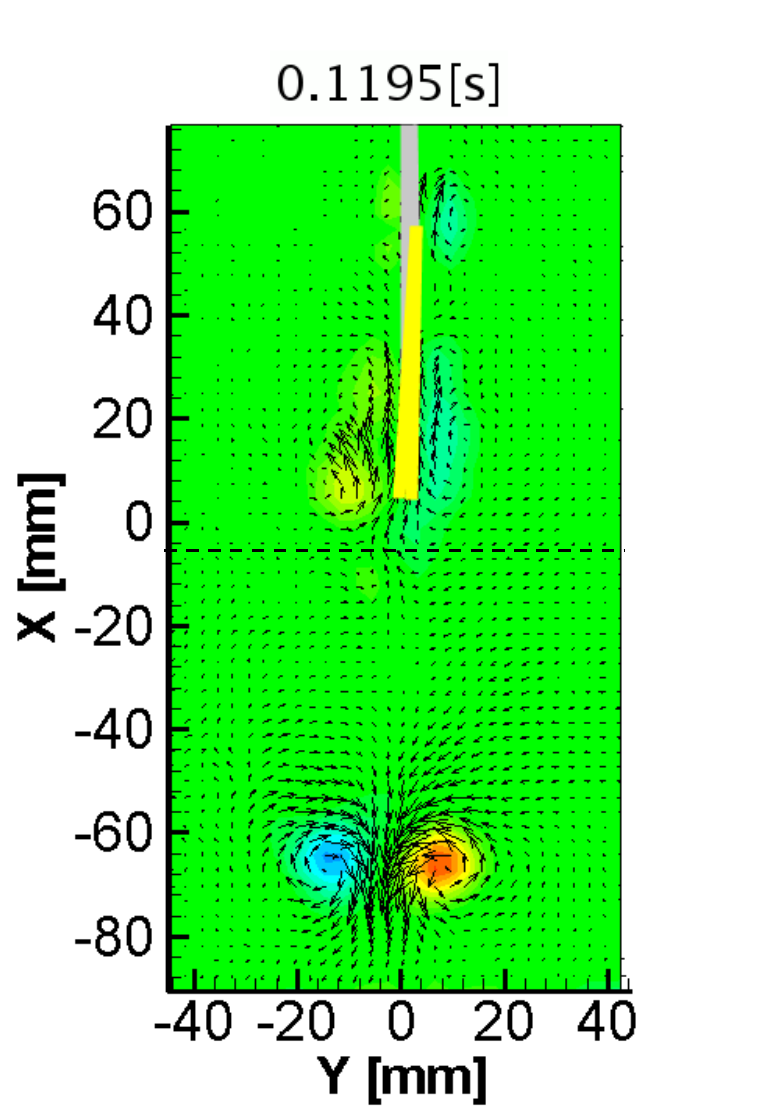}
			\caption{}
		\end{subfigure}\hspace{8mm}
		\begin{subfigure}[b]{0.28\textwidth}
			\includegraphics[width=\textwidth]{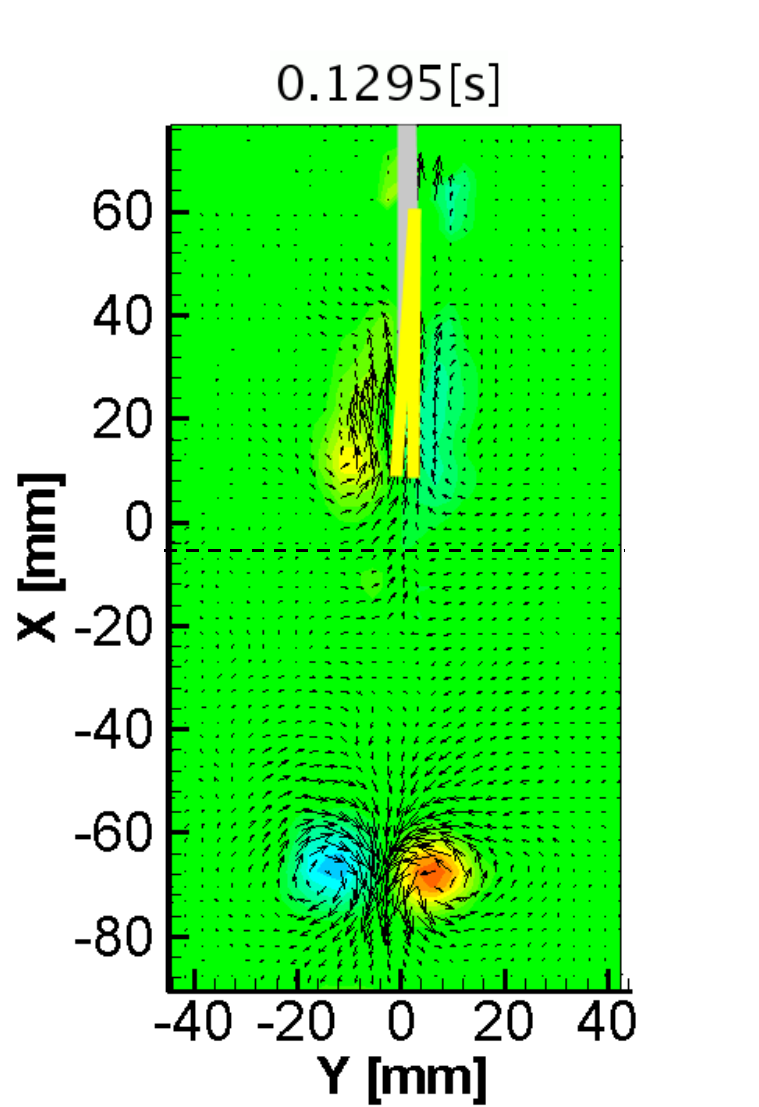}
			\caption{}
		\end{subfigure}\hspace{8mm}
		\begin{subfigure}[b]{0.28\textwidth}
			\includegraphics[width=\textwidth]{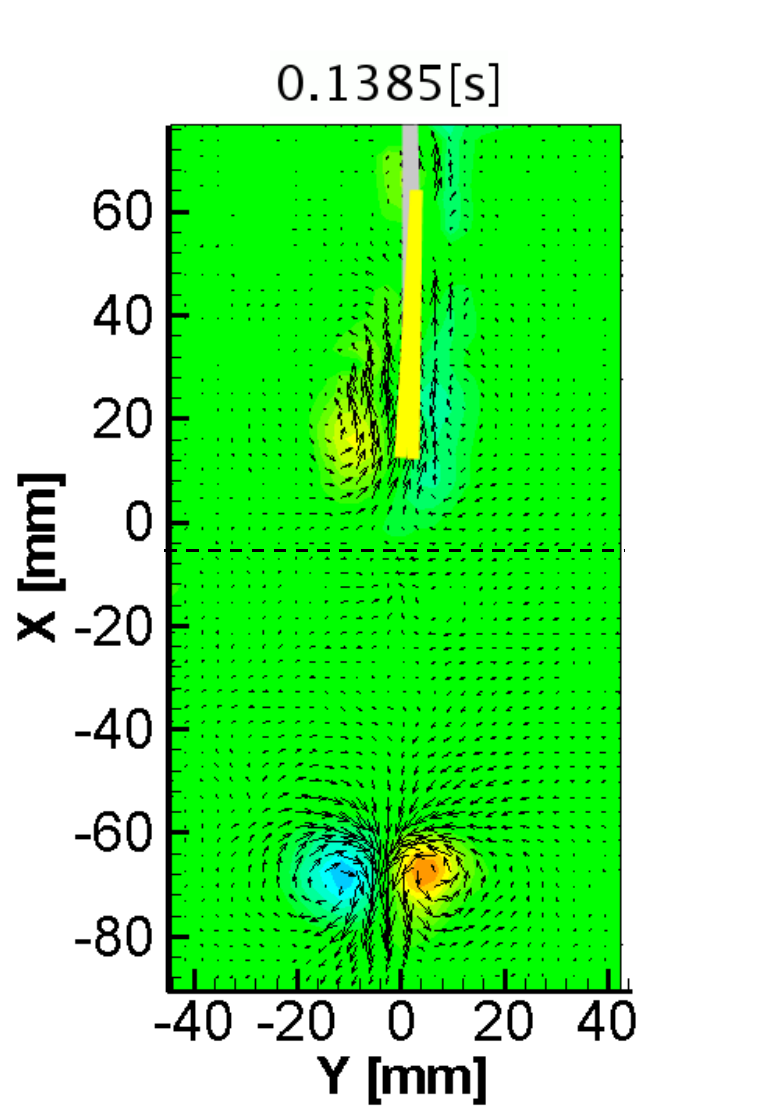}
			\caption{}
		\end{subfigure}
	
		\caption{The Z-component vorticity field for the dynamic case for $d^*$= 0.5. The dashed line indicates the initial position of the hinge point. The color bar representing vorticity levels is the same as in figure \hyperref[fig:VortZ_S_StatDyn]{8}.}\label{fig:VortZ_D_StatDyn}
	\end{figure}

	\begin{figure}
		\centering
		\begin{subfigure}[b]{0.3\textwidth}
			\includegraphics[width=\textwidth]{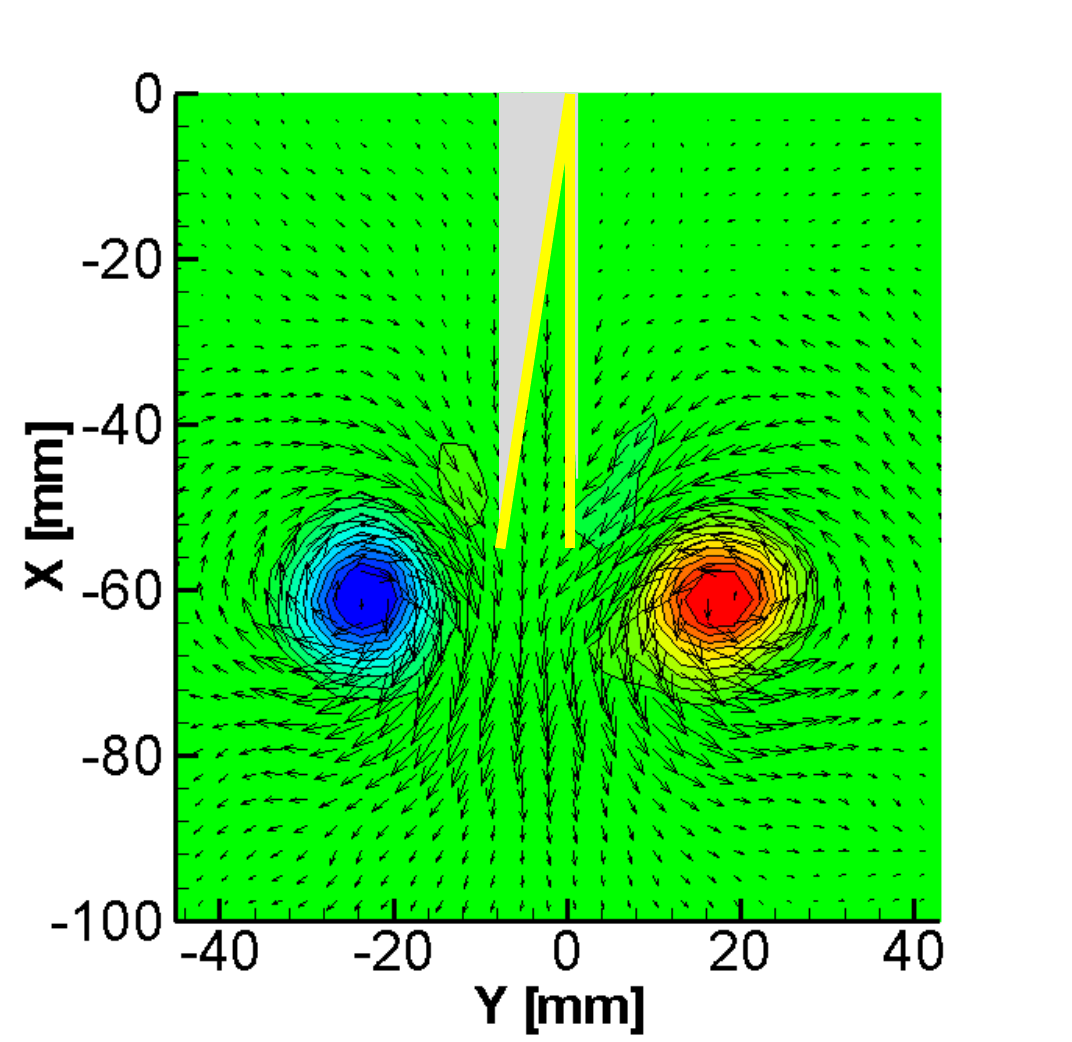}
			\caption{}
			\label{fig:VortZ_AR067_Stat}
		\end{subfigure}\hspace{2mm}
		\begin{subfigure}[b]{0.3\textwidth}
			\includegraphics[width=\textwidth]{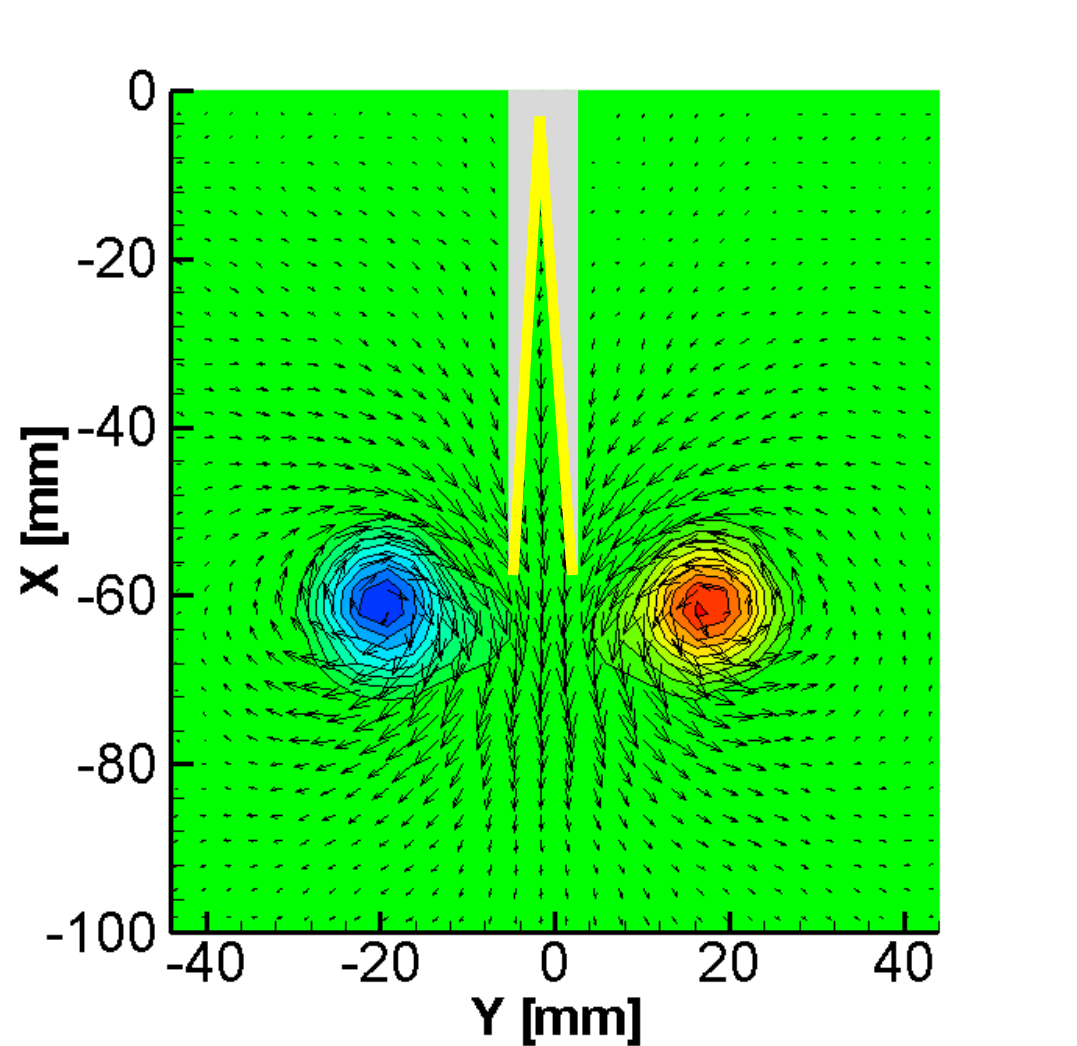}
			\caption{}
			\label{fig:VortZ_AR100_Stat}
		\end{subfigure}\hspace{2mm}
		\begin{subfigure}[b]{0.3\textwidth}
			\includegraphics[width=\textwidth]{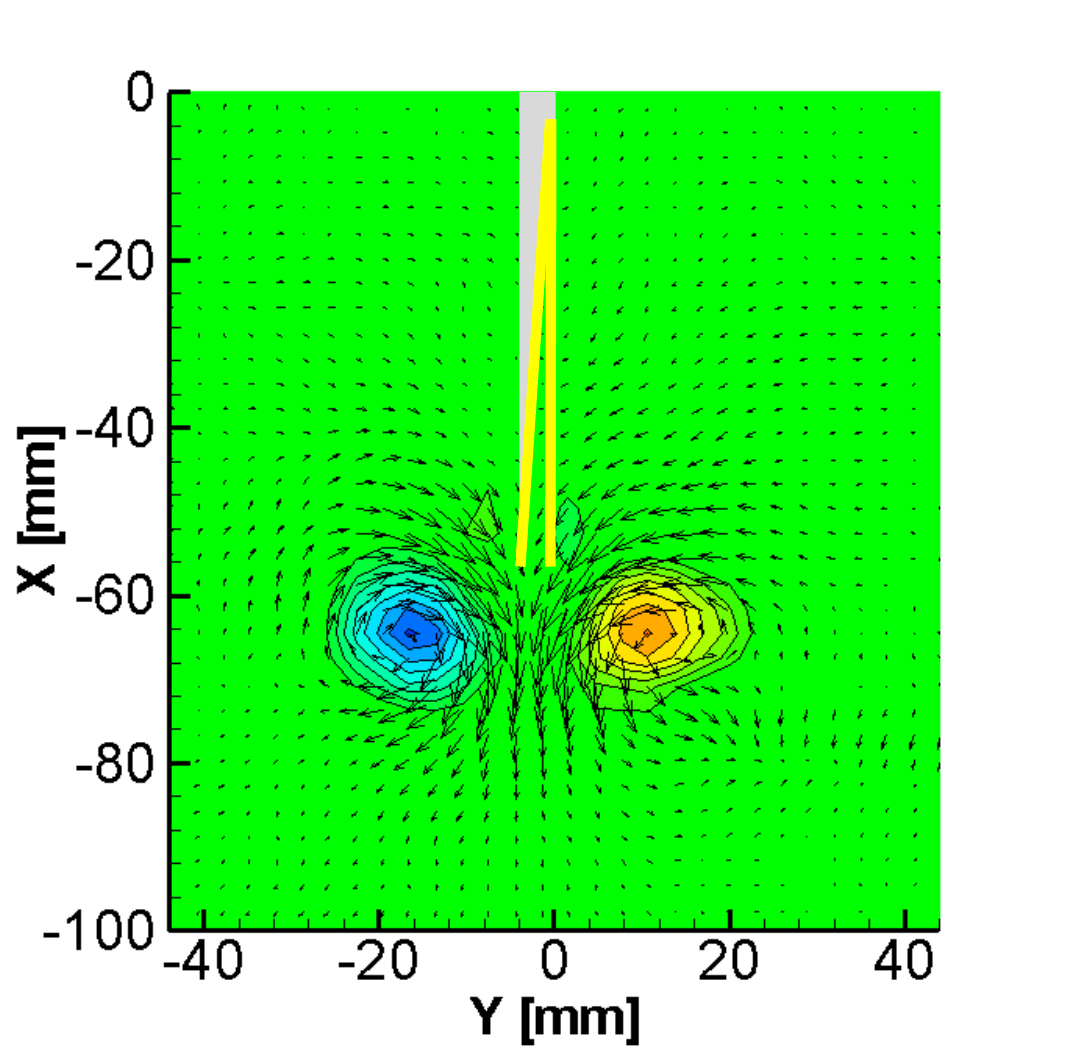}
			\caption{}
			\label{fig:VortZ_AR200_Stat}
		\end{subfigure}\vspace{1mm}
		\begin{subfigure}[b]{0.3\textwidth}
			\includegraphics[width=\textwidth]{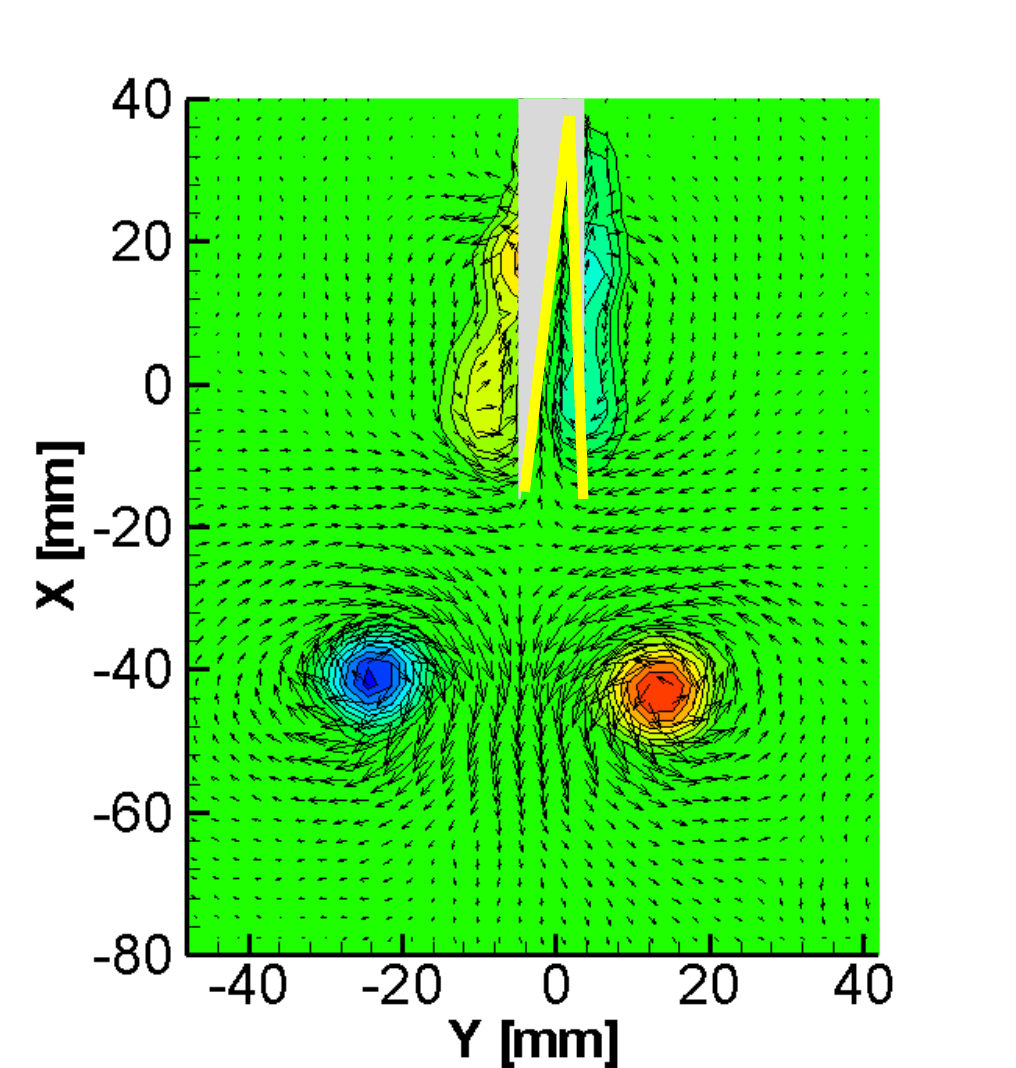}
			\caption{}
			\label{fig:VortZ_AR067_Dyn}
		\end{subfigure}\hspace{2mm}
		\begin{subfigure}[b]{0.3\textwidth}
			\includegraphics[width=\textwidth]{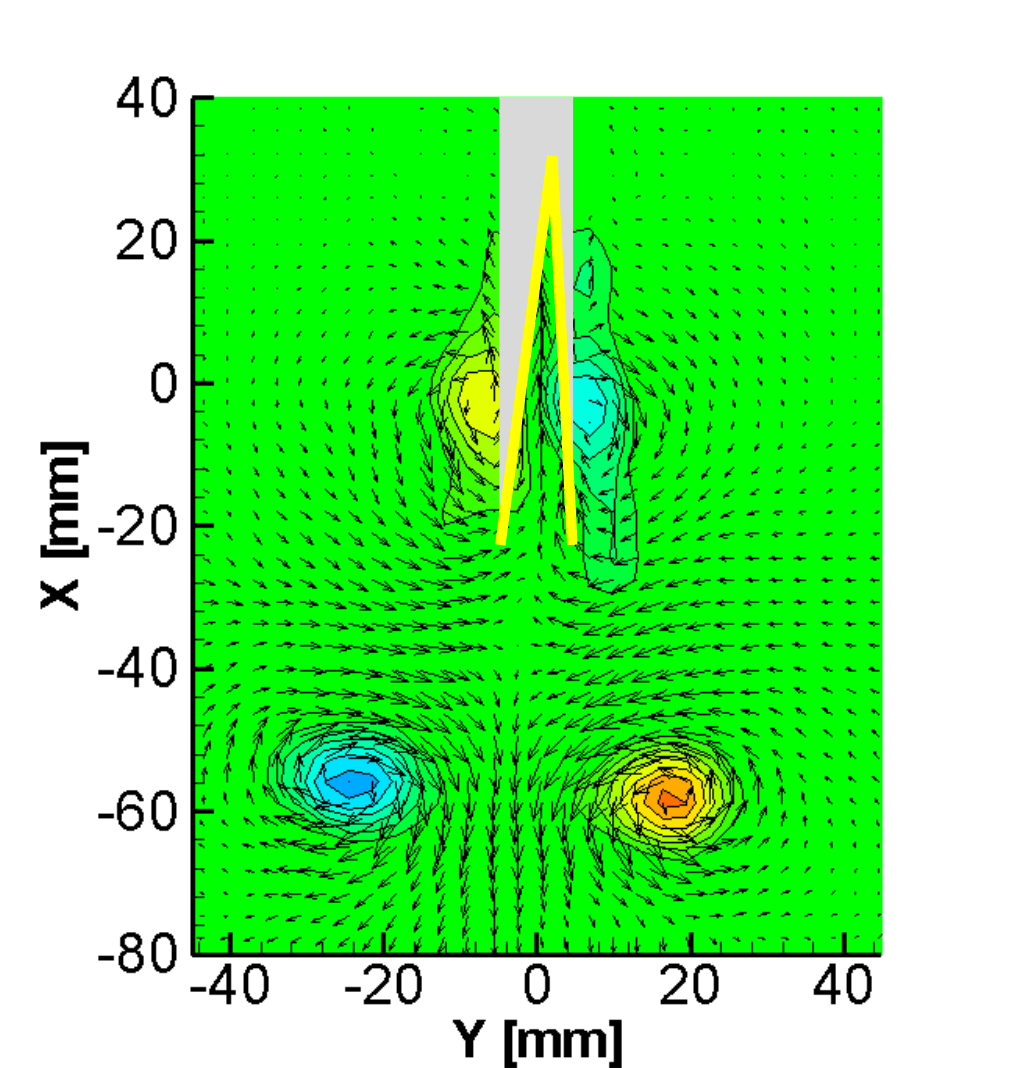}
			\caption{}
			\label{fig:VortZ_AR100_Dyn}
		\end{subfigure}\hspace{2mm}
		\begin{subfigure}[b]{0.3\textwidth}
			\includegraphics[width=\textwidth]{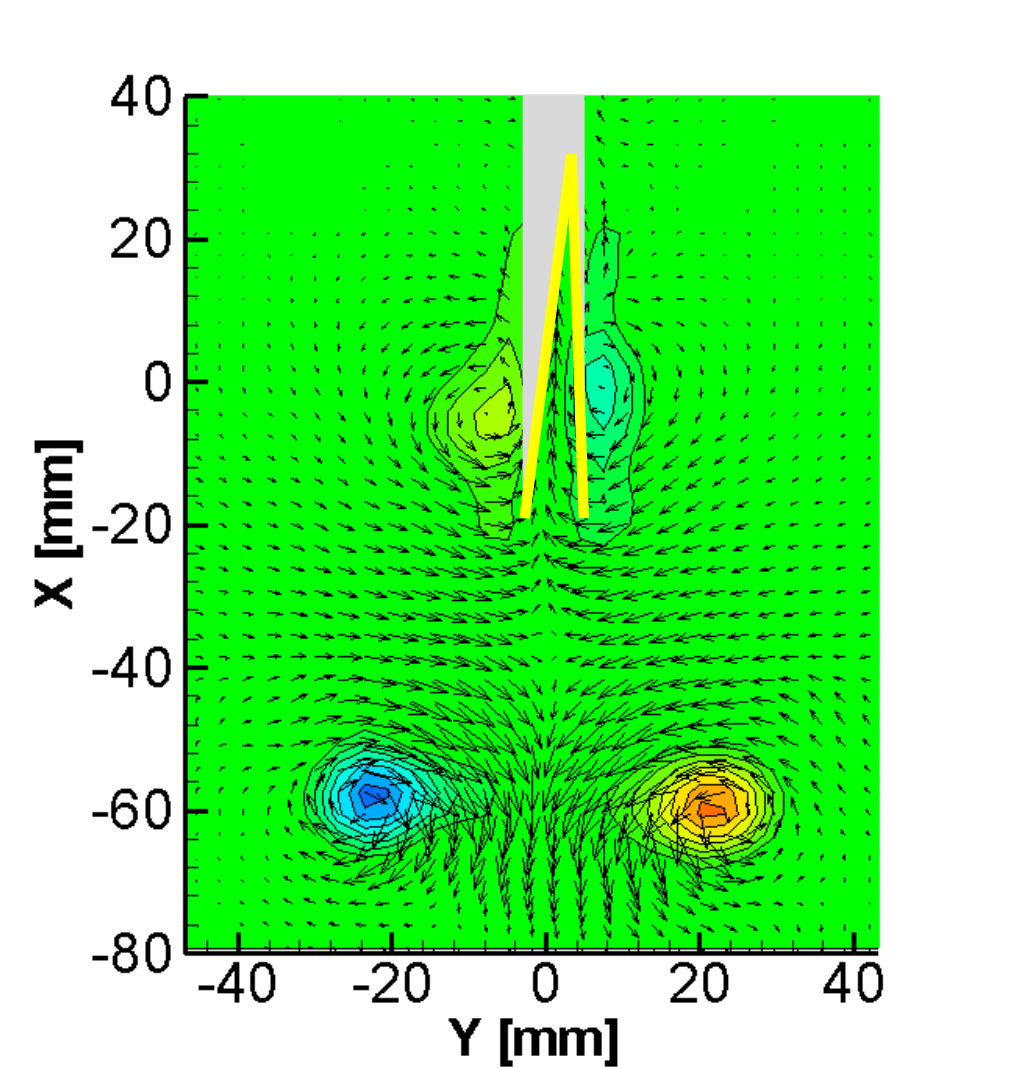}
			\caption{}
			\label{fig:VortZ_AR200_Dyn}
		\end{subfigure}
		\begin{subfigure}[b]{0.4\textwidth}
			\includegraphics[width=\textwidth]{FigDatStat/VortZ_Stat_Dyn_legend}
		\end{subfigure}
		\caption{The Z-component vorticity field towards the end of clapping motion. Figures (a), (b), and (c) show the stationary clapping cases at 100 ms for $d^* =$ 1.5, 1.0, and 0.5. Figures (d), (e), and (f) show the dynamic clapping cases at 60 ms for $d^* =$ 1.5, 1.0, and 0.5.}\label{fig:Vorticity-Z_StatDyn}
	\end{figure}
  	\begin{figure}
	\centering
	\begin{subfigure}[b]{0.3\textwidth}
		\includegraphics[width=\textwidth]{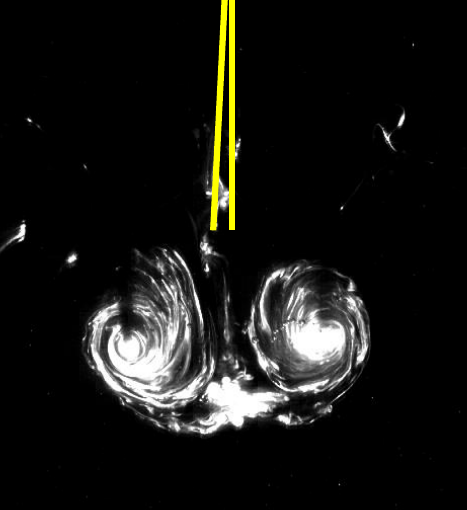}
		\caption{}
		\label{fig:FlowVis_AR067_Stat}
	\end{subfigure}\hspace{3mm}
	\begin{subfigure}[b]{0.3\textwidth}
		\includegraphics[width=\textwidth]{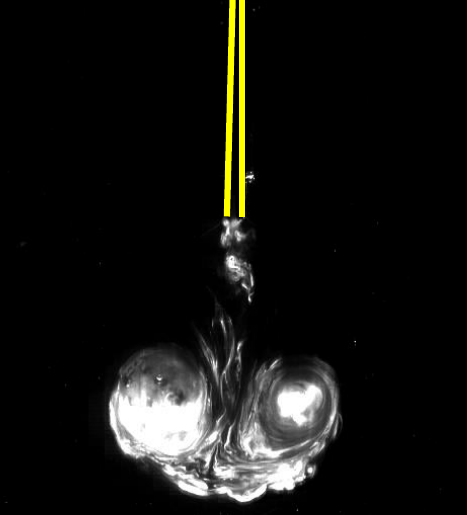}
		\caption{}
		\label{fig:FlowVis_AR100_Stat}
	\end{subfigure}\hspace{3mm}
	\begin{subfigure}[b]{0.3\textwidth}
		\includegraphics[width=\textwidth]{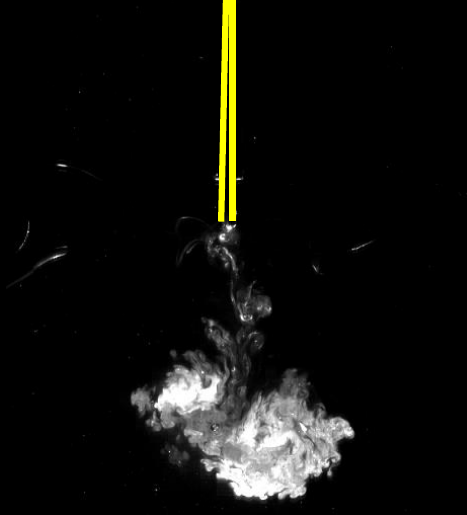}
		\caption{}
		\label{fig:FlowVis_AR200_Stat}
	\end{subfigure}\vspace{03mm}
	\begin{subfigure}[b]{0.3\textwidth}
		\includegraphics[width=\textwidth]{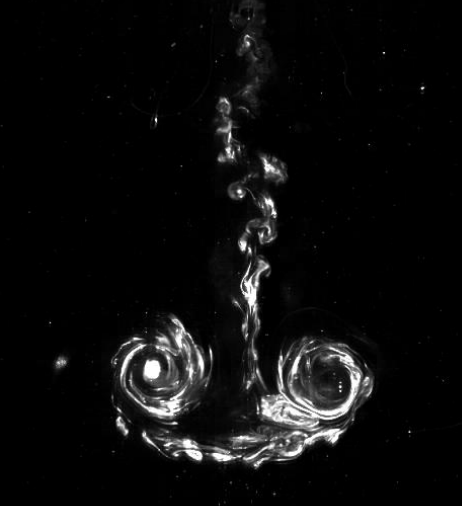}
		\caption{}
		\label{fig:FlowVis_AR067_Dyn}
	\end{subfigure}\hspace{3mm}
	\begin{subfigure}[b]{0.3\textwidth}
		\includegraphics[width=\textwidth]{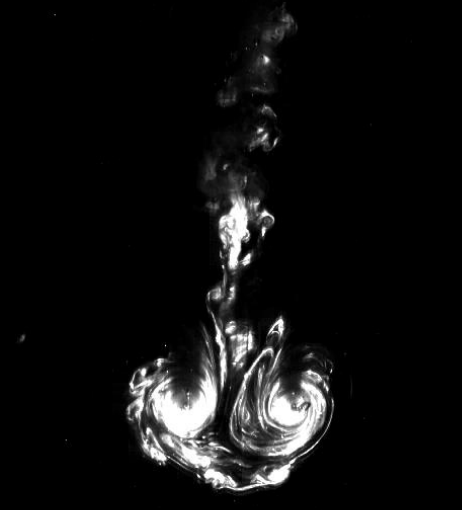}
		\caption{}
		\label{fig:FlowVis_AR100_Dyn}
	\end{subfigure}\hspace{3mm}
	\begin{subfigure}[b]{0.3\textwidth}
		\includegraphics[width=\textwidth]{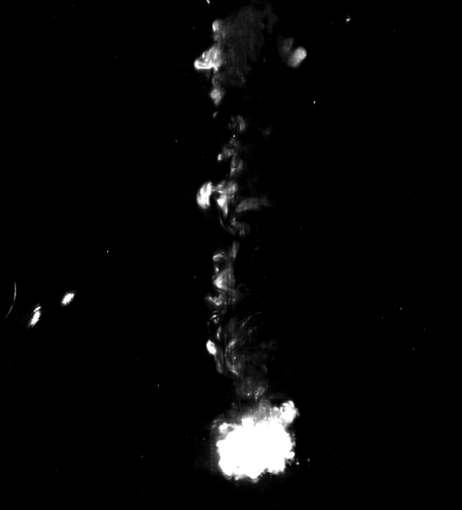}
		\caption{}
		\label{fig:FlowVis_AR200_Dyn}
	\end{subfigure}
	\caption{PLIF wake visualization in the XY plane (Z=0) at 400ms. Figures (a), (b), and (c) are for the stationary bodies with $d^*$= 1.5, 1.0, and 0.5, respectively. Figures (d), (e), and (f) are for the dynamic bodies with $d^*$= 1.5, 1.0, and 0.5, respectively.}\label{fig:Dye_visualization_StatDyn}
\end{figure}
\begin{figure}
	\centering
	\begin{subfigure}[b]{0.3\textwidth}
		\includegraphics[width=\textwidth]{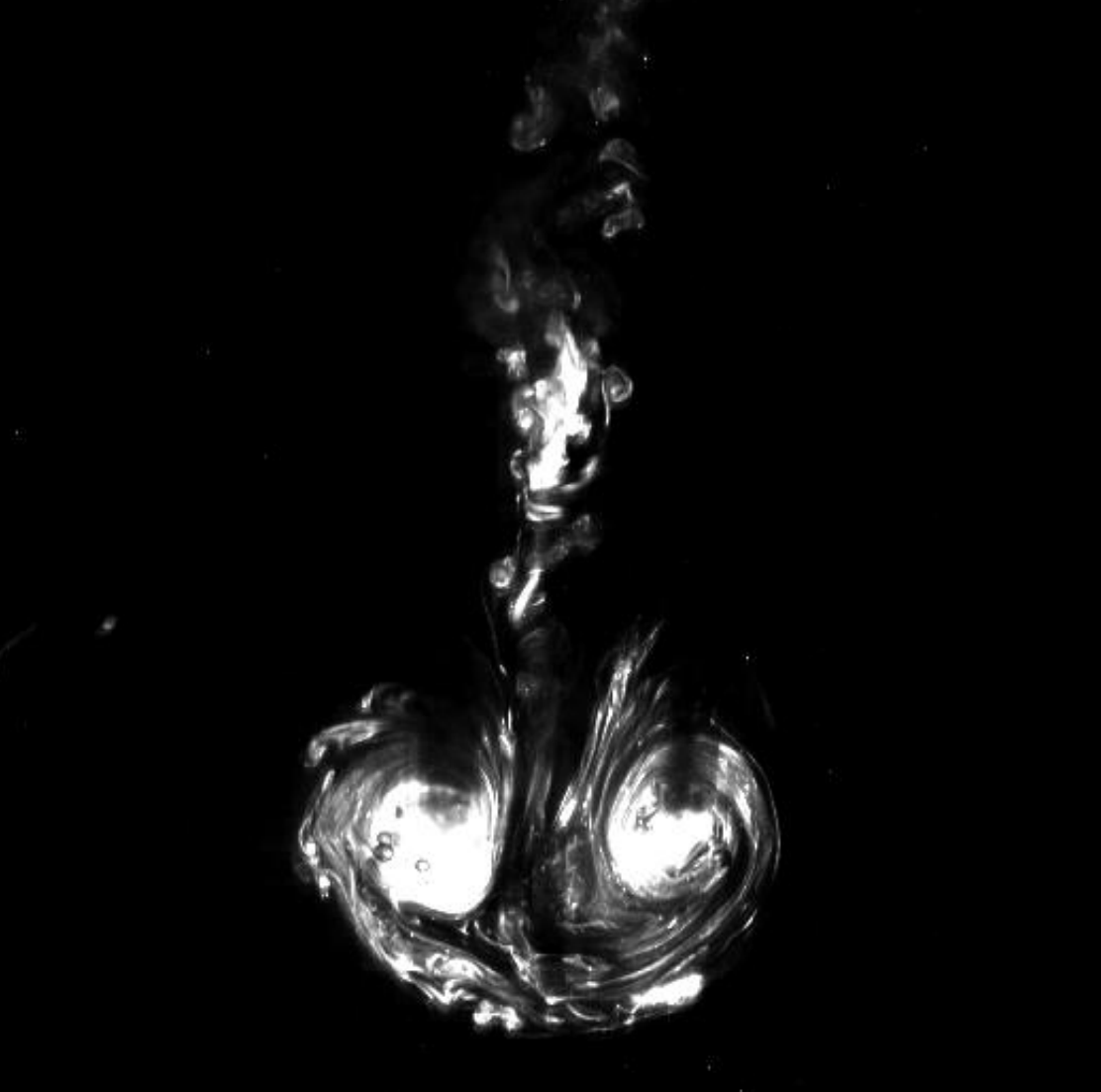}
		\caption{}
		\label{fig:VortexBreakDown_Stat_Dyn}
	\end{subfigure}\hspace{10mm}
	\begin{subfigure}[b]{0.3\textwidth}
		\includegraphics[width=\textwidth]{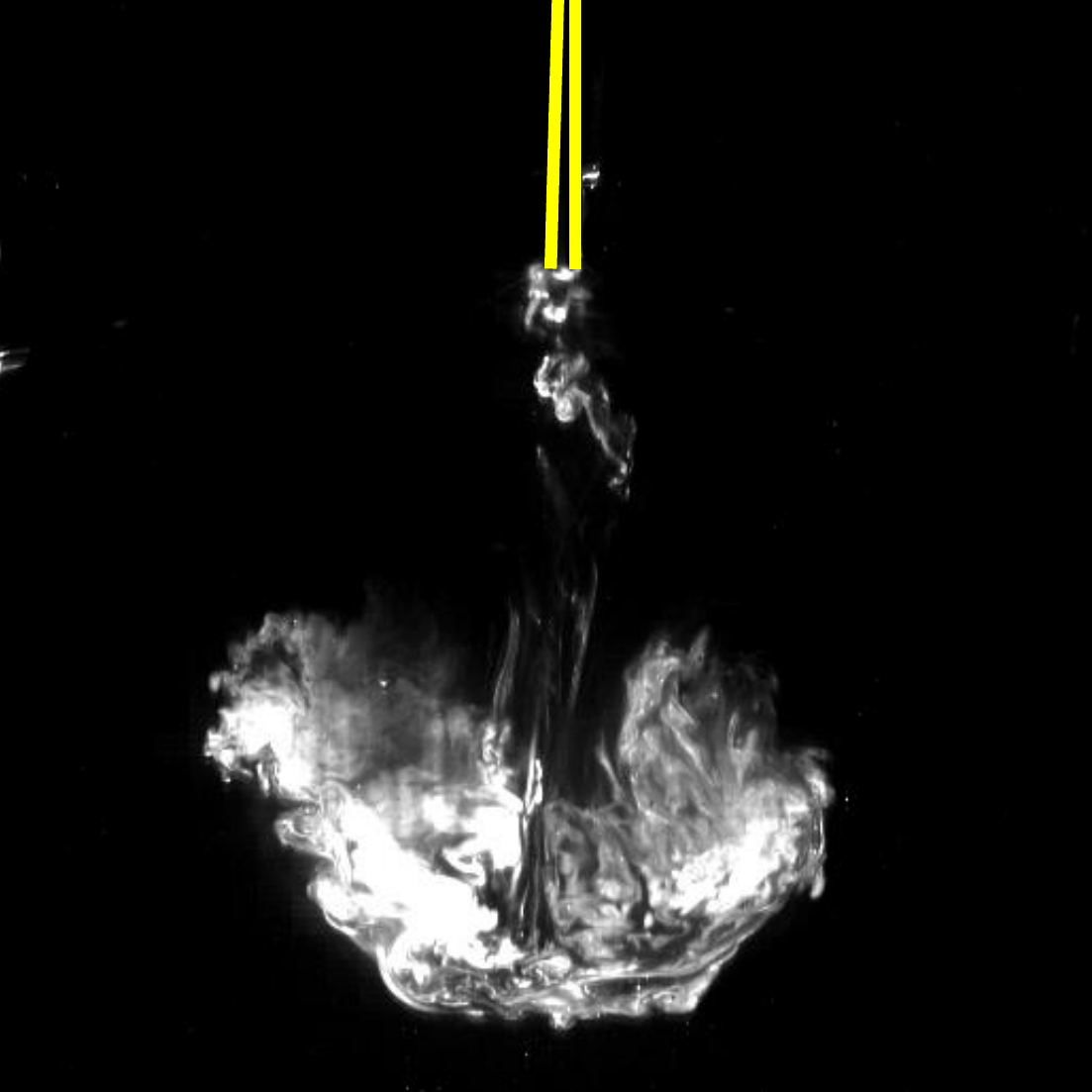}
		\caption{}
		\label{fig:VortexBreakDown_Stat}
	\end{subfigure}
	\caption{ PLIF visualization of the starting vortex at 500 ms in the wake of (a) a freely moving clapping body and (b) a forward motion constrained body, with $d^*$ = 1.0. The breakdown of the vortex occurs earlier in the stationary case, at about 340 ms.}\label{fig:VortexBreakDown_StatDyn}
\end{figure}
	Figure \hyperref[fig:VortZ_S_StatDyn]{8} shows the velocity vectors and the vorticity, $\omega_Z$, fields in the XY plane (Z=0) at different time instants after initiation of clapping for the stationary case  with $d^*=$ 0.5. The clapping action is complete in approximately 100 ms. The regions with non-zero vorticity are depicted by the red and blue areas, while the green color represents regions with approximately zero vorticity. The shadow regions are shown in gray and the two thick yellow lines show the rear parts of the plates extrapolated to a virtual pivot point. The vorticity values only above 0.05$\omega_{max}$ are shown. The clapping action produces a transient jet and counter-rotating vortex pair. The vortices form near the tips of the plates and move toward each other as the clapping motion progresses. The vortex pair seen in the XY plane is a cross-section of a three-dimensional vortex loop. As discussed below, the nature of the vortex loop changes with $d^*$. \par
	
	Figure \hyperref[fig:VortZ_D_StatDyn]{9} shows the flow fields for the dynamic case for the same $d^*$ value (= 0.5) as for the stationary case. The formation of the starting vortex pair is similar to the stationary case, but in the dynamic case, the body rapidly moves away due to the high pressure produced in the inter-plate cavity by the clapping action. The clapping motion, in this case, ends around 80 ms. After the acceleration phase ($t  \sim$ 0 - 42 ms), the body decelerates to zero velocity ($t  \sim$ 52 - 1000 ms) due to the drag force acting on it (figures \hyperref[fig:VortZ_D_StatDyn]{9e} - \hyperref[fig:VortZ_D_StatDyn]{9i}). We also observe that the fluid near the outside surfaces of both plates and in the near wake is carried forward by the body. Note during the clapping motion, the velocity vectors indicate an important difference between the stationary and dynamic bodies. In the dynamic case, the fluid in the upper part of the cavity (figure \hyperref[fig:VortZ_D_StatDyn]{9b}, $t$ = 0.0195 s) moves with the body, and the rest of the fluid moves out of the body, whereas in the stationary case (figure \hyperref[fig:VortZ_S_StatDyn]{8b}) at the same time all the fluid is moving out of the body. At 138.5 ms, the pair of starting vortices in the stationary case traversed a distance of 20 mm (figure \hyperref[fig:VortZ_S_StatDyn]{8i}); in the dynamic case, it traversed a distance of 12 mm, and the body traversed a distance of 70 mm (figure \hyperref[fig:VortZ_D_StatDyn]{9i}).\par
	
  	Figure \hyperref[fig:Vorticity-Z_StatDyn]{10} shows the velocity vectors and vorticity fields in the XY plane for the three $d^*$ values for the stationary cases (figures \hyperref[fig:Vorticity-Z_StatDyn]{10a, b, c}) and the dynamic cases (figures \hyperref[fig:Vorticity-Z_StatDyn]{10d, e, f}) near the end phase of the clapping. One major difference between stationary and dynamic cases is the existence of vorticity in the latter on the outside surfaces of the plates due to the forward motion of the body.\par
  	 	
  	The flow visualization pictures, in the XY plane, obtained using PLIF are shown for forward motion-constrained clapping cases in figures \hyperref[fig:Dye_visualization_StatDyn]{11a, b, c}, and for self-propelling cases in figures \hyperref[fig:Dye_visualization_StatDyn]{11d, e, f}. In these figures, the white dye represents the path traced by fluid ejected from the inner edges of the clapping plates. The yellow lines indicate the body’s position in the stationary cases, whereas in the dynamic cases, the body is out of the images. In figures \hyperref[fig:Dye_visualization_StatDyn]{11}(a, b, d, e), corresponding to $d^*$ = 1.5 and 1.0, the white circular patches represent part of the starting vortex, the initial formation process of which was discussed above. After the initial growth, the subsequent evolution, as we will discuss below, strongly depends on $d^*$. At 400 ms, we observed that vortices corresponding to $d^*$ = 1.5 and 1.0 cases still retain their identity (figure \hyperref[fig:Dye_visualization_StatDyn]{11}(a, b, d, e)), while for the $d^*$ = 0.5 cases, the starting vortices approach each other and undergo breakdown (figures \hyperref[fig:Dye_visualization_StatDyn]{11}(c, f)).
	
	Wake visualization reveals two significant differences between stationary and dynamic cases. The first is the presence of shear layers in the dynamic cases, that connect the edges of the two plates with the starting vortices; see figures \hyperref[fig:Dye_visualization_StatDyn]{11d, e, f}. The second is an earlier breakdown of starting vortices in the stationary cases; see figure \hyperref[fig:VortexBreakDown_StatDyn]{12}.

  \end{subsubsection}

   \begin{subsubsection}{Flow fields in the XZ plane}
			\begin{figure}
			\centering
			\begin{subfigure}[b]{0.28\textwidth}
				\includegraphics[width=\textwidth]{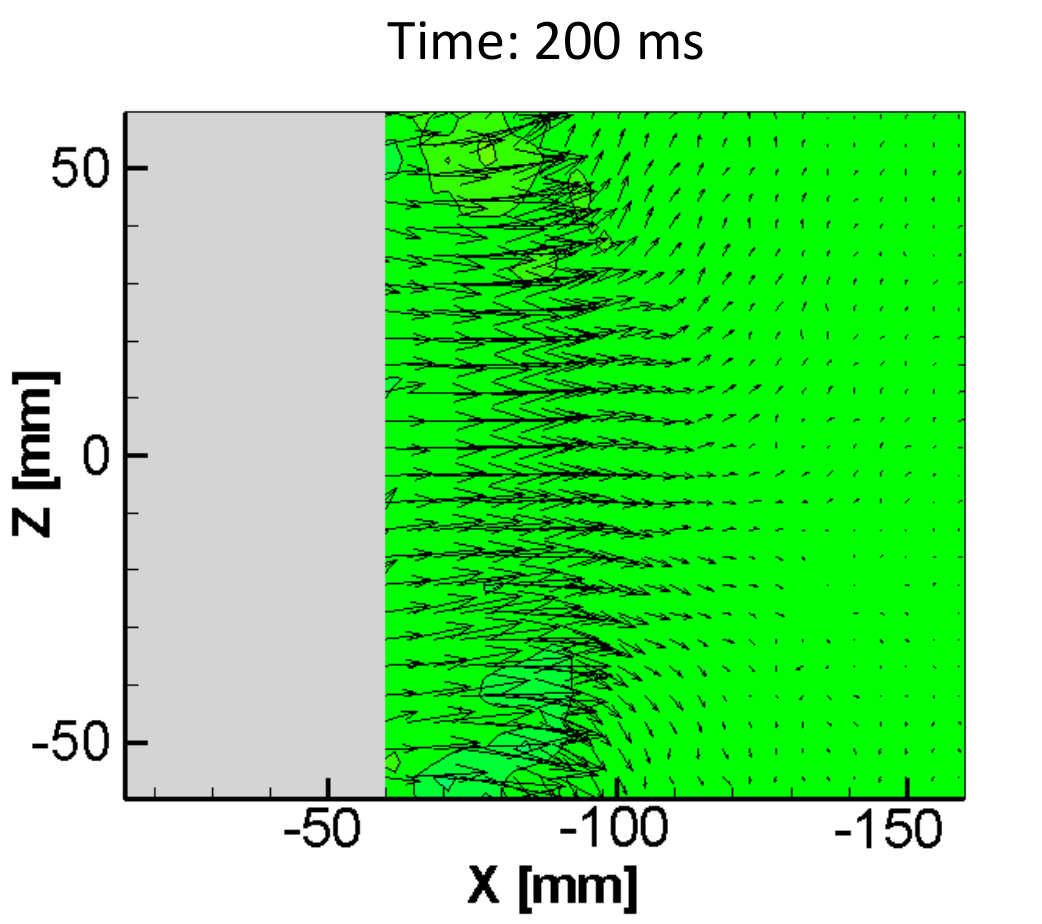}
				\caption{}
				\label{fig:VortY_AR067_Stat_01}
			\end{subfigure}\hspace{2mm}
			\begin{subfigure}[b]{0.28\textwidth}
				\includegraphics[width=\textwidth]{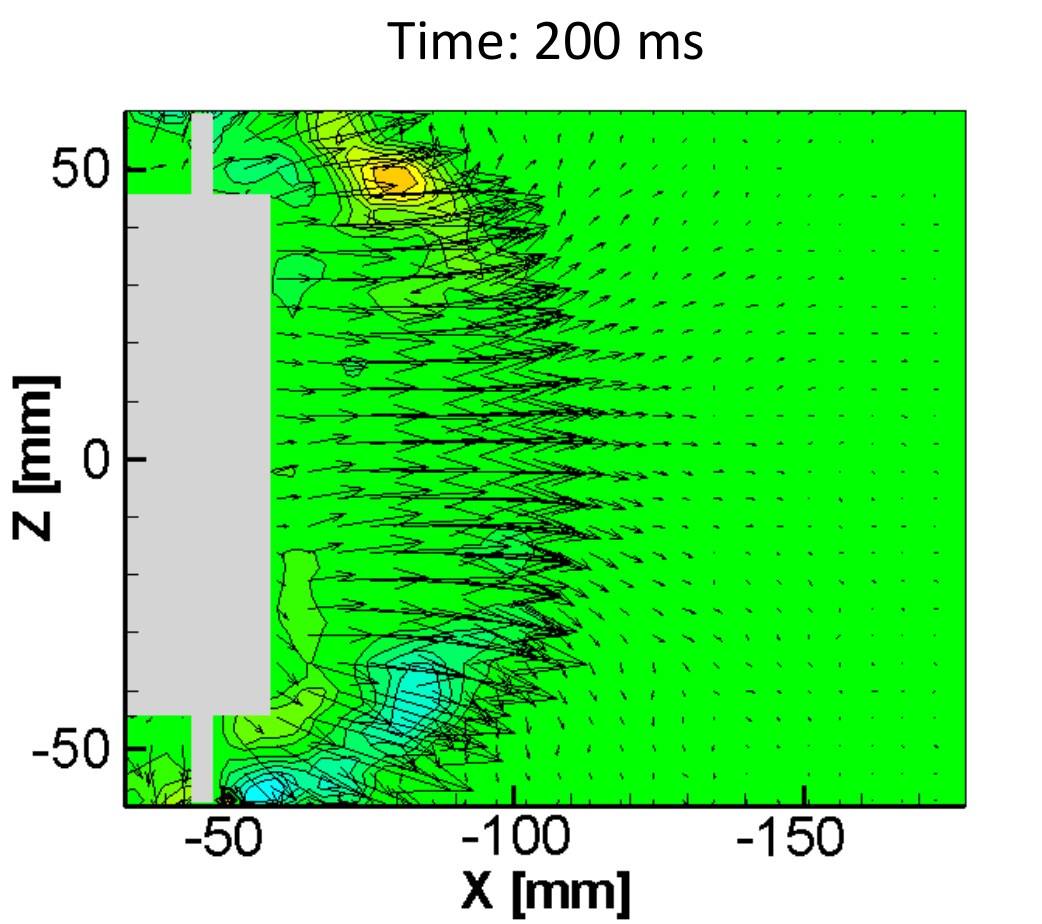}
				\caption{}
				\label{fig:VortY_AR100_Stat_01}
			\end{subfigure}\hspace{2mm}
			\begin{subfigure}[b]{0.28\textwidth}
				\includegraphics[width=\textwidth]{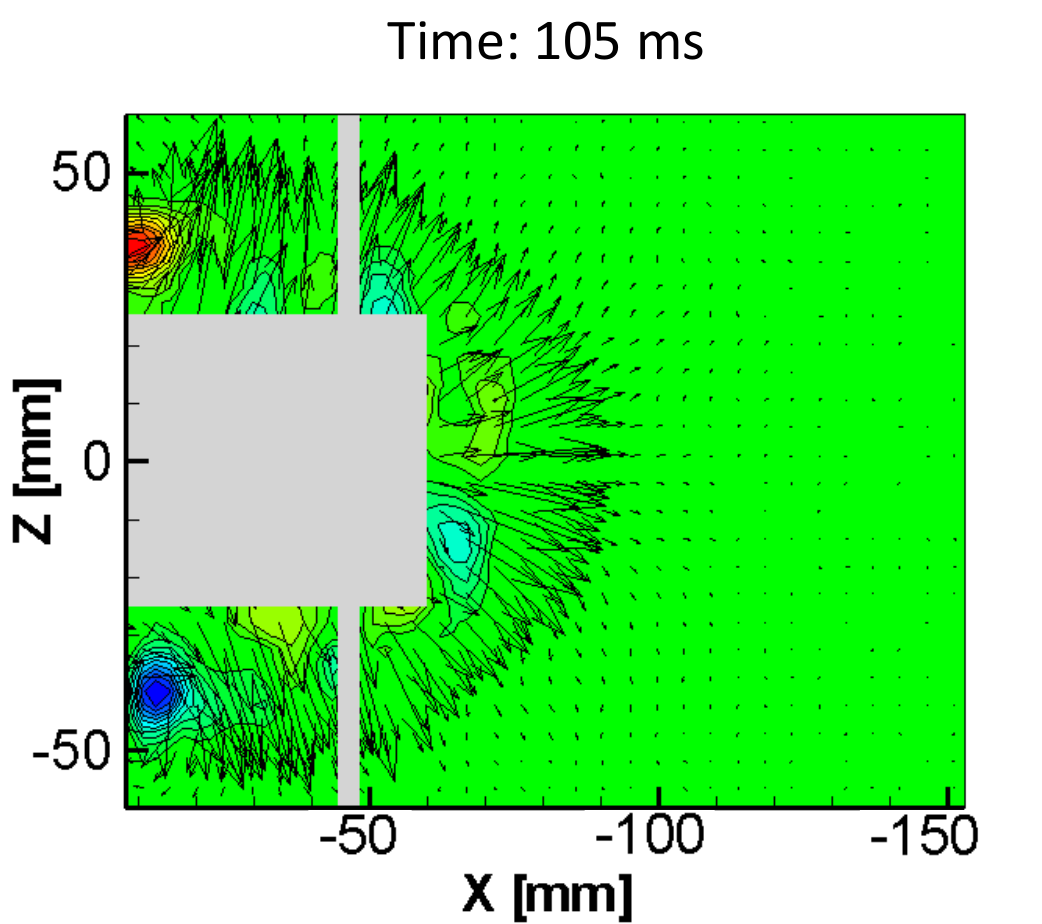}
				\caption{}
				\label{fig:VortY_AR200_Stat_01}
			\end{subfigure}\vspace{2mm}
			\begin{subfigure}[b]{0.28\textwidth}
				\includegraphics[width=\textwidth]{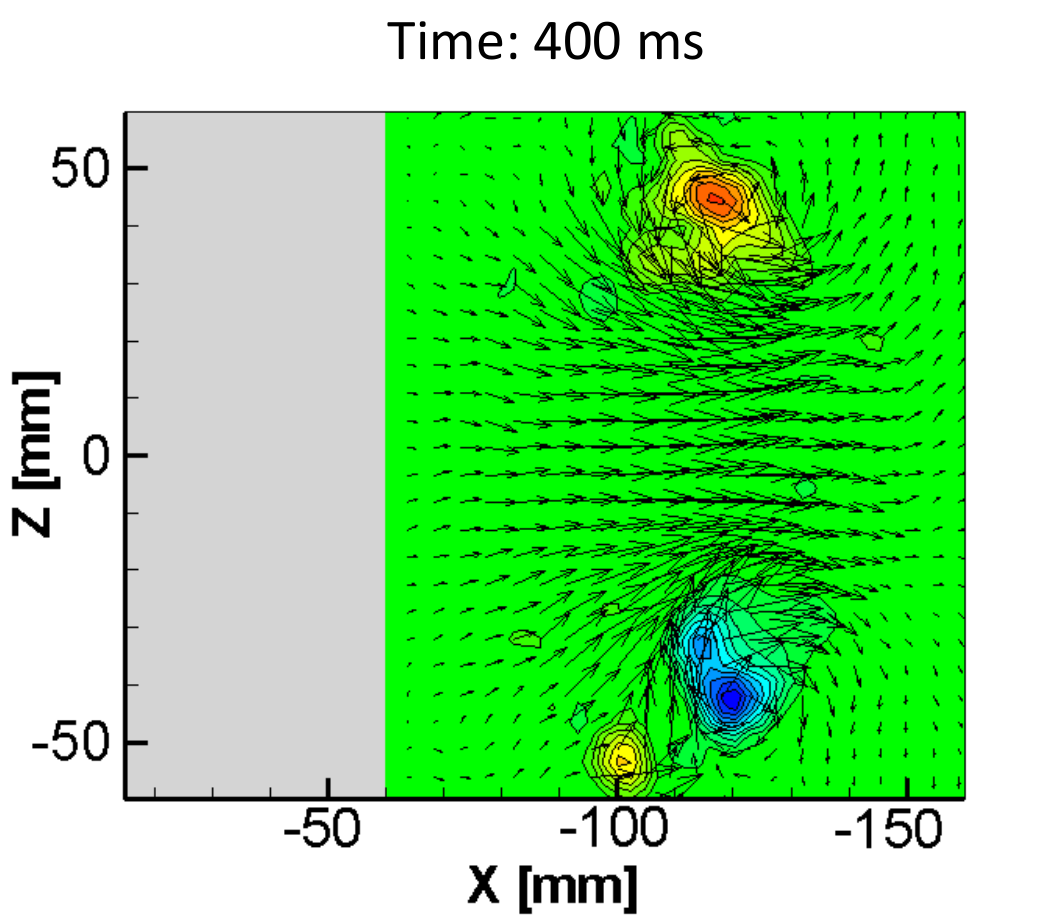}
				\caption{}
				\label{fig:VortY_AR067_Stat_02}
			\end{subfigure}\hspace{2mm}
			\begin{subfigure}[b]{0.28\textwidth}
				\includegraphics[width=\textwidth]{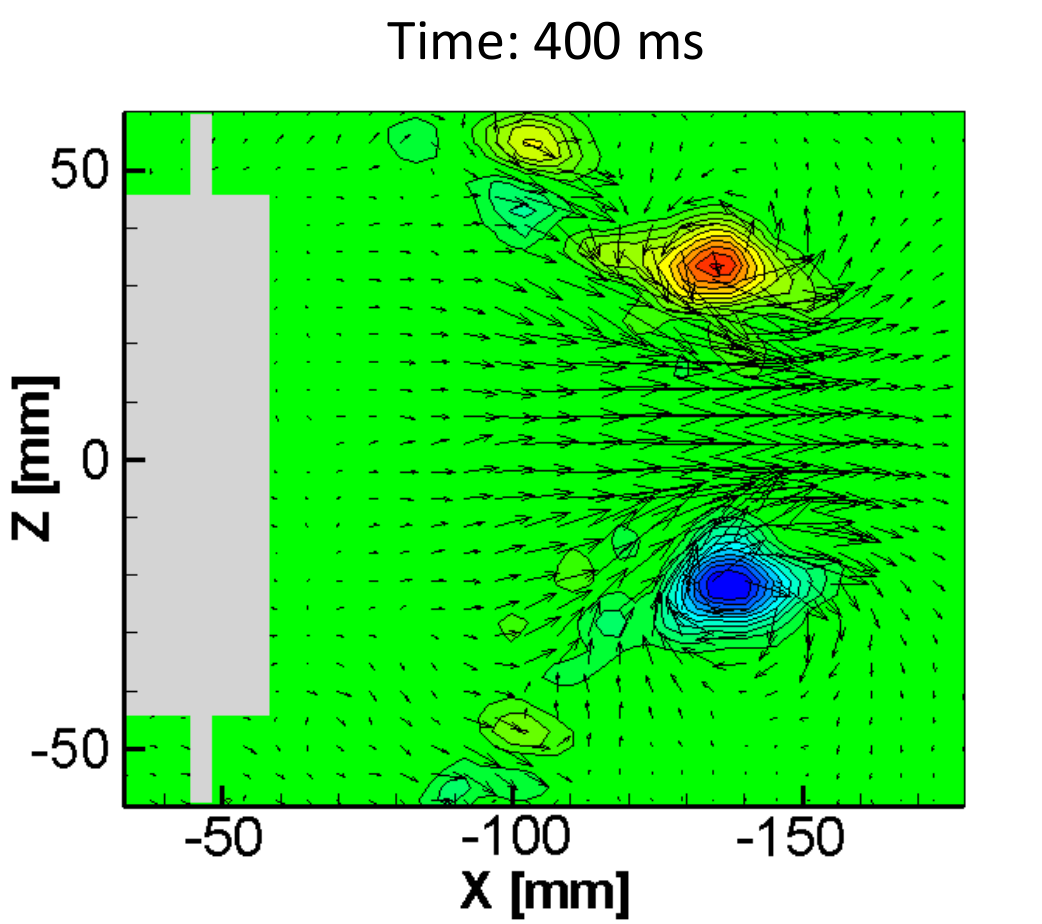}
				\caption{}
				\label{fig:VortY_AR100_Stat_02}
			\end{subfigure}\hspace{2mm}
			\begin{subfigure}[b]{0.28\textwidth}
				\includegraphics[width=\textwidth]{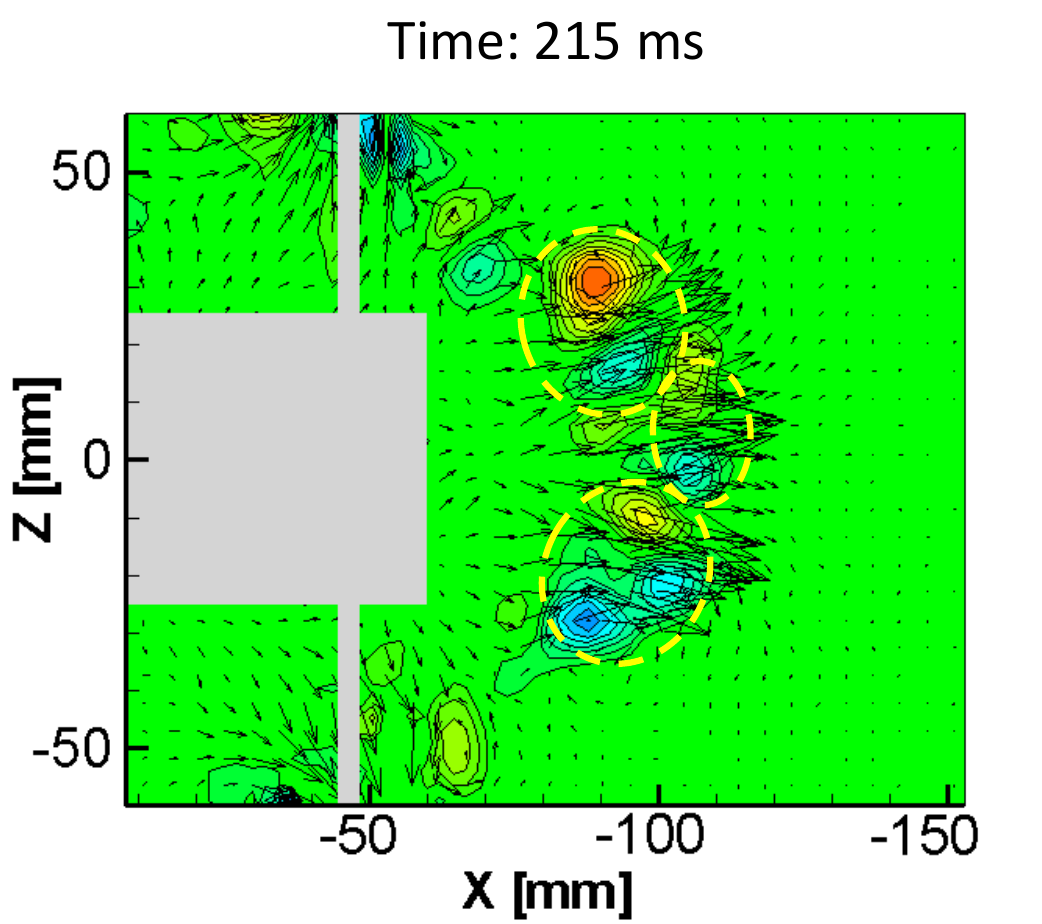}
				\caption{}
				\label{fig:VortY_AR200_Stat_02}
			\end{subfigure}\vspace{2mm}
			\begin{subfigure}[b]{0.35\textwidth}
				\includegraphics[width=\textwidth]{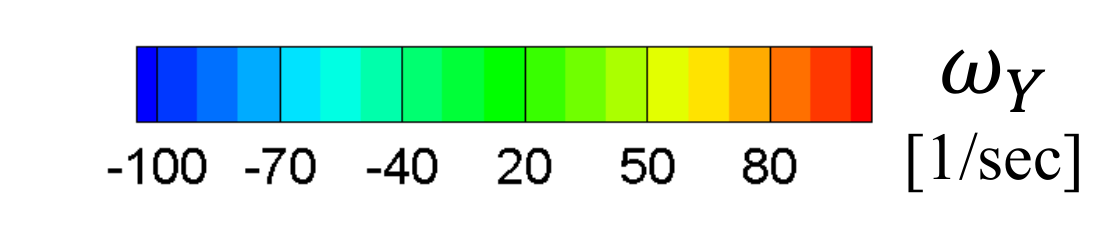}
			\end{subfigure}
			\caption{The flow field in the XZ plane for the sationary case for $d^* = $ 1.5 at (a) 200 ms and (d) 400ms; $d^* = $ 1.0 at (b) 200 ms and (e) 400ms; $d^* = $ 0.5 at (c) 105 ms and (f) 215 ms. The vertical gray line represents the shadow of the 6mm-thick rigid bar mounted on the release stand, and the gray rectangle indicates the clapping plates.}\label{fig:Vorticity-Y_S_StatDyn}
		\end{figure}
		\begin{figure}
			\centering
			\begin{subfigure}[b]{0.3\textwidth}
				\includegraphics[width=\textwidth]{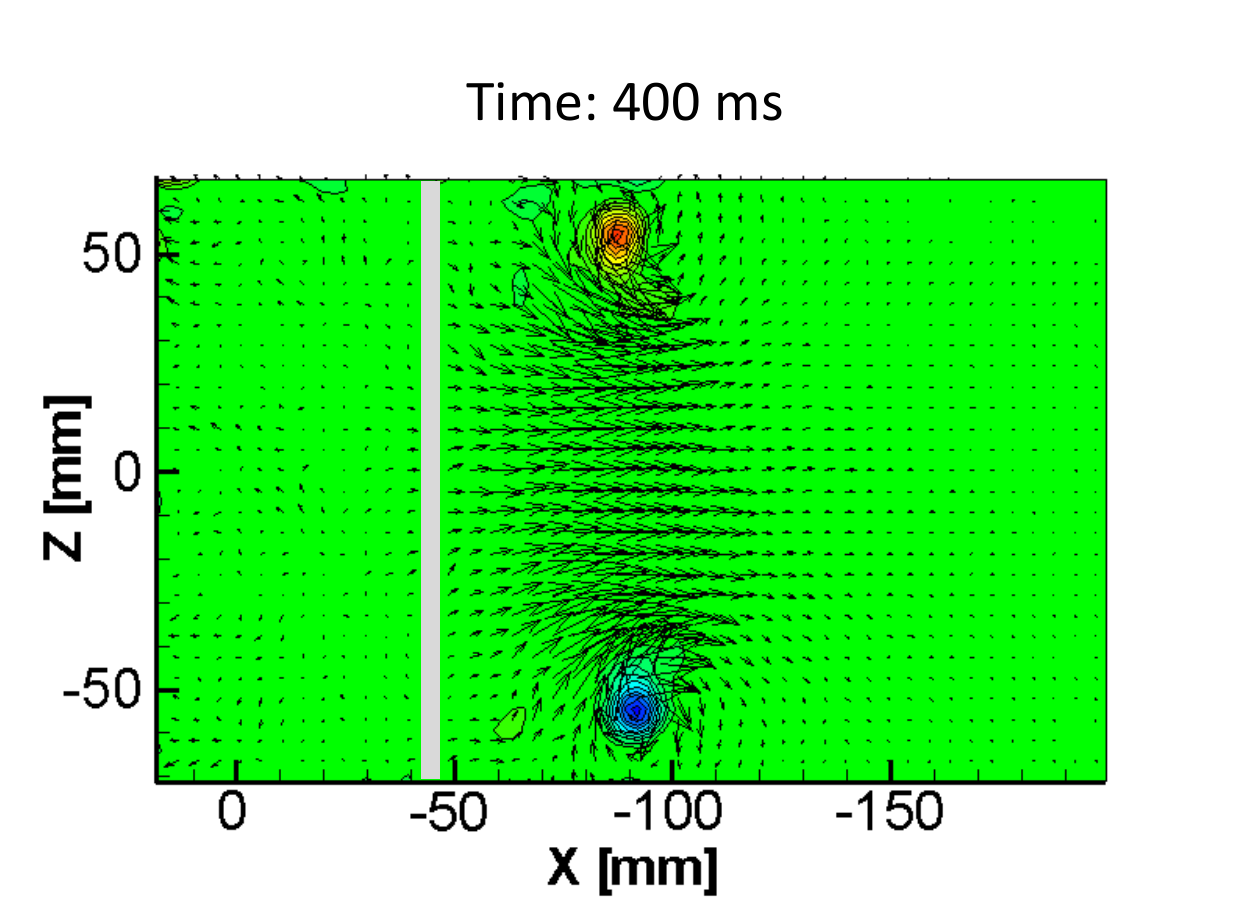}
				\caption{}
				\label{fig:VortY_AR067_Dyn_Stat}
			\end{subfigure}
			\begin{subfigure}[b]{0.3\textwidth}
				\includegraphics[width=\textwidth]{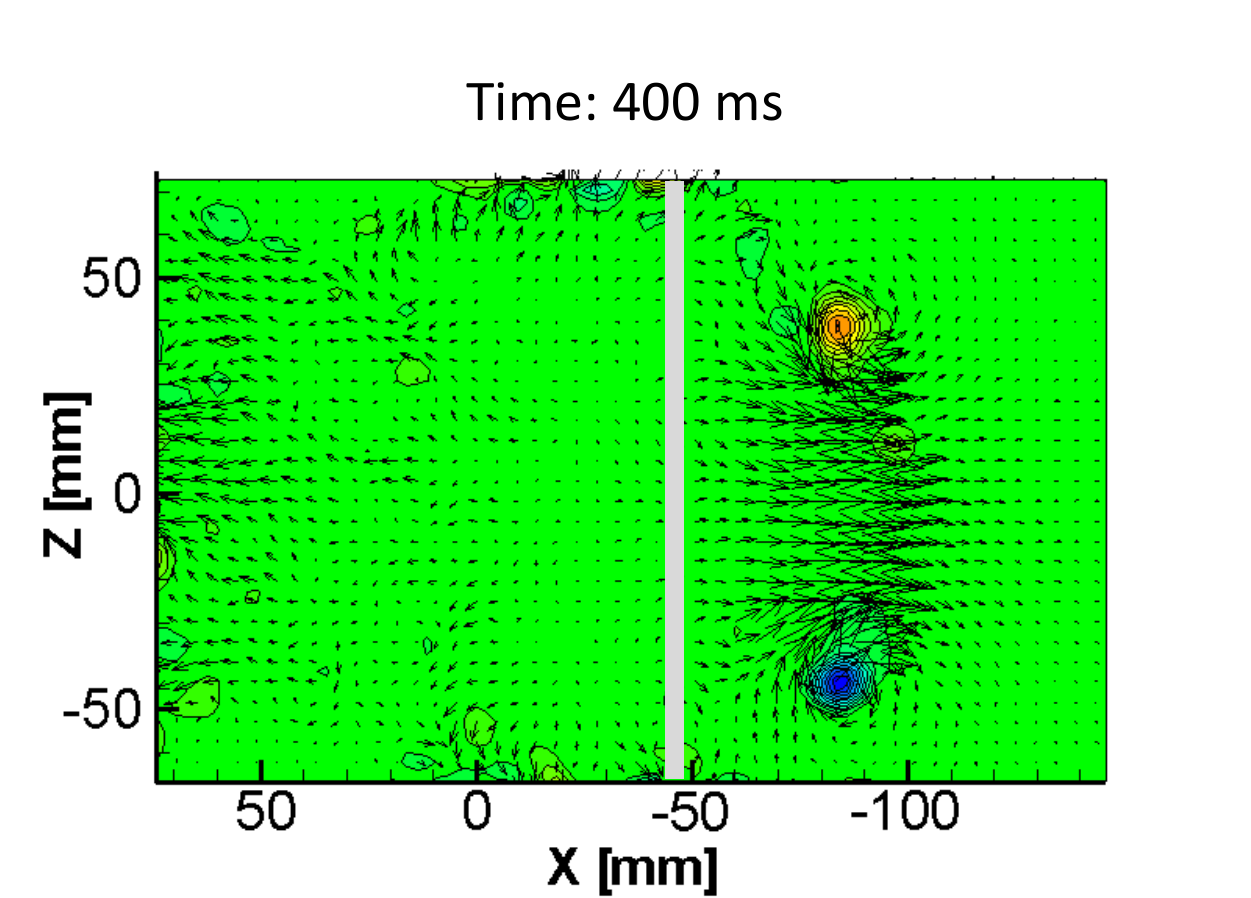}
				\caption{}
				\label{fig:Dye_AR100_Dyn_Stat}
			\end{subfigure}
			\begin{subfigure}[b]{0.3\textwidth}
				\includegraphics[width=\textwidth]{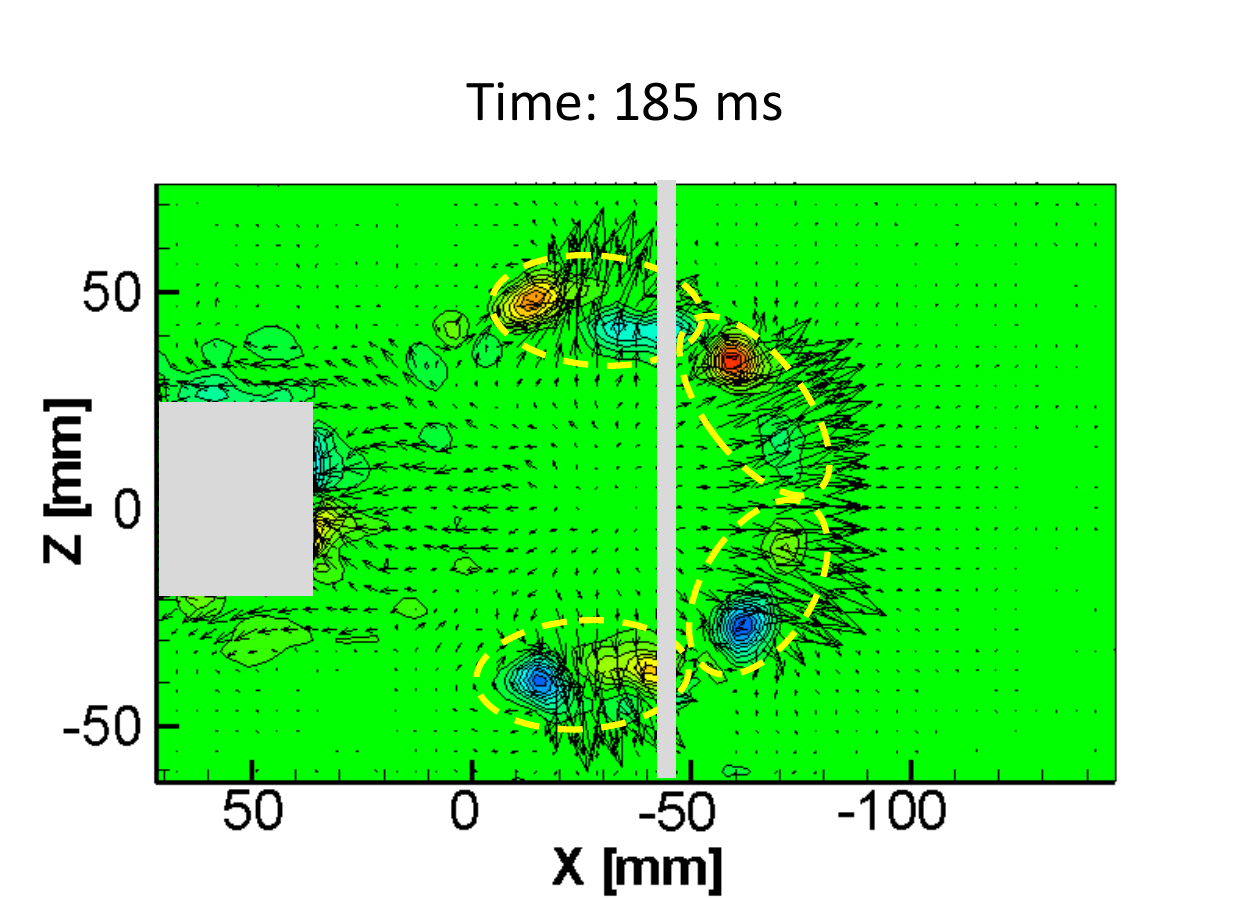}
				\caption{}
				\label{fig:Dye_AR200_Dyn_Stat}
			\end{subfigure}\vspace{2mm}
			\begin{subfigure}[b]{0.35\textwidth}
				\includegraphics[width=\textwidth]{FigDatStat/Vort_Y_Stat_Dyn_legend}
			\end{subfigure}
			\caption{The flow field in the XZ plane for the dynamic case: a) $d^* =$ 1.5 at 400 ms; b) $d^* =$ 1.0 at 400 ms; c) $d^* =$ 0.5 at 185 ms.}\label{fig:Vorticity-Y_D_StatDyn}
		\end{figure}
		
		\begin{figure}
			\centering
			\begin{subfigure}[b]{0.42\textwidth}
				\includegraphics[width=\textwidth]{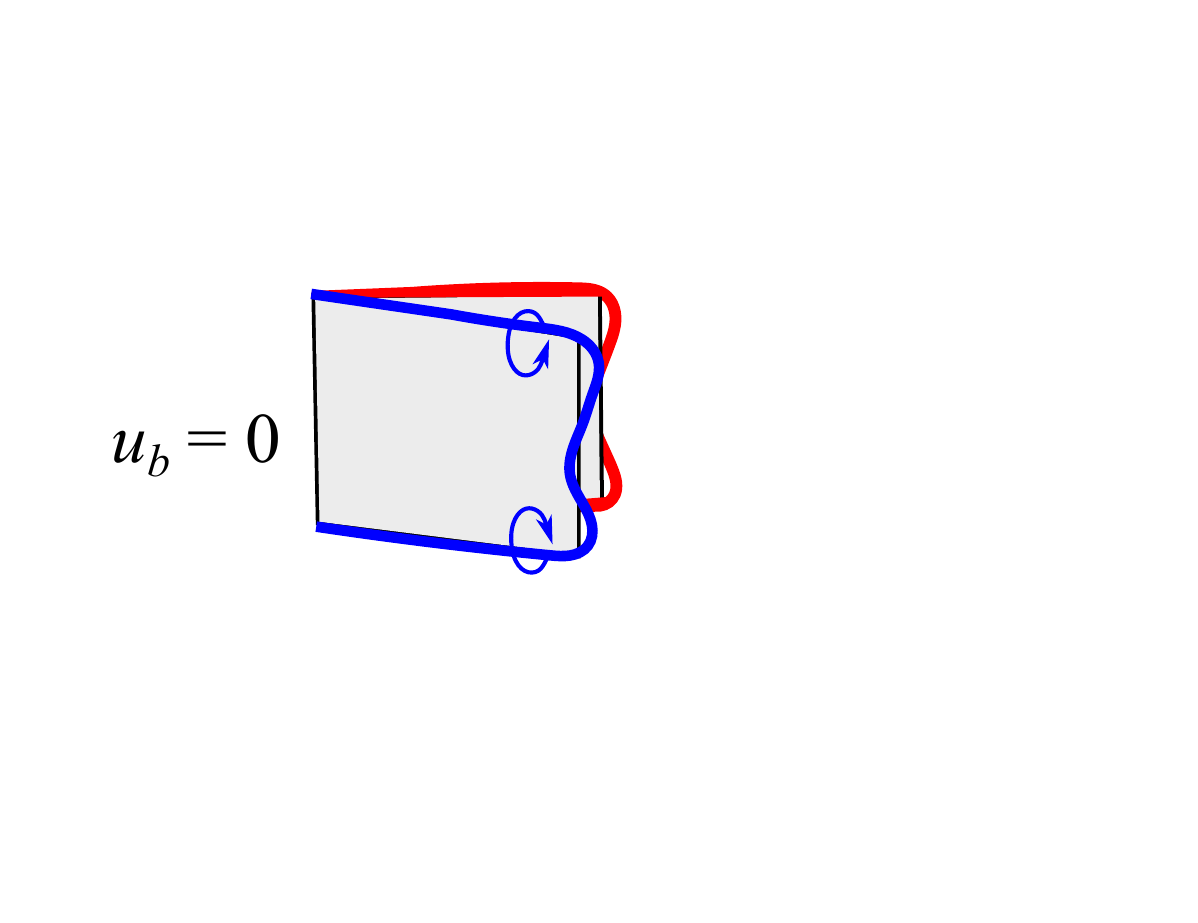}
				\caption{}
				\label{fig:3DVort_AR200_STAT_40ms}
			\end{subfigure}\hspace{5mm}
			\begin{subfigure}[b]{0.4\textwidth}
				\includegraphics[width=\textwidth]{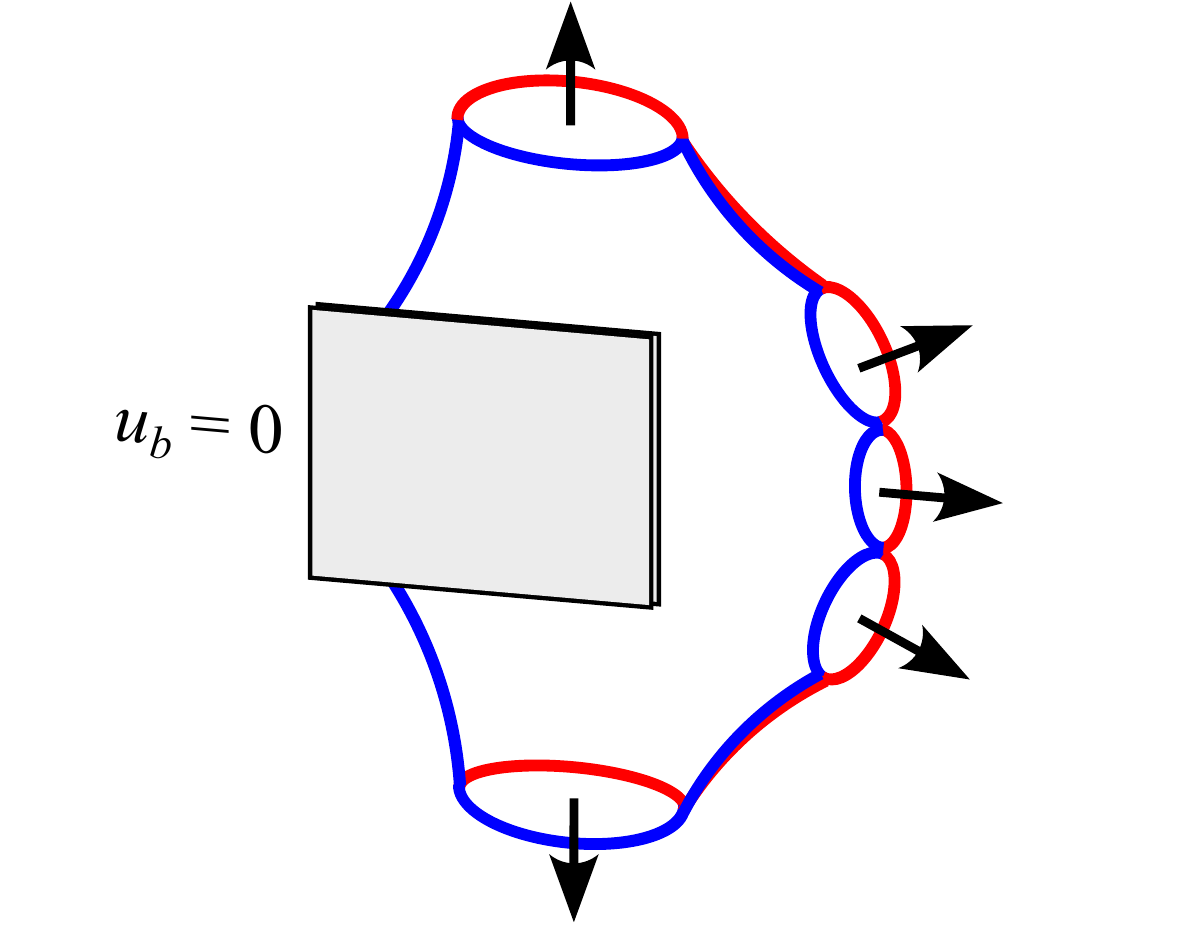}
				\caption{}
				\label{fig:3DVort_AR200_STAT_200ms}
			\end{subfigure}\vspace{2mm}
			\begin{subfigure}[b]{0.42\textwidth}
				\includegraphics[width=\textwidth]{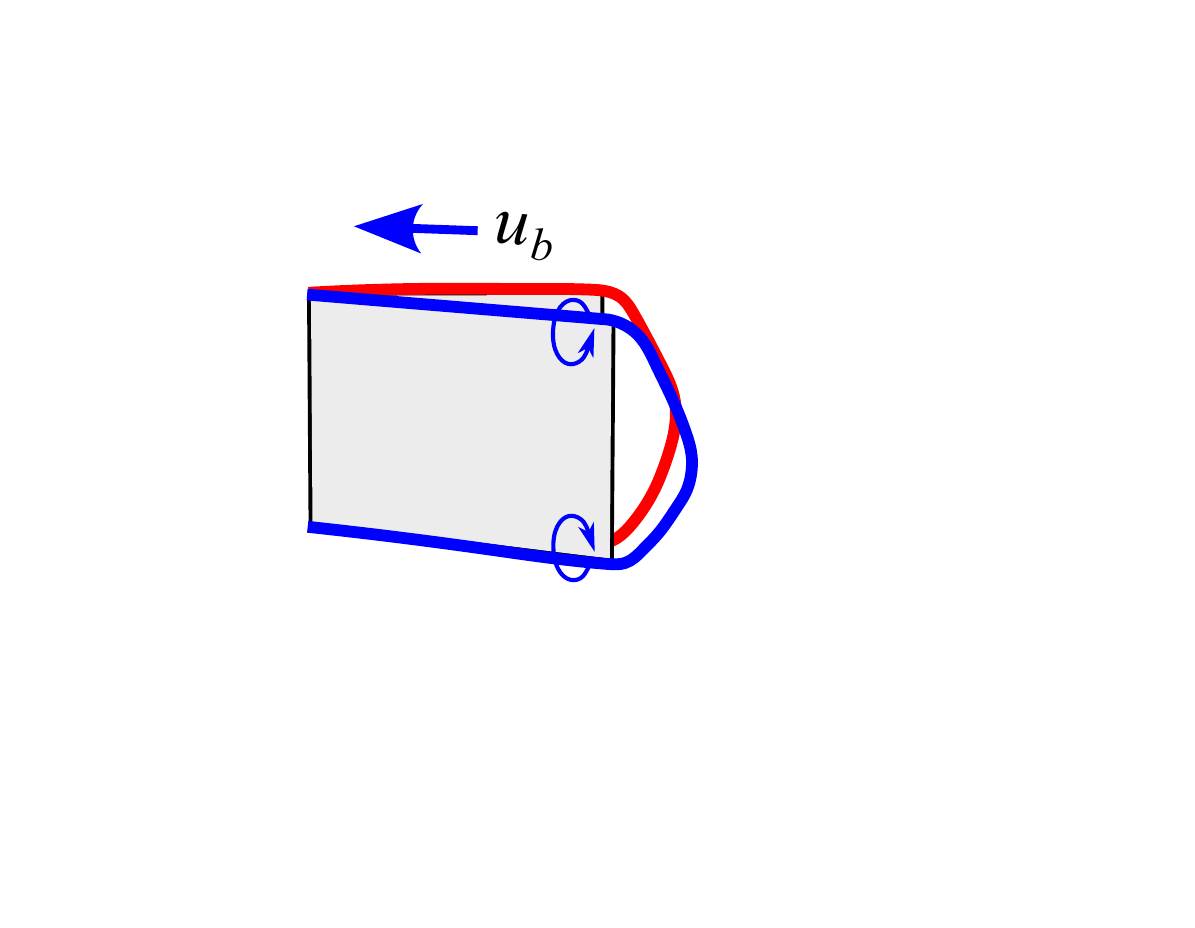}
				\caption{}
				\label{fig:3DVort_AR200_DYN_40ms}
			\end{subfigure}\hspace{5mm}
			\begin{subfigure}[b]{0.4\textwidth}
				\includegraphics[width=\textwidth]{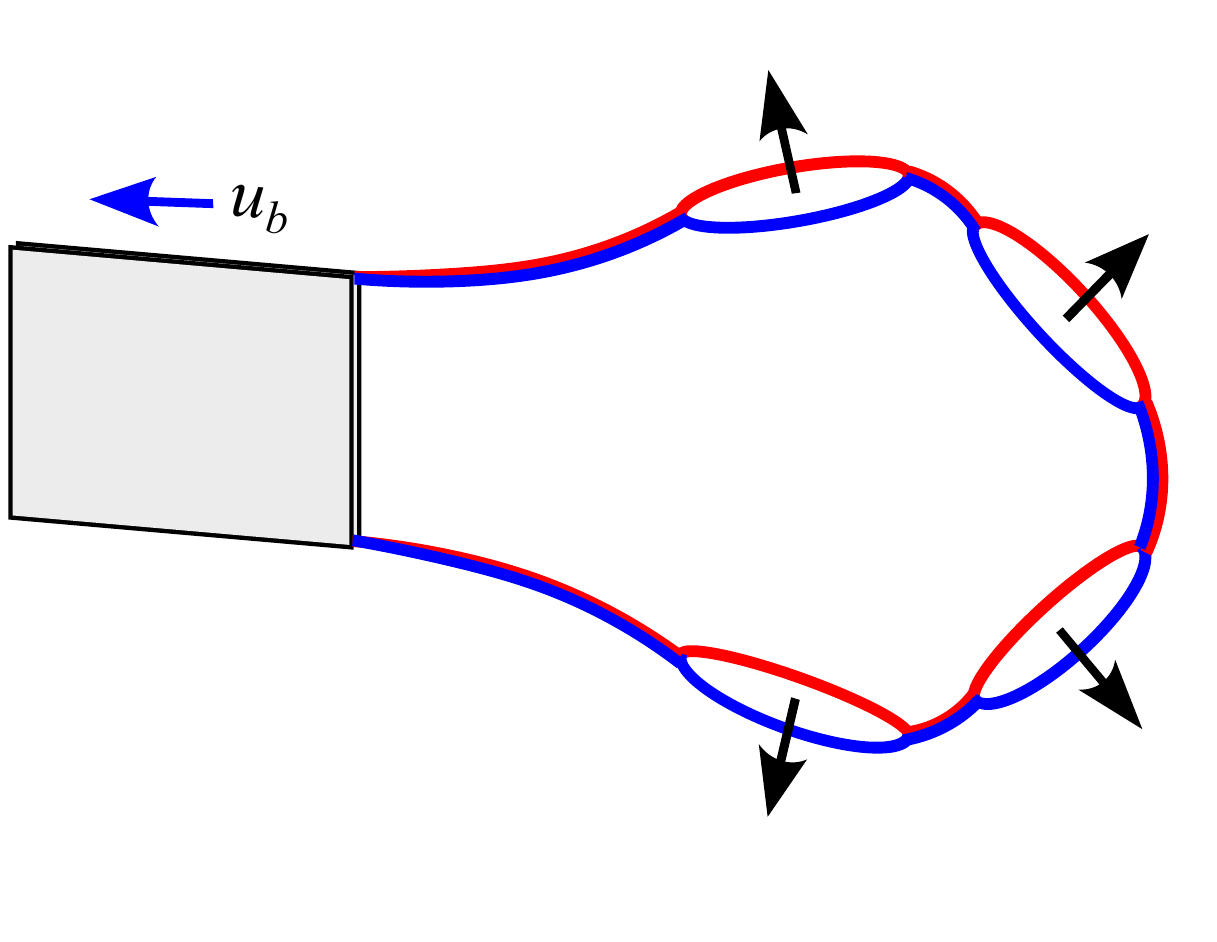}
				\caption{}
				\label{fig:3DVort_AR200_DYN_200ms}
			\end{subfigure}
			\caption{Schematics of the vortex loops for  $d^* =$ 0.5 in the stationary case at (a) 40 ms and (b) 200 ms and in the dynamic case at (c) 40 ms and (d) 200 ms.
			}\label{fig:3D Vortex loop_StatDyn}
		\end{figure}
	
	Since all three bodies have  aspect ratios of the order of one, we expect three–dimensional vortex loops to form on the three edges of each plate. The starting vortices are cross-sections of these loops in the XY plane (figures \hyperref[fig:VortZ_S_StatDyn]{8}, \hyperref[fig:VortZ_D_StatDyn]{9}). The flow fields in the XZ plane shown in figure \hyperref[fig:Vorticity-Y_S_StatDyn]{13}, for the stationary cases, and in figure \hyperref[fig:Vorticity-Y_D_StatDyn]{14}, for the dynamic cases may be used to deduce the structures of these vortex loops. The pressure generated during clapping not only forces fluid motion in the downstream (- ve X) direction but also sideways along the $\pm$ Z directions, most clearly visible for the $d^*$ = 1.0 and 0.5 bodies (figure \hyperref[fig:Vorticity-Y_S_StatDyn]{13b, c}). Due to this sideway flow, parts of the vortex loops at the top and bottom edges of the clapping plates will have vorticity predominantly in the streamwise direction. Qualitatively, the vortex generation mechanisms are the same in the stationary and dynamic cases; there are some differences due to the rapid forward motion of the body during the clapping for the latter case, that we will discuss below. As the clapping motion starts, radially outward flow develops in the XZ plane around the body (figure \hyperref[fig:Vorticity-Y_S_StatDyn]{13c}). At later times, for both stationary and dynamic cases, there are complex interactions and reconnections of the different components of the initial vortex loop, leading to single elliptical vortex rings for the $d^*$ = 1.0 and 1.5 bodies (figures \hyperref[fig:Vorticity-Y_S_StatDyn]{13d, e} and \hyperref[fig:Vorticity-Y_D_StatDyn]{14a, b}), and multiple ringlets for the shortest ($d^*$ = 0.5) body (figures \hyperref[fig:Vorticity-Y_S_StatDyn]{13f} and \hyperref[fig:Vorticity-Y_D_StatDyn]{14c}).  As the plates approach each other, splitting and reconnection (Kida et al.\cite{Kida89}, Melander et al.\cite{Melander89}) of the original vortex loops leads to the formation of the new multiple vortex loops. \par 
	
	A schematic of the vortex loop, consistent with flowfield observations in the XY and XZ planes, is shown for the forward motion-constrained clapping body with $d^*$ = 0.5 case in figures \hyperref[fig:3D Vortex loop_StatDyn]{15(a, b)}, and for a freely moving body in figures \hyperref[fig:3D Vortex loop_StatDyn]{15(c, d)}. The blue and red line shows oppositely signed vortex filaments. In the stationary cases, the vortex filament follows the rotating plate till the end of clapping motion. In the initial phase of the clapping motion, the portions of both vortex filaments in the mid-depth region lags behind the plate; this has been observed across all $d^*$ cases (figures \hyperref[fig:3D Vortex loop_StatDyn]{15a} and \hyperref[fig:VortZ_S_StatDyn]{8c}). At the end of the clapping motion, the two loops approach each other in some locations and interconnect. For the $d^*$ = 0.5 body, we observe three vortex ringlets in the wake of the body, in addition to the two ringlets at the top and bottom sides of the body (figures \hyperref[fig:3D Vortex loop_StatDyn]{15b} and \hyperref[fig:Vorticity-Y_S_StatDyn]{13f}). The top and bottom ringlets are outside the view of the velocity field shown in the figure \hyperref[fig:Vorticity-Y_S_StatDyn]{13f}. The multiple ringlet formation is possibly similar to that observed along the circumference after the head-on collision of two circular vortex rings (Lim et al.\cite{Lim92}, Cheng et al.\cite{MCheng18}). In the wake of the stationary bodies with $d^* =$ 1.5 and 1.0, after the end of the clapping motion, we observe a single elliptical vortex ring, whose cross-section can be seen in figures \hyperref[fig:Vorticity-Y_S_StatDyn]{13(d, e)}, besides the two smaller vortex rings at the top and at the bottom sides of the plate. Approximately at 500 ms, the vortex structure breaks down (figure \hyperref[fig:VortexBreakDown_StatDyn]{12b}).\par
	
	For the freely moving clapping body, the initial vortex loop formation mechanisms are similar to those for the stationary case. However, later, due to the translation, the vortex filaments near the center detach from the trailing edges of the plates and are left behind the propelling body (see figures \hyperref[fig:3D Vortex loop_StatDyn]{15c} and \hyperref[fig:VortZ_D_StatDyn]{9c}); this is observed in all $d^*$ cases. In the stationary cases, the vortices remain closer to the rotating plate (figure \hyperref[fig:3D Vortex loop_StatDyn]{15a}). The later evolutions of the vortex loops depend on the $d^*$ values in both stationary and dynamic cases. In the dynamic case for the $d^*$ = 0.5 body, we observe four circumferentially connected ringlets in the wake (figure \hyperref[fig:3D Vortex loop_StatDyn]{15d} and \hyperref[fig:Vorticity-Y_D_StatDyn]{14c}),  whereas in the stationary cases, we see three ringlets in the wake, apart from two rings at the top and bottom sides (figure \hyperref[fig:3D Vortex loop_StatDyn]{15b}). In both the stationary and dynamic cases for $d^* =$ 1.5 and 1.0, a single large elliptical ring, with major axes aligned in the depth direction, and two rings at the top and bottom sides, are observed. The cross-sections of the elliptic rings is seen in figures \hyperref[fig:Vorticity-Y_S_StatDyn]{13d, e} and \hyperref[fig:Vorticity-Y_D_StatDyn]{14a, b} for the stationary and dynamics cases,  respectively. For the dynamic cases, the elliptical rings switch axes as they translate downstream, around 800 ms in the $d^* =$ 1.5 case and around 1300 ms in the $d^* =$ 1.0 case; axis switching is not observed in the stationary cases due to early vortex breakdown (figure \hyperref[fig:VortexBreakDown_StatDyn]{12b}). Cheng et al.\cite{MCheng16} and  Dahank et al.\cite{Dhanak81} have provided a detailed discussion on axis switching.\par
	\end{subsubsection}

	%%%%%%%%%%%%%%%%%%%%%%%%%%%%%%%%%%%%%%%%%%%%%%%%%%%%%%%%%%%%%%%%%%%%%%%%%%%%%%%%%%%
	\begin{subsubsection}{Core separation}
	\label{sec:coreSep_StatDyn}
		\begin{figure}
		\centering
		\begin{subfigure}[b]{0.45\textwidth}
			\includegraphics[width=\textwidth]{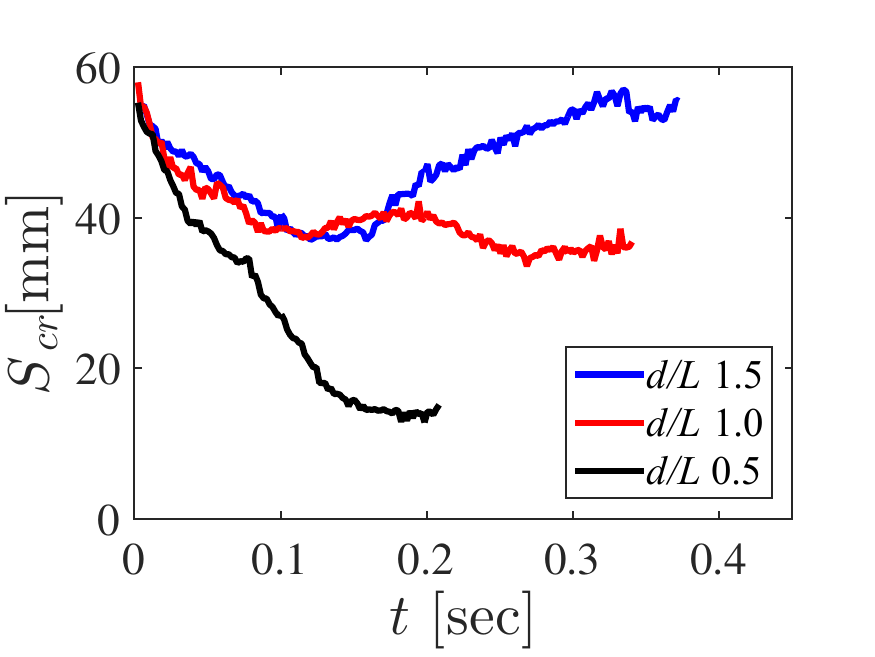}
			\caption{}
			\label{fig:MeanCircualtion_Stat}
		\end{subfigure}\hspace{02mm}
		\begin{subfigure}[b]{0.45\textwidth}
			\includegraphics[width=\textwidth]{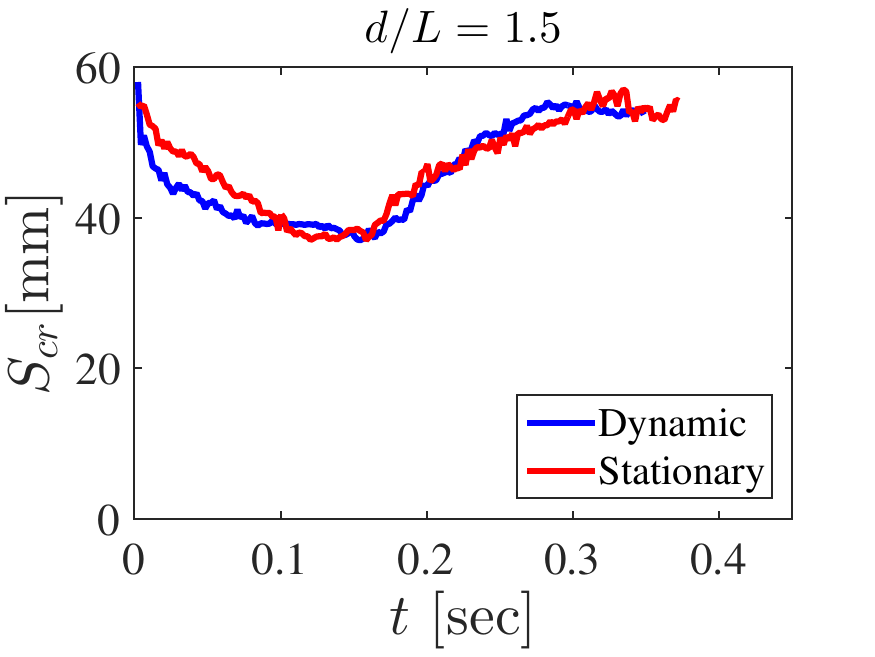}
			\caption{}
			\label{fig:CoreSep_AR067_Stat_Dyn}
		\end{subfigure}\vspace{02mm}
		\begin{subfigure}[b]{0.45\textwidth}
			\includegraphics[width=\textwidth]{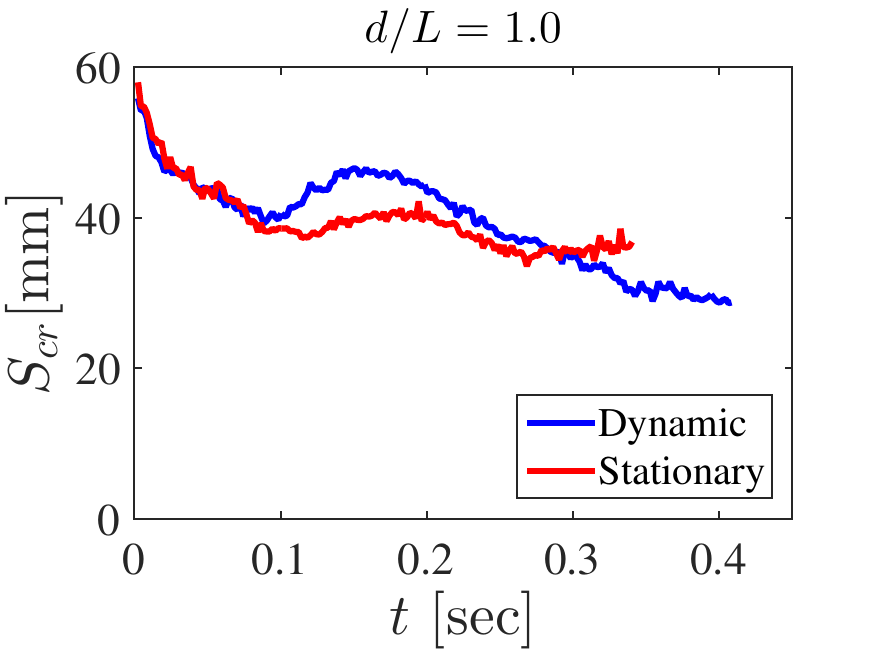}
			\caption{}
			\label{fig:CoreSep_AR100_Stat_Dyn}
		\end{subfigure}\hspace{02mm}	
		\begin{subfigure}[b]{0.45\textwidth}
			\includegraphics[width=\textwidth]{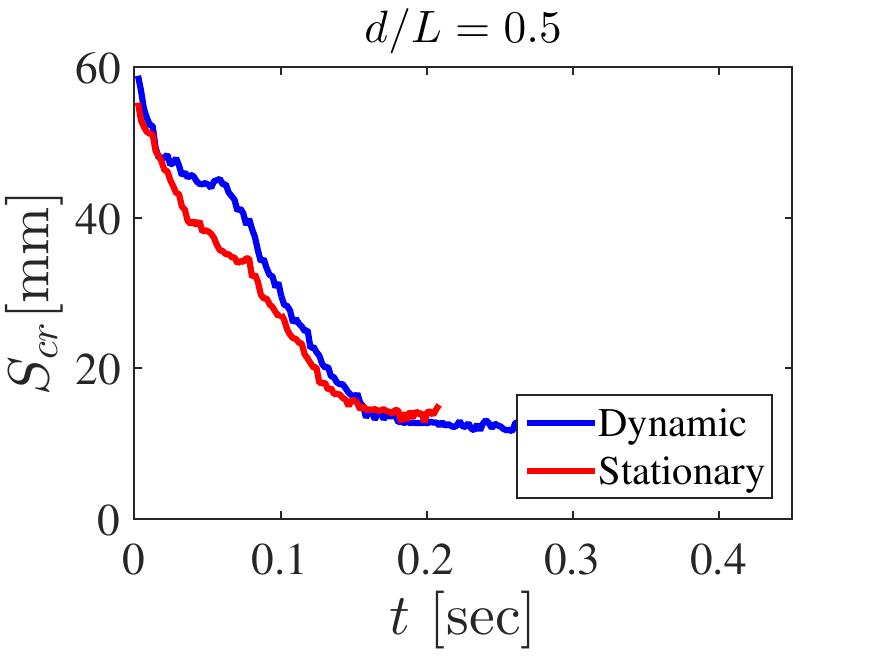}
			\caption{}
			\label{fig:CoreSep_AR200_Stat_Dyn}
		\end{subfigure}\vspace{02mm}
		\begin{subfigure}[b]{0.45\textwidth}
			\includegraphics[width=\textwidth]{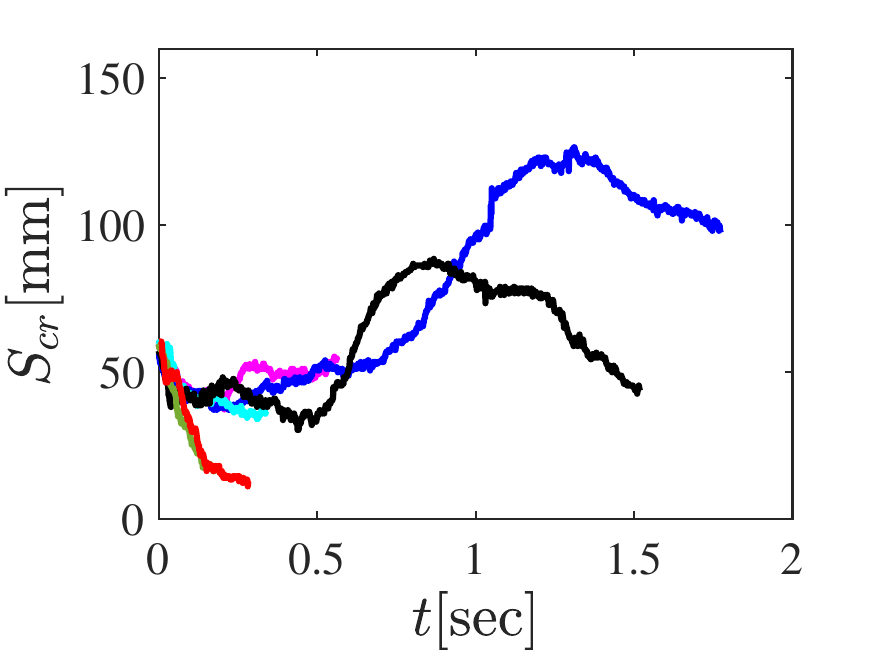}
			\label{fig:WakeVis_CoreSep_Stat_Dyn}
		\end{subfigure}
		\begin{subfigure}[b]{0.55\textwidth}
			\includegraphics[width=\textwidth]{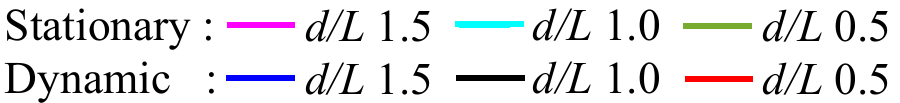}
			\caption{}
		\end{subfigure}
		\caption{(a) Time variation in the separation distance, $S_{cr}$, between the vortex cores in the XY plane for the stationary cases corresponding to the three $d^*$ values. Comparison of $S_{cr}$ versus time plots for the stationary and dynamic cases for (b) $d^* =$ 1.5, (c) $d^* =$ 1.0, and (d) $d^* =$ 0.5. (e) $S_{cr}$ versus time curves for stationary and dynamic cases obtained from PLIF visualization.}\label{fig:CoreSeparation_StatDyn}
	\end{figure}

	The variation of the spacing of the starting vortices in the XY plane with time gives some insight into the initial formation and later evolution of the wake vortex loops. Figure \hyperref[fig:CoreSeparation_StatDyn]{16a} shows the core separation, $S_{cr}$, with time for the three aspect ratios for the stationary body. In all three cases, the initial core separation value is approximately 60 mm, comparable to the initial spacing between the tips of both plates, after which there is a rapid reduction in $S_{cr}$ till the end of the clapping motion. The process of the initial variation in $S_{cr}$ is understood by looking at the velocity fields, an example of which is seen in figure \hyperref[fig:VortZ_S_StatDyn]{8} for $d^* =$ 0.5 body. The later evolution of $S_{cr}$ is strongly dependent on the vortex reconnection process and thus strongly dependent on $d^*$. The variations of $S_{cr}$ with time, as shown in figures \hyperref[fig:CoreSeparation_StatDyn]{16a,b,c,d}, are obtained from the PIV velocity field. There is not much difference in the initial vortex spacing time evolutions between the stationary and dynamic cases, as shown in figures \hyperref[fig:CoreSeparation_StatDyn]{16b,c,d}. In both stationary and dynamic cases, after the end of the clapping motion, $S_{cr}$ shows rapid reduction in $d^* =$ 0.5 cases, whereas, after 200 ms, it gradually increases in $d^* =$ 1.5 cases and remains approximately constant in the $d^* =$ 1.0 cases. The average of standard deviations over time in $S_{cr}$ is less than 7\%  of the initial spacing between the tips of both plates.\par  
	
	The vortex core spacing data obtained for longer times from dye visualization are shown in figure \hyperref[fig:CoreSeparation_StatDyn]{16e} for both stationary and dynamic cases across the three $d^*$ values. At 400 ms, dye visualization images show almost equal core separations in stationary and dynamic cases for $d^* =$ 1.5 and 1.0 cases (see also figure \hyperref[fig:Dye_visualization_StatDyn]{11}). The subsequent variations in $S_{cr}$ for the dynamic cases with $d^*$ = 1.0 and 1.5 are consistent with axis switching in the elliptic vortex rings. For $d^*$ = 1.5, initially $S_{cr}$(= 60 mm) corresponds to  the minor axis, which is in the  XY plane when the major axis is approximately equal to depth (133 mm) along the Z direction; at around, 1300 ms, $S_{cr}$ increases to 126 mm (figure \hyperref[fig:CoreSeparation_StatDyn]{16e}) corresponding to the major axis, now in the Y direction, indicating the axes switching. For the $d^*$ = 1.0 case the switch in axes happens earlier at around 700 ms. Due to early breakdown (figure \hyperref[fig:VortexBreakDown_StatDyn]{12b}), the axis switching is not observed in stationary cases, although the initial variations in the core separations are similar to the dynamic cases.\par
	
	\end{subsubsection}
%%%%%%%%%%%%%%%%%%%%%%%%%%%%%%%%%%%%%%%%%%%%%%%%%%%%%%%%%%%%%%%%%%%%%%%%%%%%%%%%%%%%%%%%%%%%
	\begin{subsubsection}{Momentum and circulation in the wake}
	\label{sec:WakeMom_Gamma_StatDyn}	
		\begin{figure}
		\centering
		\begin{subfigure}[b]{0.32\textwidth}
			\includegraphics[width=\textwidth]{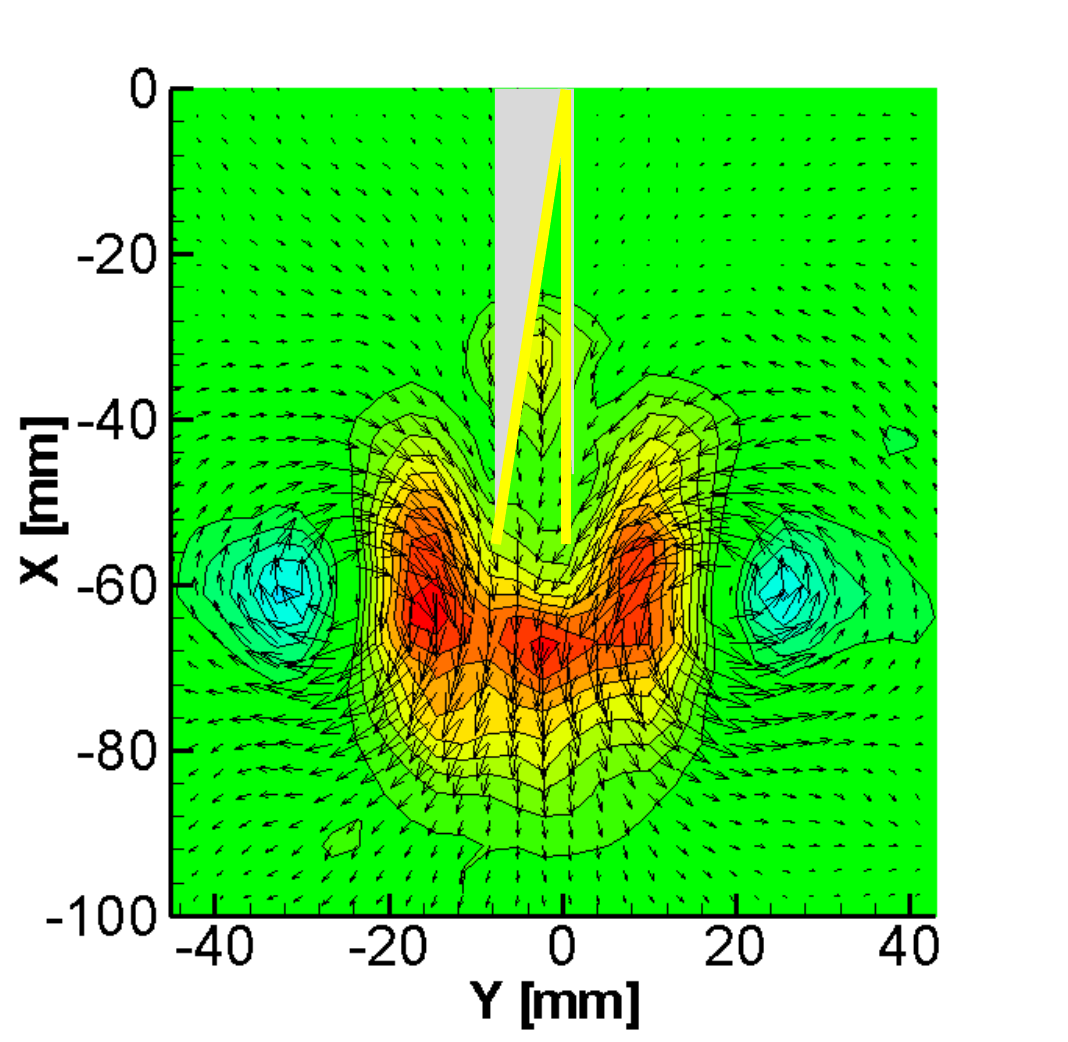}
			\caption{}
			\label{fig:Mflx_AR067_Stat}
		\end{subfigure}\hspace{1mm}
		\begin{subfigure}[b]{0.32\textwidth}
			\includegraphics[width=\textwidth]{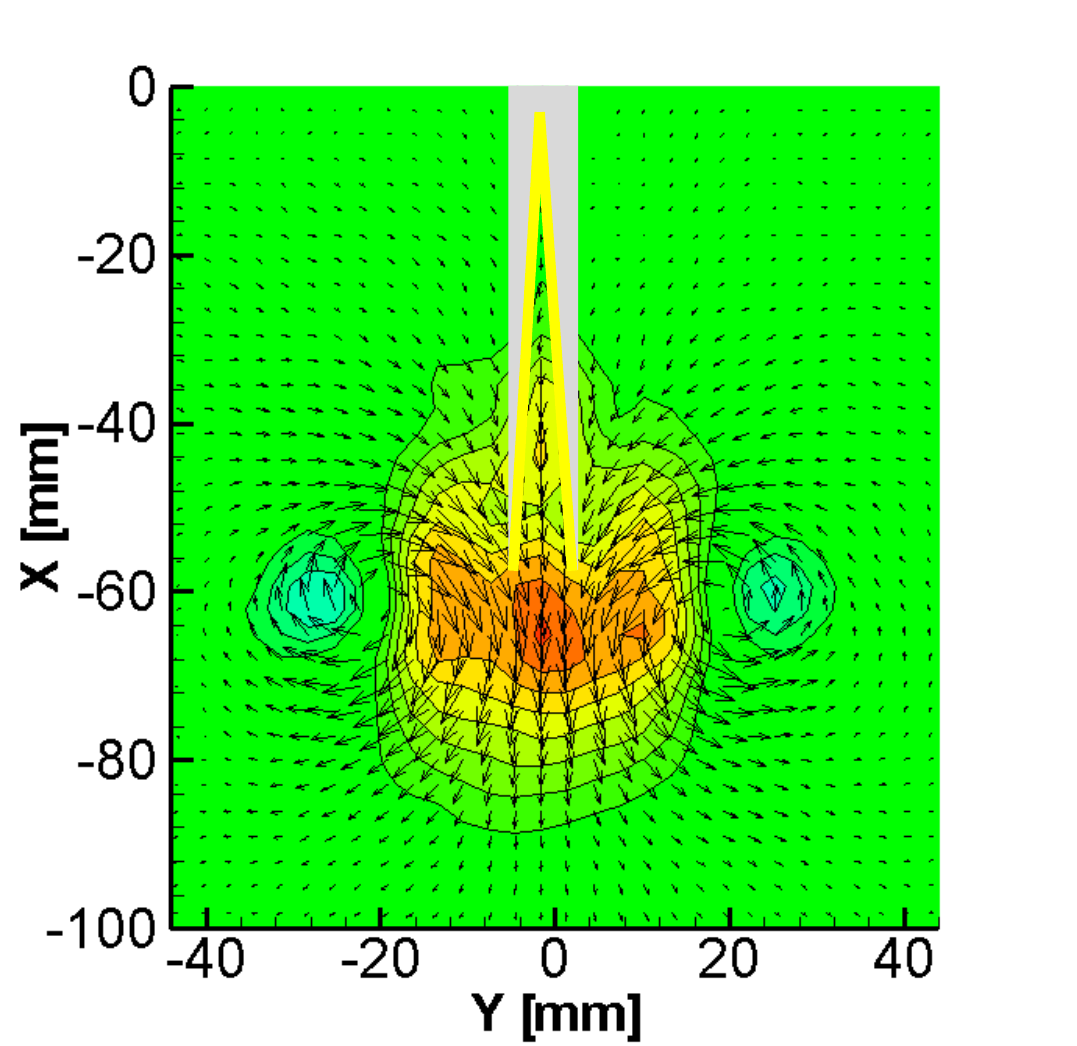}
			\caption{}
			\label{fig:Mflx_AR100_Stat}
		\end{subfigure}\hspace{1mm}
		\begin{subfigure}[b]{0.32\textwidth}
			\includegraphics[width=\textwidth]{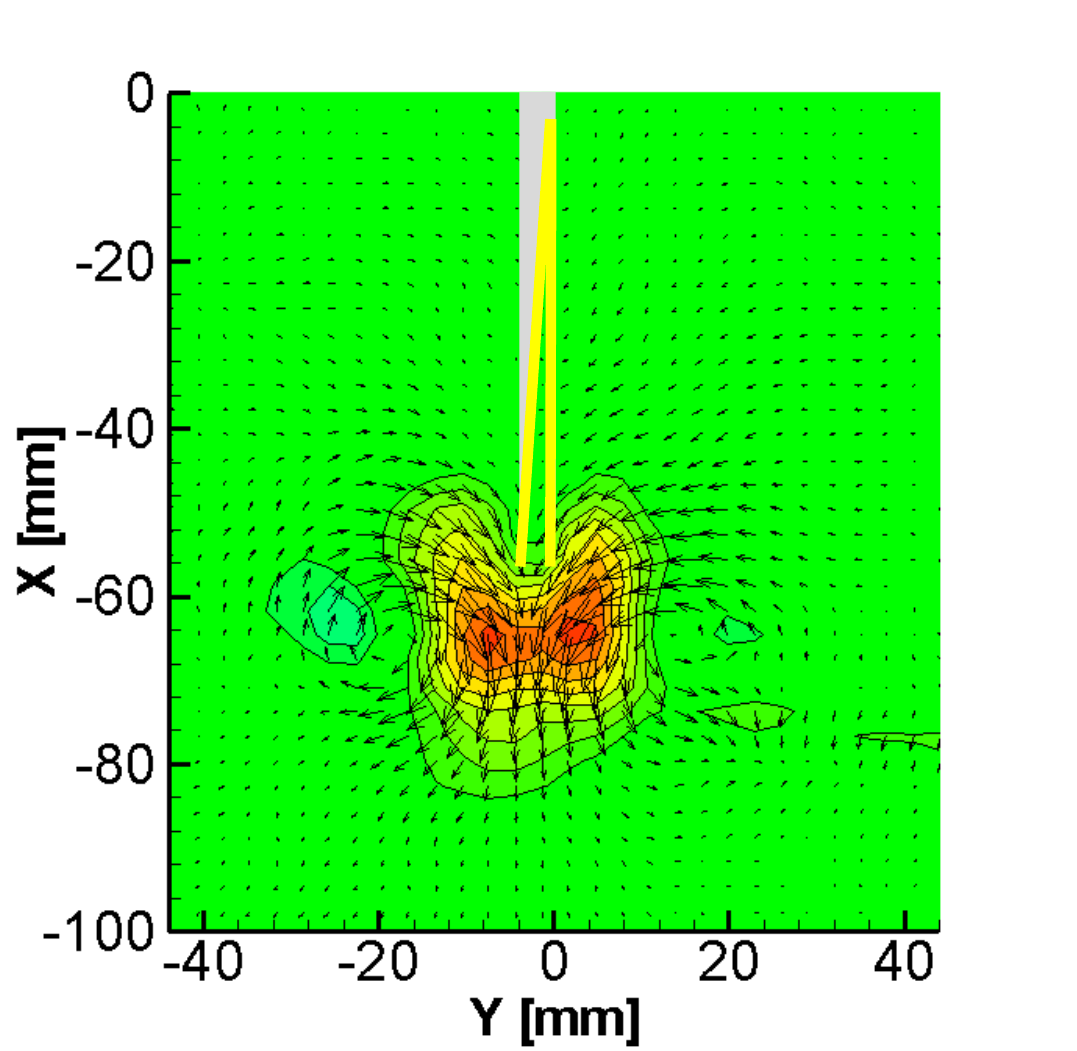}
			\caption{}
			\label{fig:Mflx_AR200_Stat}
		\end{subfigure}\vspace{5mm}
		\begin{subfigure}[b]{0.32\textwidth}
			\includegraphics[width=\textwidth]{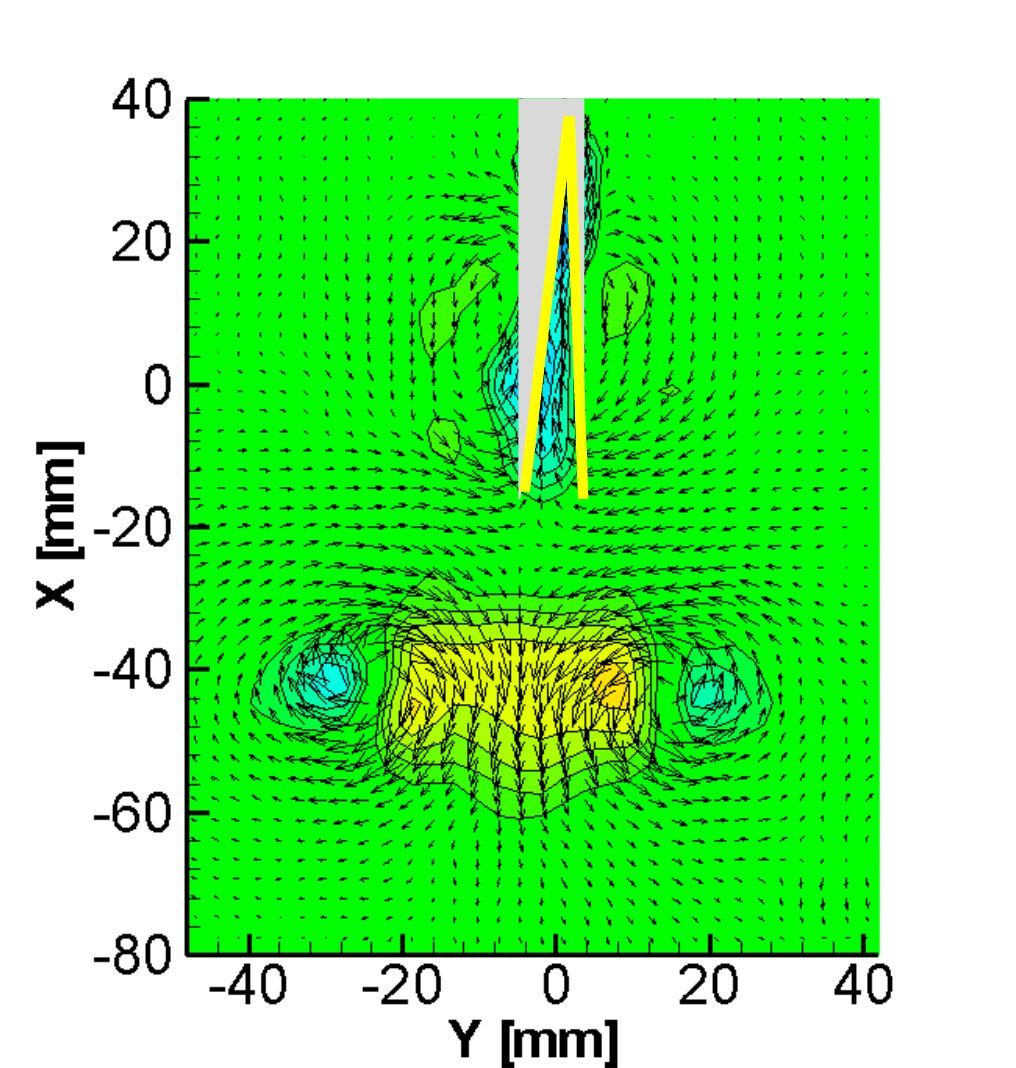}
			\caption{}
			\label{fig:Mflx_AR067_Dyn}
		\end{subfigure}\hspace{1mm}
		\begin{subfigure}[b]{0.32\textwidth}
			\includegraphics[width=\textwidth]{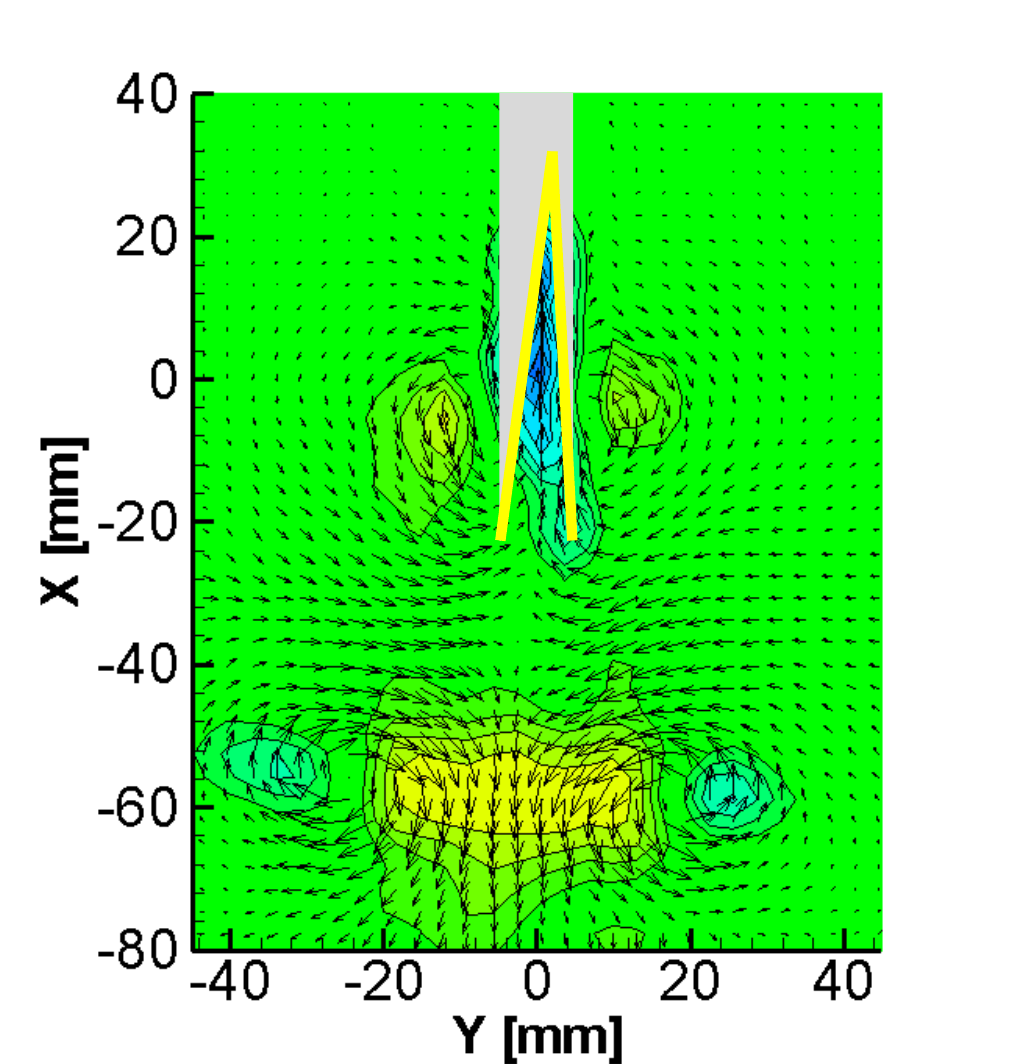}
			\caption{}
			\label{fig:Mflx_AR100_Dyn}
		\end{subfigure}\hspace{1mm}
		\begin{subfigure}[b]{0.32\textwidth}
			\includegraphics[width=\textwidth]{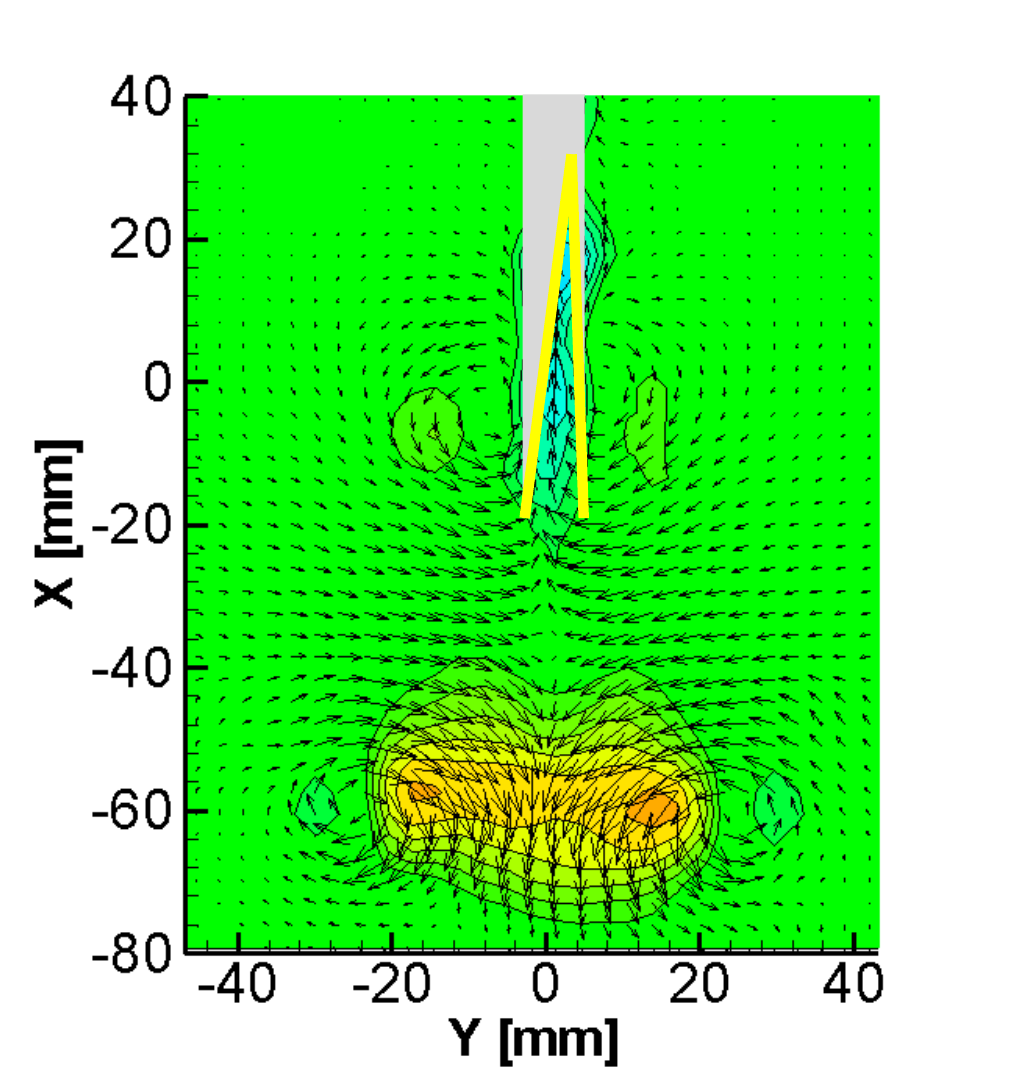}
			\caption{}
			\label{fig:Mflx_AR200_Dyn}
		\end{subfigure}
		\begin{subfigure}[b]{0.35\textwidth}
			\includegraphics[width=\textwidth]{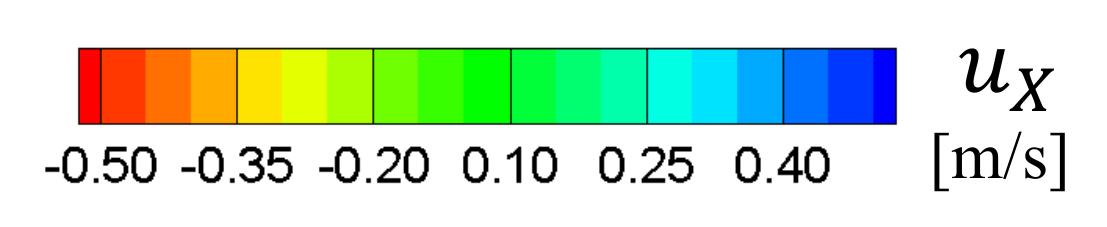}
		\end{subfigure}
		\caption{The flow field in the XY plane displays contours of the X-component of fluid velocity, $u_X$, at the end of the clapping motion. Figures (a), (b), and (c) illustrate the flow fields at 100 ms for the stationary cases with $d^* = $ 1.5, 1.0, and 0.5. Figures (d), (e), and (f) showcase the flow fields at 60 ms for the dynamic cases with with $d^* = $ 1.5, 1.0, and 0.5. Yellow lines denote the positions of the clapping body, and gray regions indicate the shadow regions.}\label{fig:U-velo_stat_dyn}
	\end{figure}
	For the stationary body, the momentum of the fluid in the -ve X-direction will be equal to the impulse on the support holding the body. For the body which is free to translate, the X-momentum in the body plus that in the fluid being carried along with it will be equal to the X-momentum of the fluid moving in the –ve X direction. The flow velocity in the X-direction, $u_X$, characterizes the wake momentum. In figures \hyperref[fig:U-velo_stat_dyn]{17(a-c)}, which show the X-direction velocity magnitude contours for the constrained body, it is obvious that momentum (per unit depth) in the wake is maximum for $d^*$ = 1.5 body, minimum for $d^*$ = 0.5, and a value in between $d^*$ = 1.0. For the freely moving bodies, the velocity distribution for each $d^*$ in the wake region exhibits lower values (figures \hyperref[fig:U-velo_stat_dyn]{17d, e, f}), indicating lower momentum compared to the corresponding stationary case.\par

	Most of the momentum in the wake is expected to be in the vortex structures (figure \hyperref[fig:Vorticity-Z_StatDyn]{10}) and related to the circulation (Sullivan et al.\cite{Sullivan08}). Figure \hyperref[fig:Gamma_fix_stat_Dyn]{18a} shows the variation in circulation, $\Gamma$, with time for the stationary body with $d^*$ = 1.5. The circulation values for the vortex patches related to the two plates are plotted separately and are nearly identical, showing that the clapping is symmetrical. The initial steep and nearly linear increase in circulation for about 100 ms is during the vortex formation phase. After reaching the maximum, the circulation reduces slowly with time. The average of standard deviations in $\Gamma$ over time is less than 7.5\% of the maximum circulation, $\Gamma_m$. The nature of variation of circulation with time is similar for the three aspect ratio bodies, although the magnitudes are different, with $d^*$ = 1.5 body having the largest and the $d^*$ = 0.5 body having the least (figure \hyperref[fig:Gamma_fix_stat_Dyn]{18b} and Table \hyperref[tab:stat_wake]{2}). Also, the lowest aspect ratio body shows a steeper decline in $\Gamma$ after the peak is reached. In this case, both vortices come very close to each other, resulting in the vorticity cancellation (Melander et al.\cite{Melander89}). A previous study on stationary clapping plates by Kim et al.\cite{Kim13} also reported an increase in $\Gamma$ with an increase in the depth of the plate. \par	
	\begin{table}
		\centering
		\begin{tabular}{cccccccccccc}
			&       &       &       &       &       &       &       &       &       &       &  \\
			\cmidrule{2-2}\cmidrule{4-5}\cmidrule{7-8}\cmidrule{10-11}    \multicolumn{1}{c}{} & \multicolumn{1}{c}{\multirow{2}[4]{*}{$d^*$}} & \multicolumn{1}{c}{} & \multicolumn{2}{c}{$\Gamma_m$ [cm$^2$ /s]} & \multicolumn{1}{c}{} & \multicolumn{2}{c}{$R_v$ [mm]} & \multicolumn{1}{c}{} & \multicolumn{2}{c}{$I_v$ [gm.m/s]} & \multicolumn{1}{c}{} \\
			\cmidrule{4-5}\cmidrule{7-8}\cmidrule{10-11}    \multicolumn{1}{c}{} & \multicolumn{1}{c}{} & \multicolumn{1}{c}{} & \multicolumn{1}{c}{Stationary } & \multicolumn{1}{c}{Dynamic} & \multicolumn{1}{c}{} & \multicolumn{1}{c}{Stationary } & \multicolumn{1}{c}{Dynamic} & \multicolumn{1}{c}{} & \multicolumn{1}{c}{Stationary } & \multicolumn{1}{c}{Dynamic} & \multicolumn{1}{c}{} \\
			\cmidrule{2-2}\cmidrule{4-5}\cmidrule{7-8}\cmidrule{10-11}    \multicolumn{1}{c}{} & \multicolumn{1}{c}{1.5} & \multicolumn{1}{c}{} & \multicolumn{1}{c}{267} & \multicolumn{1}{c}{137} & \multicolumn{1}{c}{} & \multicolumn{1}{c}{57.4} & \multicolumn{1}{c}{57.3} & \multicolumn{1}{c}{} & \multicolumn{1}{c}{65.7} & \multicolumn{1}{c}{33.6} & \multicolumn{1}{c}{} \\
			\multicolumn{1}{c}{} & \multicolumn{1}{c}{1.0} & \multicolumn{1}{c}{} & \multicolumn{1}{c}{206} & \multicolumn{1}{c}{138} & \multicolumn{1}{c}{} & \multicolumn{1}{c}{50.0} & \multicolumn{1}{c}{50.4} & \multicolumn{1}{c}{} & \multicolumn{1}{c}{38.7} & \multicolumn{1}{c}{26.3} & \multicolumn{1}{c}{} \\
			\multicolumn{1}{c}{} & \multicolumn{1}{c}{0.5} & \multicolumn{1}{c}{} & \multicolumn{1}{c}{158} & \multicolumn{1}{c}{139} & \multicolumn{1}{c}{} & \multicolumn{1}{c}{39.8} & \multicolumn{1}{c}{40.6} & \multicolumn{1}{c}{} & \multicolumn{1}{c}{18.8} & \multicolumn{1}{c}{17.1} & \multicolumn{1}{c}{} \\
			\cmidrule{2-2}\cmidrule{4-5}\cmidrule{7-8}\cmidrule{10-11}          &       &       &       &       &       &       &       &       &       &       &  \\
		\end{tabular}%
		\caption{The circulation, $\Gamma$, equivalent radius, $R_v$, and impulse of the wake vortex, $I_v$.}
		\label{tab:stat_wake}%
	\end{table}%

	\begin{figure}
		\centering
		\begin{subfigure}[b]{0.45\textwidth}					        				\includegraphics[width=\textwidth]{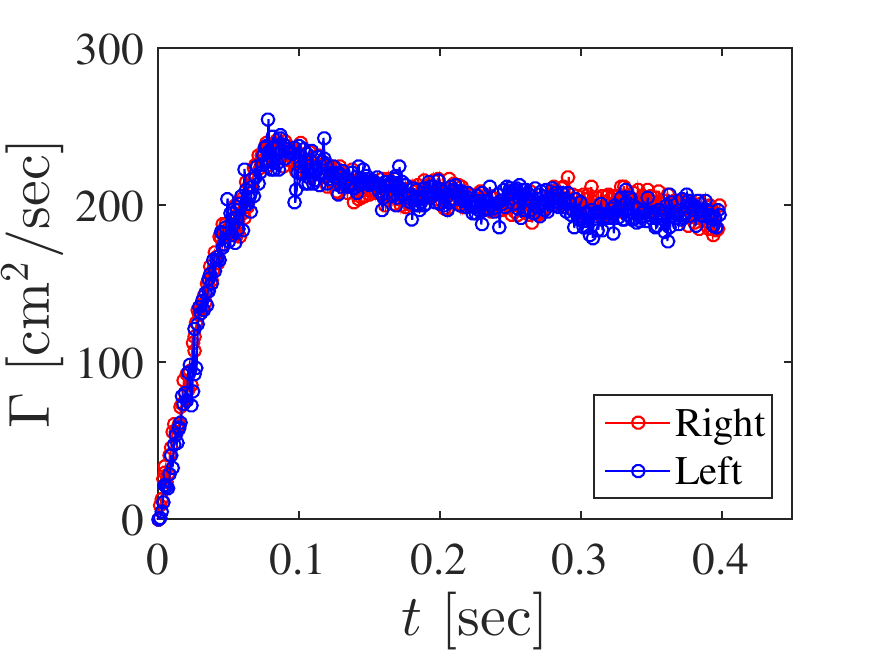}	
			\caption{}
			\label{fig:Circulation_fix_RL_StatDyn}
		\end{subfigure}\hspace{10mm}
		\begin{subfigure}[b]{0.45\textwidth}
			\includegraphics[width=\textwidth]{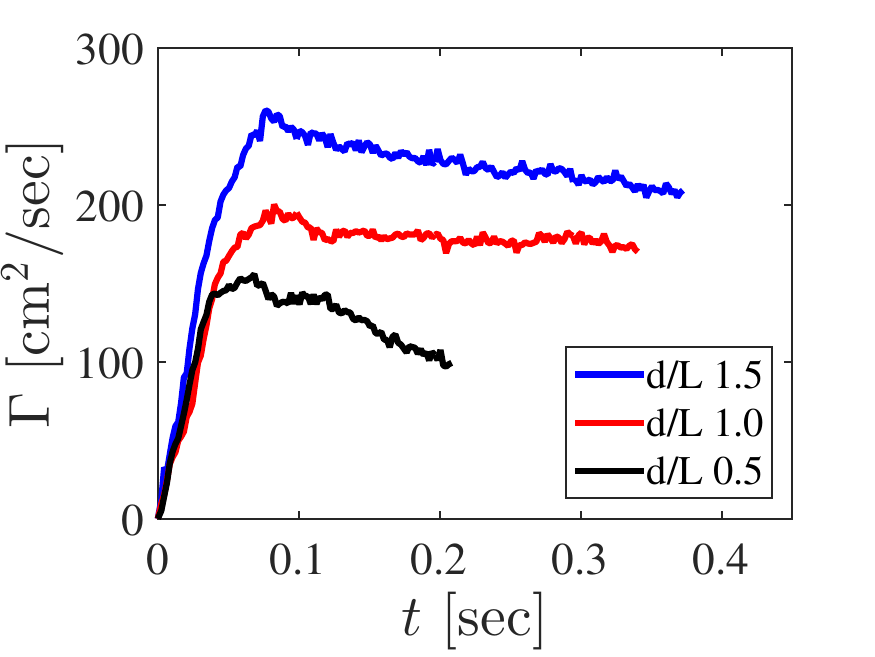}
			\caption{}
			\label{fig:Circulation_fix_d*_statDyn}
		\end{subfigure}
		\caption{(a) The circulation, $\Gamma$, in the starting vortices for the left and right clapping plates of the stationary clapping body with $d^*$ = 1.5. (b) Variations of $\Gamma$ with time for the stationary case with three different $d^*$ values.}\label{fig:Gamma_fix_stat_Dyn}
	\end{figure}
	\begin{figure}
		\centering
		\begin{subfigure}[b]{0.45\textwidth}
			\includegraphics[width=\textwidth]{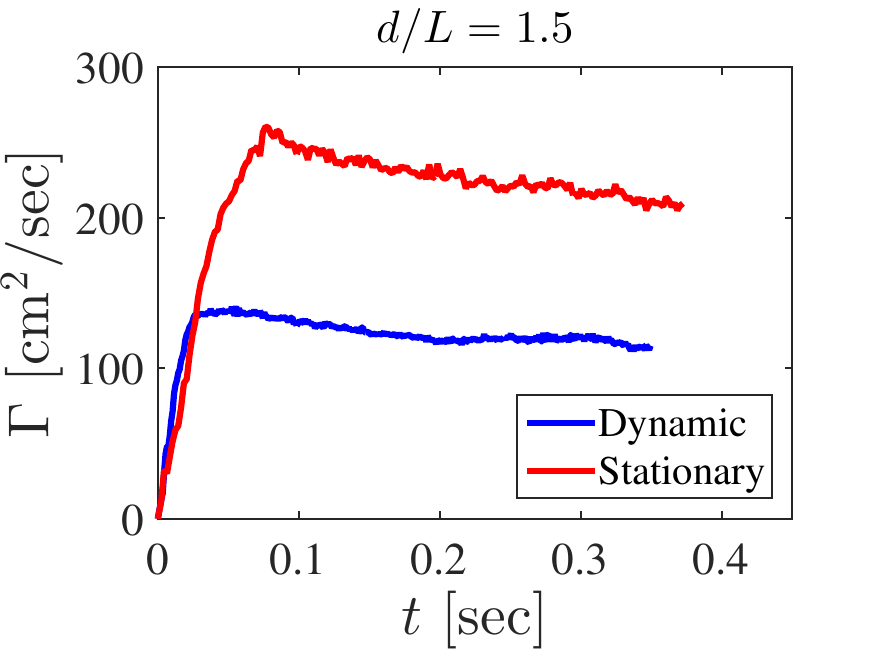}
			\caption{}
			\label{fig:Circulation_free_fix_AR067_StatDyn}
		\end{subfigure}\hspace{10mm}
		\begin{subfigure}[b]{0.45\textwidth}
			\includegraphics[width=\textwidth]{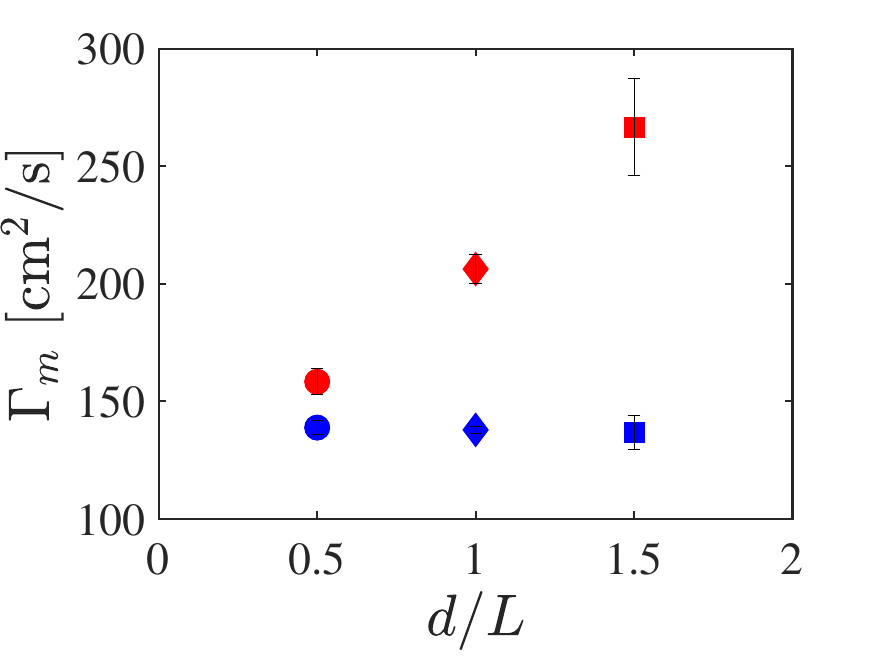}
			\caption{}
			\label{fig:Circulation_AR_d*_Stat_Dyn}
		\end{subfigure}
		\caption{(a) Variations of circulation with time for the body with $d^* =$ 1.5 in both stationary and dynamic cases. (b) Maximum values of circulation, $\Gamma_m$, versus $d^*$ for stationary cases (red) and dynamic cases (blue). Circle, rhombus, and square symbols represent $d^* =$ 0.5, 1.0, and 1.5.}\label{fig:Gamma_free_fix_StatDyn}
	\end{figure}

	\begin{figure}
		\centering
		\begin{subfigure}[b]{0.45\textwidth}
			\includegraphics[width=\textwidth]{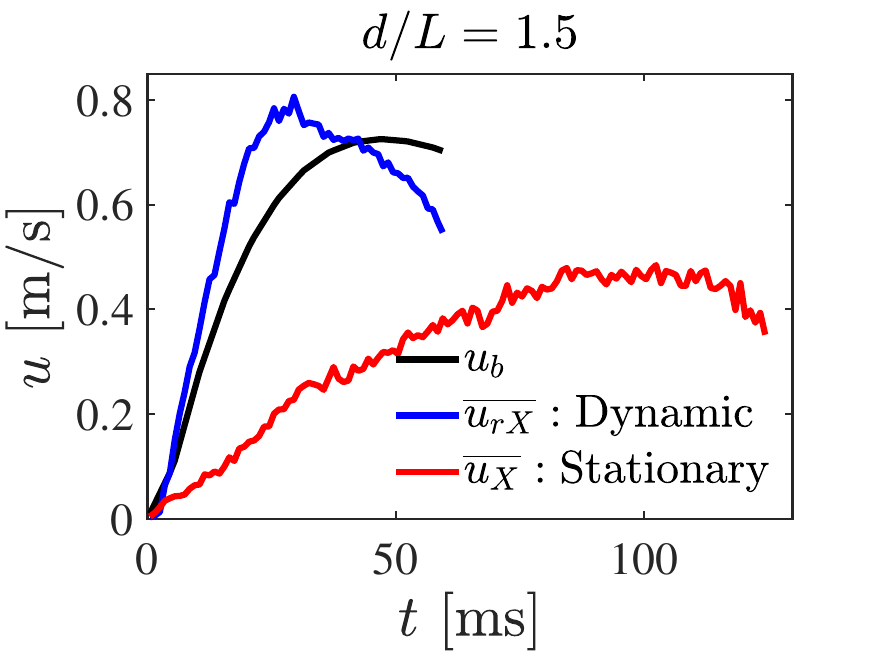}
			\caption{}
			\label{fig:Mean_lab_ux}
		\end{subfigure}\hspace{10mm}
		\begin{subfigure}[b]{0.45\textwidth}
			\includegraphics[width=\textwidth]{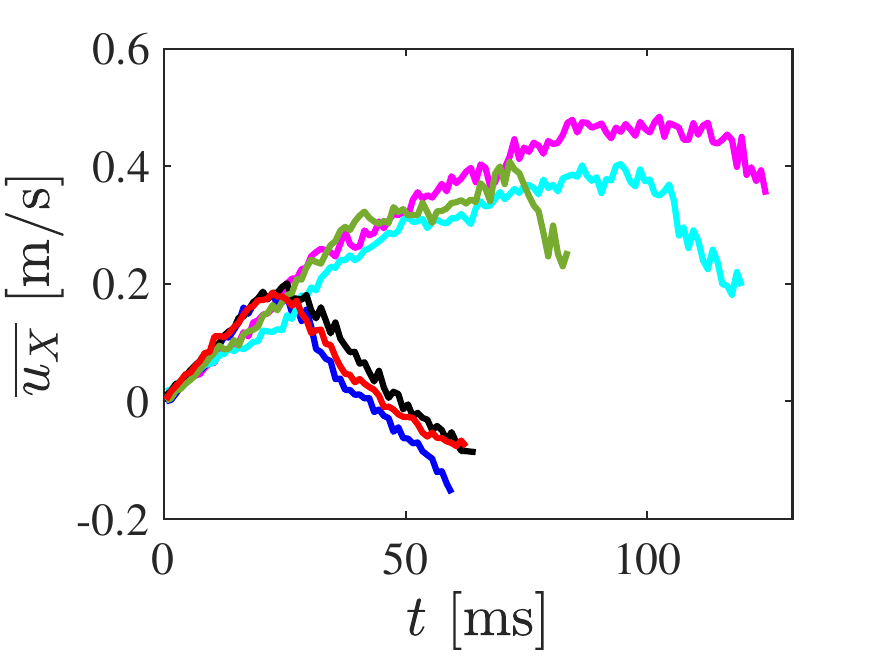}
			\caption{}
			\label{fig:Mean_ux}
		\end{subfigure}
		\caption{(a) Time variation of the mean flow velocity magnitude in the X-direction, $\overline{u_X}$, for the stationary case with $d^*=$ 1.5. Time variation of the mean flow velocity magnitude, $\overline{u_{rX}} (= \overline{u_X}-u_b)$, relative to the body and of the body velocity $u_b$ (black) for the dynamic case with $d^* =$ 1.5. (b) Time variation of $\overline{u_X}$ for stationary and dynamic cases for the three $d^*$ values. The data legends are same as given in figure \hyperref[fig:CoreSeparation_StatDyn]{16e}.}\label{fig:Mean_ux_lab_body}
	\end{figure}

	For the freely moving body, the circulation value is lower compared to the corresponding stationary case (Table \hyperref[tab:stat_wake]{2}), although the overall nature of circulation variation over time, for a particular $d^*$ case, remains nearly the same; see figure \hyperref[fig:Gamma_free_fix_StatDyn]{19a}. In dynamic cases, the body quickly moves away from its starting vortices. As a result, the vorticity, generated by the rotation of the plates, is supplied to these vortices for shorter durations compared to stationary cases, leading to lower circulation values. As discussed below, another reason for the lower circulation in the dynamic cases is the lower exit velocity of the fluid from the clapping cavity. For the stationary cases, the maximum circulation, $\Gamma_m$, increases close to linearly with $d^*$, in contrast to dynamic cases, where the value remains nearly constant (figure \hyperref[fig:Gamma_free_fix_StatDyn]{19b} and Table \hyperref[tab:stat_wake]{2}). The variation in $\Gamma_m$ with respect to $d^*$ is possibly due to the flow-induced by the vortex loops surrounding its top and bottom edges of the plates (figure \hyperref[fig:3D Vortex loop_StatDyn]{15a,b}). Another view of this phenomenon is the influence of `leakage' from the top and bottom edges that becomes increasingly pronounced as $d^*$ reduces. We may expect the value of $\Gamma_m$ to stop increasing with $d^*$ and level off to the 2-D value beyond perhaps $d^*$ of 5. However, for the  dynamic cases, it is intriguing that $\Gamma_m$ remains nearly constant with $d^*$, at least in the range currently covered. The rapid body movement during the clapping phase (figure \hyperref[fig:ub_disp_StatDyn]{7}) seems to reduce the aspect ratio effect; the body has moved already moved about 0.3 body lengths in about 50 ms. Another consequence of the body movement is that even though the plate angular velocity is higher for the dynamic cases, the circulation values are lower compared to the stationary cases. To better understand the difference in circulation values between the stationary and dynamic cases, we plot in figure \hyperref[fig:Mean_lab_ux]{20a}, for $d^* =$ 1.5, the body velocity, $u_b$, the X component of fluid velocity relative to moving body averaged over the exit distance, $\overline{u_{rX}}$, for the dynamic case, and the average fluid velocity at the exit in lab reference frame, $\overline{u_X}$, for the stationary case; exit refers to the line joining the edges of the clapping plates in the XY plane, and the velocities are obained from PIV data. The rapid closing of the plates for the dynamic case results in much higher relative fluid velocities at the exit compared to those for the stationary case. But it is the difference in the relative fluid velocity and body velocity, $\overline{u_{rX}} - u_b (= \overline{u_X})$, that contributes to the circulation in the starting vortex. $u_X$ is essentially the X component of velocity in the lab reference frame, which is what was measured and shown in the figures (for example, figure \hyperref[fig:VortZ_D_StatDyn]{9}). Figure \hyperref[fig:Mean_ux]{20b} shows the variations of $\overline{u_X}$ for both the dynamic and stationary cases for the three $d^*$ values, and we see for the dynamic cases the maximum $\overline{u_X}$ ($\sim$ 0.2 m/s) values are about half compared to the corresponding stationary values ($\sim$ 0.4 m/s). It is interesting to note that the initial variations (till about 20 ms) of $\overline{u_X}$ with time for all $d^*$ values in both dynamic and stationary cases are nearly identical. After 20 ms, for all three dynamic cases, there is a sharp linear drop in $\overline{u_X}$, and beyond about 40 ms, it becomes negative, corresponding to the body velocity being higher than the fluid relative velocity. The negative velocity implies that the fluid at the exit is being carried forward by the body. These differences in the exit velocities explain the observation that maximum circulation values are higher for the stationary cases compared to those for the dynamic cases. The flow field relative to body in dynamic case is related to the study by Krueger et al.\cite{Kruger06}, where it was found that the presence of co-flow reduced the circulation in the vortex rings generated using a piston-cylinder arrangement. However, in the present study, both body and fluid velocities are transient\par
	
	\begin{figure}
		\centering
		\begin{subfigure}[b]{0.45\textwidth}
			\includegraphics[width=\textwidth]{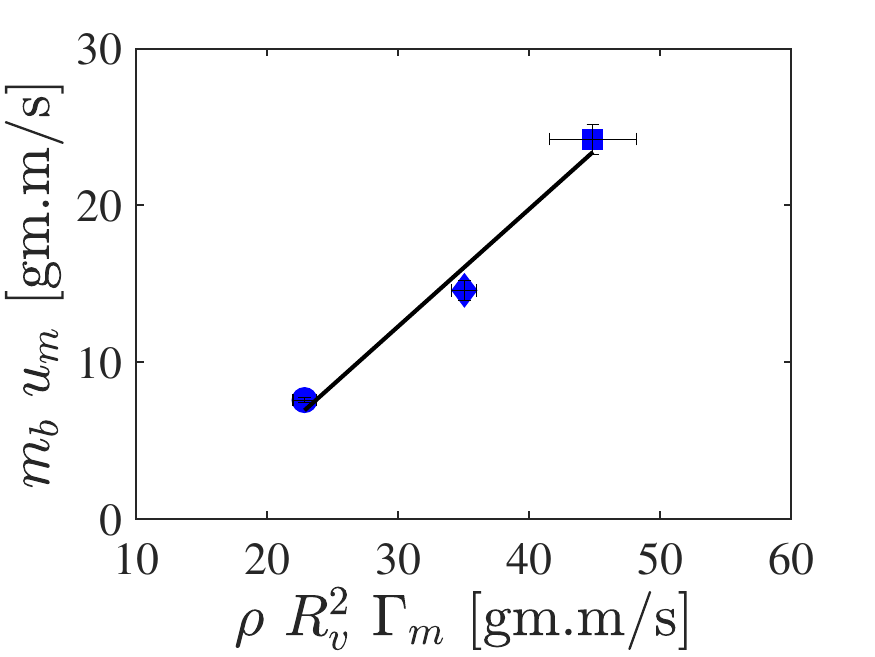}
			\caption{}
			\label{fig:ImpulseCoeff_Stat_Dyn}
		\end{subfigure}\hspace{10mm}
		\begin{subfigure}[b]{0.45\textwidth}
			\includegraphics[width=\textwidth]{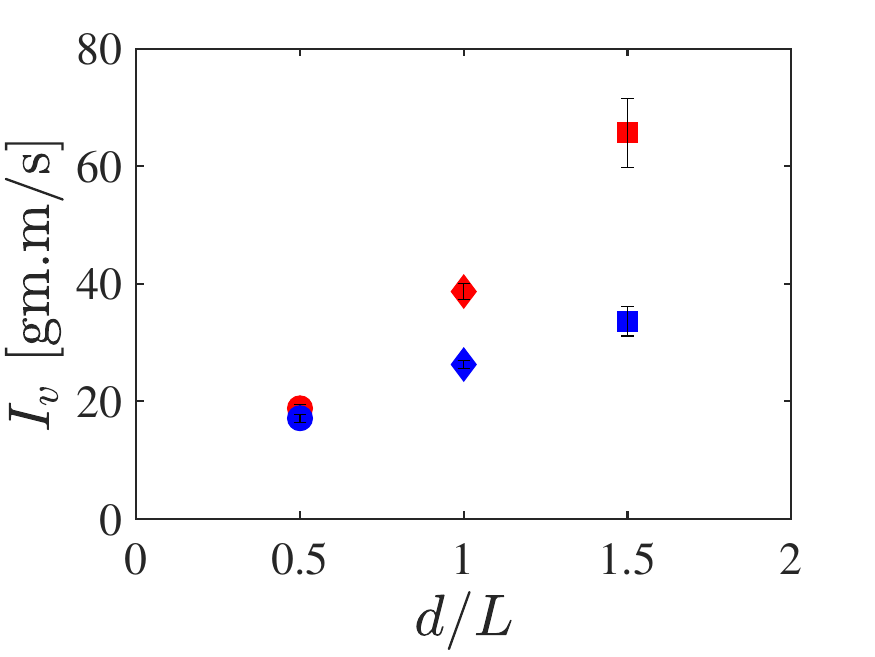}
			\caption{}
			\label{fig:Impulse_Stat_Dyn}
		\end{subfigure}\hspace{10mm}
		\caption{(a) Plot showing the dependence between momentum of the freely moving clapping body and momentum in the ‘equivalent’ vortex ring. (b) Plot of vortex impulse $I_v$ versus $d^*$  for stationary cases (red), and  for dynamic cases (blue).}\label{fig:VortImpulse_StatDyn}
		\end{figure}

		\begin{figure}
			\centering
			\begin{subfigure}[b]{0.45\textwidth}
				\includegraphics[width=\textwidth]{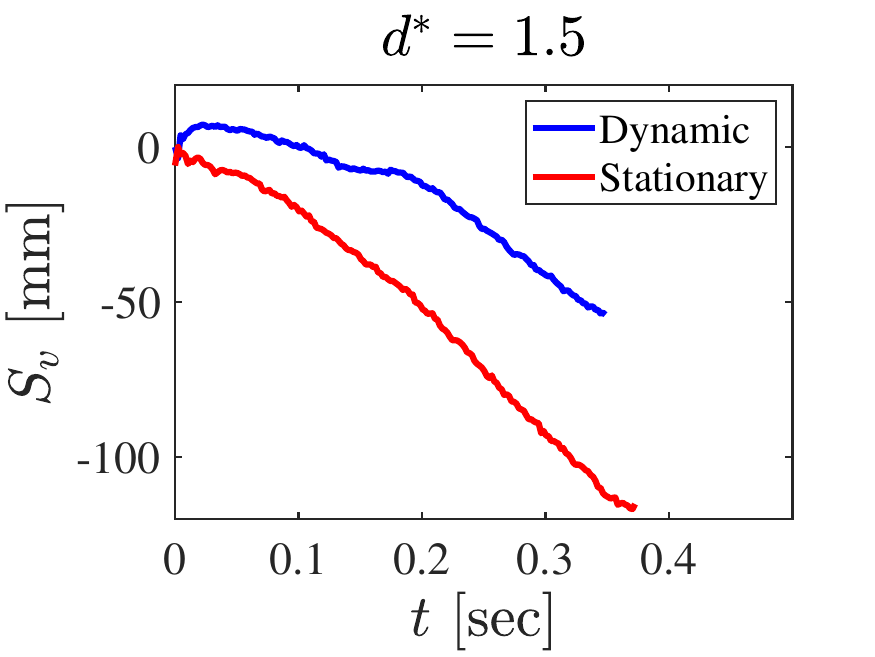}
				\caption{}
				\label{fig:VortDisp_Stat_Dyn}
			\end{subfigure}\hspace{10mm}
			\begin{subfigure}[b]{0.42\textwidth}
				\includegraphics[width=\textwidth]{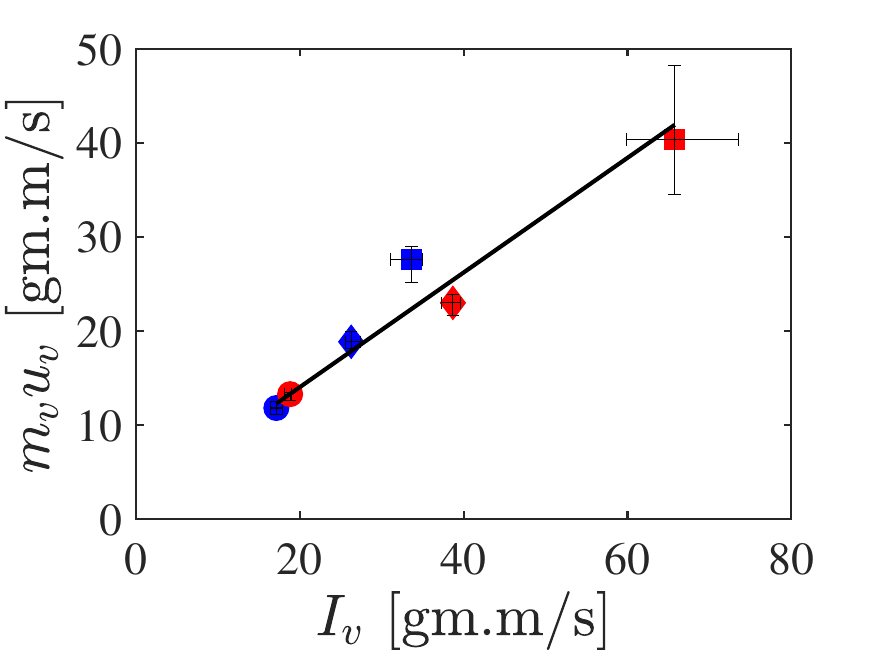}
				\caption{}
				\label{fig:ImpulseMom_Stat_Dyn}
			\end{subfigure}
			\caption{(a) X-displacement of the wake vortex in stationary and dynamic cases with $d^*$ = 1.5. (b) Plot showing the linear dependence between $m_v u_v$ and $I_v$ for stationary cases (red) and dynamic cases (blue). Circle, rhombus, and square symbols respectively represent $d^* =$ 0.5, 1.0, and 1.5.}
			\label{fig:DispX_Stat_Dyn}
		\end{figure}
	\begin{table}
		\centering
		\begin{tabular}{rrrrrrrrrrrr}
			&       &       &       &       &       &       &       &       &       &       &  \\
			\cmidrule{2-2}\cmidrule{4-5}\cmidrule{7-8}\cmidrule{10-11}    \multicolumn{1}{c}{} & \multicolumn{1}{c}{\multirow{2}[4]{*}{$d^*$}} & \multicolumn{1}{c}{} & \multicolumn{2}{c}{$u_v$ [m/s]} & \multicolumn{1}{c}{} & \multicolumn{2}{c}{$t_c$ [ms]} & \multicolumn{1}{c}{} & \multicolumn{2}{c}{$\overline{C_T}$} &  \\
			\cmidrule{4-5}\cmidrule{7-8}\cmidrule{10-11}    \multicolumn{1}{c}{} & \multicolumn{1}{c}{} & \multicolumn{1}{c}{} & \multicolumn{1}{c}{Stationary } & \multicolumn{1}{c}{Dynamic} & \multicolumn{1}{c}{} & \multicolumn{1}{c}{Stationary } & \multicolumn{1}{c}{Dynamic} & \multicolumn{1}{c}{} & \multicolumn{1}{c}{Stationary } & \multicolumn{1}{c}{Dynamic} &  \\
			\cmidrule{2-2}\cmidrule{4-5}\cmidrule{7-8}\cmidrule{10-11}    \multicolumn{1}{c}{} & \multicolumn{1}{c}{1.5} & \multicolumn{1}{c}{} & \multicolumn{1}{c}{0.21} & \multicolumn{1}{c}{0.15} & \multicolumn{1}{c}{} & \multicolumn{1}{c}{125} & \multicolumn{1}{c}{60} & \multicolumn{1}{c}{} & \multicolumn{1}{c}{3.15} & \multicolumn{1}{c}{0.98} &  \\
			\multicolumn{1}{c}{} & \multicolumn{1}{c}{1.0} & \multicolumn{1}{c}{} & \multicolumn{1}{c}{0.18} & \multicolumn{1}{c}{0.15} & \multicolumn{1}{c}{} & \multicolumn{1}{c}{119} & \multicolumn{1}{c}{65} & \multicolumn{1}{c}{} & \multicolumn{1}{c}{2.89} & \multicolumn{1}{c}{1.12} &  \\
			\multicolumn{1}{c}{} & \multicolumn{1}{c}{0.5} & \multicolumn{1}{c}{} & \multicolumn{1}{c}{0.21} & \multicolumn{1}{c}{0.18} & \multicolumn{1}{c}{} & \multicolumn{1}{c}{84} & \multicolumn{1}{c}{63} & \multicolumn{1}{c}{} & \multicolumn{1}{c}{1.96} & \multicolumn{1}{c}{1.15} &  \\
			\cmidrule{2-2}\cmidrule{4-5}\cmidrule{7-8}\cmidrule{10-11}          &       &       &       &       &       &       &       &       &       &       &  \\
		\end{tabular}%
		\caption{Velocity of the vortex, $u_v$, clapping time, $t_c$,  and mean thrust coefficient, $\overline{C_T}$.}
		\label{tab:stat_wake_ct}%
	\end{table}%

	An accurate estimation of the impulse acting on a clapping body requires 3-D wake field information, which is unavailable. Therefore, we modeled the wake as an equivalent circular vortex ring, which can be analyzed using the data obtained from 2-D PIV. The radius of the ring, $R_v \ (\sim \forall^{1/3})$, is scaled as a cube root of the initial interplate cavity volume, $\forall$, defined as $0.5 d R_c^2 \sin(2\theta_o)$, where $R_c$ represents the radius of rotation as shown by the yellow line in figure \hyperref[fig:Fabricated clapping body_StatDyn]{2b}. The impulse of a vortex ring, $I_v$, is given as
	\begin{gather}
	\label{Iv}
	 I_v = C_1 \ \rho \ R_v^2 \ \Gamma_m,
	\end{gather} 
	 where $C_1$ is a proportionality constant. While this constant is analytically determined for a steadily translating circular ring (Sullivan et al.\cite{Sullivan08}), it isn't directly applicable to the complex 3-D wake of the clapping body. To address this, we used a momentum balance between the forward-moving clapping body and backward-moving vortex ring, given as:
	 \begin{gather}
	 \label{mom_stat}
	  (m_b + m_{add}) \ u_m =  C_1 \ \rho \ R_v^2 \ \Gamma_m ,
	 \end{gather}
	where the LHS represents the momentum of the body with mass $m_b$, carrying an additional mass of fluid $m_{add}$, when it reaches maximum velocity $u_m$ approximately at the end the clapping motion, and  RHS shows the impulse of the vortex when it reaches maximum circulation, as given by \eqref{Iv}. At maximum velocity, when both plates touch and the body becomes almost streamlined, we can assume the added mass to be negligible. The $m_b u_m$  shows linear fit with $\rho R_v^2 \Gamma_m$ expressed as $m_b u_m = 0.75 \ \rho R_v^2 \Gamma_m - 10.2$, see figure \hyperref[fig:VortImpulse_StatDyn]{21a}.  The fit has an $R^2$ value of 0.98 and the slope of the fit gives $C_1$ (= 0.75). For the calculation of $I_v$ using \eqref{Iv}, we assumed that both stationary and dynamic cases share the same $C_1$, based on the observed similarity in the wake structure for $d^*$ = 1.5 and 1.0; see figures \hyperref[fig:Vorticity-Z_StatDyn]{10}, \hyperref[fig:Vorticity-Y_S_StatDyn]{13} and \hyperref[fig:Vorticity-Y_D_StatDyn]{14}. $I_v$ values are higher in stationary cases than in dynamic cases, and they both show a linear increase with $d^*$, although the rate of increase is higher in the former (figure \hyperref[fig:VortImpulse_StatDyn]{21b} and table \hyperref[tab:stat_wake]{2}). These findings align with the observations in the $u_X$ field as discussed earlier (figure \hyperref[fig:U-velo_stat_dyn]{17}). \par
	
	An alternative way to calculate vortex impulse is by multiplying the mass of vortex $m_v$ ( $= \rho R_v^3$) by its velocity $u_v$, which is determined from vortex displacement data in the XY plane obtained from PIV measurements. In figure \hyperref[fig:DispX_Stat_Dyn]{22a}, the vortex displacement, $S_v$, in the X direction is shown for both stationary and dynamic cases with $d^*$ = 1.5. In dynamic cases, vortices initially travel in the direction of the body (+ve X direction) for the first 20 ms, followed by a reversal to the -ve X direction. In stationary cases, they consistently move only in the -ve X direction. These observations hold true across all $d^*$ values. After the clapping motion ceases, both vortices primarily travel in the -ve X direction, and after 200 ms, the displacement of vortices in the Y direction becomes nearly zero. During this phase, we determined the vortex patch velocities by calculating the slopes of the $S_v - t$ curves (figure \hyperref[fig:DispX_Stat_Dyn]{22a}). The vortex velocity in stationary cases is higher than that in dynamic cases for all $d^*$ values, as specified in table \hyperref[tab:stat_wake_ct]{3}. Interestingly, $m_v u_v$ and $I_v$ are approximately linearly related: $m_v u_v \approx 0.61 I_v$, as shown in the figure \hyperref[fig:DispX_Stat_Dyn]{22b}. \par

	The mean thrust coefficient, $\overline{C_T}$, for the clapping bodies is defined as:
	\begin{gather}
	\overline{C_T} = \frac{\overline{F}}{0.5 \ \rho \ \overline{u_T}^2  \ R_c d} \ ,
	\end{gather}
	where $\overline{F}$ is the mean net thrust force acting on the body, and $\overline{u_T} \ (= R_c \ \overline{\dot{\theta}})$ is the mean tip velocity of the rotating plate. The net thrust force, $F$, acting on a translating body, representing the difference between thrust and drag is determined by taking the time derivative of \eqref{mom_stat}. The mean net thrust force $\overline{F}$ is expressed as:
	\begin{gather}
	\label{mean_Ft}
	 \overline{F} = \frac{1}{t_c} \ \int_{0}^{t_c} F \ dt 
	\end{gather}
	where $t_c$ is the clapping time period. By Substituting \eqref{mom_stat} in \eqref{mean_Ft}  and assuming negligible $m_{add}$, we can simplify this to:
	\begin{gather}
	\label{mean_Iv_tc}
	\overline{F} \approx \frac{m_b u_m}{t_c} \approx \frac{I_v}{t_c}.
	\end{gather}
	In stationary cases, $\overline{C_T}$ signifies the mean thrust coefficient, while in dynamic cases, it represents the mean net thrust coefficient. The angular velocity curve of the plate shows a gradual stopping of the clapping motion, with values becoming inaccurate in the tail portion across the experiments. Therefore, we defined the clapping time period, $t_c$, as the point in time when the angular velocity of plate reaches 15\% of its maximum value. Figure \hyperref[fig:AngVelo_AR067_StatDyn]{5a} illustrates that for the stationary case, $t_c$ is 125 ms, and for the dynamic case, it is 60 ms, signifying the effective end of clapping motions. The $t_c$ values for the remaining cases are listed in table \hyperref[tab:stat_wake_ct]{3}. When we restrict the forward motion of the body, we notice that $\overline{C_T}$ increases with $d^*$ (figure \hyperref[fig:Ct_StatDyn]{23} and table \hyperref[tab:stat_wake_ct]{3}). This observation aligns with the findings from a previous study on clapping plates by Kim et al.\cite{Kim13}. In freely moving bodies, $\overline{C_T}$ values remain nearly constant across the three $d^*$ values, consistent with the observation that the maximum velocity of the bodies also remains largely unchanged across different $d^*$ values, see table \hyperref[tab:maximaAngular]{1} and \hyperref[tab:stat_wake_ct]{3}.
	 
	\begin{figure}
		\centering
		\begin{subfigure}[b]{0.45\textwidth}
			\includegraphics[width=\textwidth]{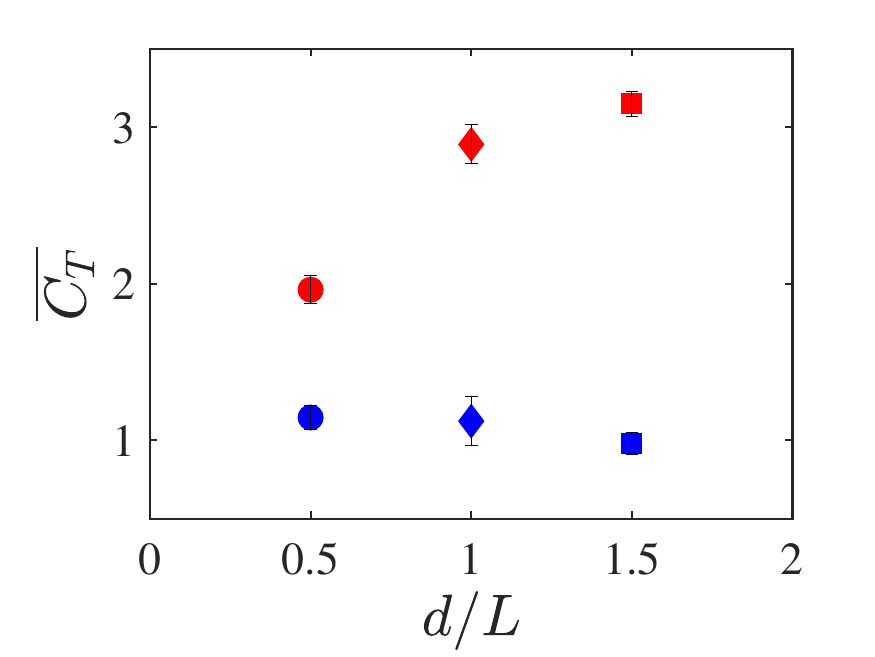}
			\label{fig:Ct_Stat_Dyn}
		\end{subfigure}
		\caption{The plot of mean thrust coefficient $\overline{C_T}$ versus $d^*$ for stationary cases (red) and dynamic cases (blue).}\label{fig:Ct_StatDyn}
	\end{figure}

\end{subsubsection}	
\end{subsection}
\end{section}
%%%%%%%%%%%%%%%%%%%%%%%%%%%%%%%%%%%%%%%%%%%%%%%%%%%%%%%%%%%%%%%%%%%%%%%%%%%%%%%%%%%%%%%%
\begin{section}{Concluding remarks}
\label{sec:conclusion_StatDyn}
	We presented a comparative study of flow dynamics between a freely moving clapping body (dynamic case) and one with constrained forward motion (stationary case). The clapping body, composed of two plates with variable density components, was designed to achieve near-neutral buoyancy in water. Initially, torque was applied to both plates to create an interplate cavity using a fine-threaded loop. Cutting the loop initiated clapping and released a fluid jet from the cavity. Under freely moving conditions, the straight-line motion of the body, driven by the thrust-producing jet, is achieved by aligning the center of mass with the buoyancy. We employed image analysis, PIV, and PLIF to compare differences in body kinematics and wake dynamics between the stationary and dynamic bodies for three values of aspect ratio,  $d^*$ = 1.5, 1.0, and 0.5. \par
	
	In both freely moving and constrained conditions, the angular velocity of the clapping plates remain nearly constant across all three $d^*$ values. The finite aspect ratios of the bodies lead to complex three-dimensional wake structures, which we have elucidated from PIV data on two perpendicular planes.The interaction of vortex loops from the three free edges of each clapping plate results in an elliptical vortex ring for the $d^* =$ 1.5 and 1.0 bodies, while three to four interconnected ringlets are observed in the wakes of the $d^* =$ 0.5 bodies (see figure \hyperref[fig:3D Vortex loop_StatDyn]{15}). Wake visualization reveals that in freely moving bodies with $d^*$ = 1.5 and 1.0, the elliptical ring undergoes axis switching but it undergoes early breakdown in stationary cases before axis switching can take place. \par
	
	There are several key differences between the dynamic and stationary bodies. Despite the spring stiffness being the same for both cases, the plates in the freely moving body clap more rapidly compared to the constrained body. The angular velocity of the clapping plate is approximately double in the former case. The circulation values of the starting vortex in the dynamic cases are lower compared to the stationary cases, even though there's a higher efflux of fluid in the dynamic cases due to rapid closure of the interplate cavity. One reason for the lower circulation values in the dynamic cases may be attributed to the rapid forward movement of the body, leading to a lower supply of vorticity to the starting vortex. The second reason is the lower mean fluid velocities , $\overline{u_X}$, ejected from the interplate cavity of fast-moving clapping bodies, see figure \hyperref[fig:Mean_ux_lab_body]{20b}. The circulation remains almost constant with aspect ratio for the dynamic case, whereas it increases with $d^*$ for the stationary case.\par
	
	Using a simple model relating cirulcation with impulse, we calculate the mean thrust coefficient, $\overline{C_T}$, for both dynamic and stationary cases. $\overline{C_T}$ is lower for the dynamic cases and remains nearly constant with $d^*$. It increases with $d^*$ for the stationary cases, which has also been observed by Kim et al.\cite{Kim13}. The increase in circulation and $\overline{C_T}$ with increases in $d^*$ for the stationary cases is due to the finite span effects, which seems to be absent for the dynamic cases. Our results show that there are significant differences, some unexpected, between a clapping body that is free to propel forward and one that is constrained from translating, that will be important when studying animals and designing vehicles based on pulsed jet propulsion.\\
	
	\textbf{Funding.}This work received support from the Department of Science and Technology, India, under the `Fund for Improvement of S\&T' (grant number: FA/DST0-16.009) and the Naval Research Board, India (grant number: NRB/456/19-20) \\
	
	\textbf{Declaration of interests.} The authors report no conflict of interest.\\
	
	\textbf{Author ORCIDs.} \\
	\orcidlink{0000-0001-8497-183X} Suyog Mahulkar \href{https://orcid.org/0000-0001-8497-183X}{https://orcid.org/0000-0001-8497-183X};\\
	\orcidlink{0009-0001-2575-0604} Jaywant Arakeri
	\href{https://orcid.org/0009-0001-2575-0604}{https://orcid.org/0009-0001-2575-0604}.

\end{section}

\bibliographystyle{jfm}
%\bibliography{jfm2esam}

\end{document}